\documentclass[twocolumn]{aastex63}
\usepackage{amsmath}
\usepackage{algorithmic}
\usepackage{fontawesome}
\definecolor{linkcolor}{rgb}{0.1216,0.4667,0.7059}
\definecolor{revcolor}{rgb}{0,0,0}
\newcommand{\rev}[1]{\textcolor{revcolor}{#1}}

\newcommand{\thetaGP}{\boldsymbol{\theta}}
\newcommand{\thetaGPdag}{\boldsymbol{\theta}_n^\dagger}
\newcommand{\phiGP}{\boldsymbol{\phi}}
\newcommand{\Ng}{\mathcal{N}}
\newcommand{\xv}{\boldsymbol{x}}
\newcommand{\zv}{\boldsymbol{z}}
\newcommand{\tvi}{\hat{\boldsymbol{t}}}
\newcommand{\uvi}{\hat{\boldsymbol{u}}}
\newcommand{\av}{\boldsymbol{a}}
\newcommand{\avj}{\hat{\boldsymbol{a}}} 
\newcommand{\avi}{\check{\boldsymbol{a}}} 
\newcommand{\yv}{\boldsymbol{y}}

\newcommand{\psiv}{\boldsymbol{\psi}}
\newcommand{\ThetaM}{\Theta_\mathrm{eq}}
\newcommand{\dv}{\boldsymbol{d}}
\newcommand{\wvj}{\hat{\boldsymbol{w}}}

\newcommand{\qv}{\boldsymbol{q}}
\newcommand{\gvv}{\boldsymbol{g}}
\newcommand{\gvvdag}{\gvv^\dagger_n}
\newcommand{\muv}{\boldsymbol{\mu}}
\newcommand{\ip}{i^\prime}
\newcommand{\jp}{j^\prime}
\newcommand{\Jp}{J^\prime}
\newcommand{\reshape}{\mathrm{reshape}}
\newcommand{\Pspin}{P_\mathrm{spin}}
\received{June 5, 2020}
\revised{July 14, 2020}
\accepted{July 23, 2020}
\shorttitle{Dynamic Mapping of an Exo-Earth}
\shortauthors{Kawahara and Masuda}
\graphicspath{{./}{figures/}}
\begin{document}

\title{Bayesian Dynamic Mapping of an Exo-Earth from Photometric Variability}
 \correspondingauthor{Hajime Kawahara}
 
\author[0000-0003-3309-9134]{Hajime Kawahara}
\email{kawahara@eps.s.u-tokyo.ac.jp}
\affiliation{Department of Earth and Planetary Science, The University of Tokyo, 7-3-1, Hongo, Tokyo, Japan}
\affiliation{Research Center for the Early Universe, 
School of Science, The University of Tokyo, Tokyo 113-0033, Japan}
\author[0000-0003-1298-9699]{Kento Masuda}
\affiliation{Department of Earth and Space Science, Osaka University, Osaka 560-0043, Japan}
 
\begin{abstract}
Photometric variability of a directly imaged exo-Earth conveys spatial information on its surface and can be used to retrieve a two-dimensional geography and axial tilt of the planet (spin-orbit tomography). 
In this study, we relax the assumption of the static geography and present a computationally tractable framework for dynamic spin-orbit tomography applicable to the time-varying geography.
First, a Bayesian framework of static spin-orbit tomography is revisited using analytic expressions of the Bayesian inverse problem with a Gaussian prior.
We then extend this analytic framework to a time-varying one through a Gaussian process in time domain, and present analytic expressions that enable efficient sampling from a full joint posterior distribution of geography, axial tilt, spin rotation period, and hyperparameters in the Gaussian-process priors. 
Consequently, it only takes 0.3~s for a laptop computer to sample one posterior dynamic map \rev{conditioned on the other parameters} with 3,072 pixels and 1,024 time grids, for a total of $\sim 3 \times 10^6$ parameters. 
We applied our dynamic mapping method on a toy model and found that the time-varying geography was accurately retrieved along with the axial-tilt and spin rotation period.
In addition, we demonstrated the use of dynamic spin-orbit tomography with a real multi-color light curve of the Earth as observed by the Deep Space Climate Observatory. We found that the resultant snapshots from the dominant component of a principle component analysis roughly captured the large-scale, seasonal variations of the \rev{clear-sky} and cloudy areas on the Earth. \href{https://github.com/HajimeKawahara/sot}{\color{linkcolor}\faGithub}

\end{abstract}


\section{Introduction}

Direct imaging from space is a key technology to characterize potentially habitable exoplanets for future exploration. 
One important task is to decode the surface of an exo-Earth observed as a pale blue dot.
A multi-band photometric light curve has been studied as a probe of the surface of such a dot \citep{2001Natur.412..885F}. Longitudial mapping along the direction of planetary rotation has also been explored \citep{2009ApJ...700..915C, 2009ApJ...700.1428O, 2010ApJ...715..866F}. 
Two-dimensional mapping proposed by \cite{2010ApJ...720.1333K}, called spin-orbit tomography, utilizes the fact that the star illuminates the planet surface from various different directions as the planet spins as well as revolves around the star.

Thus far, the spin-orbit tomography has been developed from a variety of perspectives \citep{2011ApJ...739L..62K,2012ApJ...755..101F,2016MNRAS.457..926S,2018AJ....156..146F, 2019AJ....158..246B, 2020arXiv200403941A}. 
\cite{2018AJ....156..146F} constructed a Bayesian mapping framework using a Gaussian process for the spatial regularization, and obtained posterior samples for the map and axial-tilt using a Markov chain Monte Carlo (MCMC) method. 
In \cite{2020arXiv200403941A}, a sparse modeling technique was introduced to obtain a clearer map. 
Nowadays such retrieval methods can be applied to real light curves of the Earth monitored from space \citep{2018AJ....156...26J}, as has been demonstrated using a single-band light curve from the Transiting Exoplanet Survey Satellite \citep[TESS;][]{2019arXiv190312182L} and a multi-band light curve from the Deep Space Climate Observatory \citep[DSCOVR;][]{2019ApJ...882L...1F}. Spin-orbit tomography and spectral unmixing were unified into a single retrieval method and tested using DSCOVR data \citep{Kawahara2020}.

Most of these previous studies assumed a static geography during an observation period, however this assumption does not always work. Indeed, \citet{Kawahara2020} found that a static model fails to retrieve the map of a flat spectrum component in the DSCOVR light curve, which corresponds to clouds.
In \cite{2019arXiv190312182L}, the authors first attempted a time-varying mapping using a single-band light curve of Earthshine as observed by TESS. They modeled time evolution of the surface map by injecting an orthogonal polynomial basis into the coefficients of spherical harmonic expansion of the planetary sphere map.

In this study, we develope a dynamic mapping technique for time-varying geography, leveraging the high flexibility of a Gaussian process \citep{rasmussen2003gaussian}.
The high flexibility of the Gaussian process originates from the kernel trick \citep{bishop2006pattern}: The Gaussian process models the covariance as a function of the time interval, instead of tracking time evolution using a parametrized model.
Our technique is an extention of the Bayesian mapping framework developed by \cite{2018AJ....156..146F}, who adopted a Gaussian process for spatial regulaization: we do so both in spatial and time directions.
The main difficulty in such a full modeling of the time-varying geography is an extremely large number of free parameters. There are a million fitting parameters if we adopt $10^3$ grids for the geography and $10^3$ data points for a time evolution. 
In our framework, we overcome this difficulty by utlizing analytic solutions of the Bayesian inverse problem with a (multivariate) Gaussian prior and their isomorphic representations. This significantly reduces the computational complexity and makes it possible to sample from the full posterior distribution even for a large number of spatial grids and data points.

The remainder of this paper is organized as follows. In Section 2, we first describe a Bayesian framework of static spin-orbit tomography and then develop its extension to time-varying geography.
A test of our method using a toy time-varying map of the cloudless Earth is described in Section 3,
and the method is applied to the real data collected by DSCOVR in Section 4. We summarize our findings and discuss remaining issues in Section 5. 
Derivations of the mathematical formulae used in the main body
are provided in the Appendix. 
Our code written in Python 3 is publicly available on GitHub \href{https://github.com/HajimeKawahara/sot}{\color{linkcolor}\faGithub}.

\section{Dynamic Spin-Orbit Tomography in a Bayesian Framework}

\subsection{Light curve of Planet's Reflection }
We start from the theoretical modeling of the reflected light curve. 
The flux of the reflected (or scattered) light from a planet can be approximated by the surface integral of the bidirectional reflection distribution function (BRDF) over the illuminated and visible (IV) area on the planet as follows: 
\begin{eqnarray}
\label{eq:fp}
f_p(t) = \frac{f_\star R_p^2}{\pi a_\mathrm{sp}^2} \int_{\mathrm{IV}} d \Omega  R^s(\vartheta_0,\vartheta_1,\varphi, t)  \cos{\vartheta_0} \cos{\vartheta_1},
\end{eqnarray}
where $f_\star$ is the flux of the host star, $R_p$ is the planet radius, $a_\mathrm{sp}$ is the star-planet separation (we assume a circular orbit in this study), and $d \Omega$ is the differential solid angle on the planetary sphere. The BRDF $R^s(\vartheta_0,\vartheta_1,\varphi, t)$ is a function of the solar zenith angle $\vartheta_0$,  the zenith angle between the line-of-sight direction and the surface normal vector ${\vartheta_1}$, and the relative azimuth angle between the directions toward an observer and the host star $\varphi$. In general, the BRDF is also time dependent. The derivation of Equation (\ref{eq:fp}) is given in Appendix \ref{ss:matriidentity} in \cite{Kawahara2020}.

Here we assume that the BRDF is isotropic, that is, $R^s(\vartheta_0,\vartheta_1,\varphi, t) = a(t,\Omega)$, where $a(t,\Omega)$ is the surface albedo at the spherical coordinate of $\Omega$ fixed on the planet surface. For the isotropic assumptions, we can rewrite Equation (\ref{eq:fp}) as follows:
\begin{eqnarray}
\label{eq:fredfirstjyanai}
f_p (t) = \int d \Omega  \, W (t,\Omega; \gvv) \, a(t, \Omega),
\end{eqnarray}
where we define the geometric kernel by
\begin{eqnarray}
\label{eq:weight}
W (t,\Omega; \gvv) = 
  \left\{
    \begin{array}{l}
\displaystyle{\frac{f_\star R_p^2}{\pi a_\mathrm{sp}^2}  \cos{\vartheta_0} \cos{\vartheta_1} \mbox{ for $\cos{\vartheta_0}, \cos{\vartheta_1}>0$}}\\
\\
\displaystyle{ 0 \mbox{\,\, otherwise.} }
    \end{array}
  \right.
\end{eqnarray}
The geometric kernel depends on the spin parameter $\gvv = (\zeta, \ThetaM, \Pspin)$, where $\zeta$ is the axial tilt (obliquity), $\ThetaM$ is the orbital phase at equinox, and $\Pspin$ is the spin rotation period.

For the above explanation, we took a single-band photometric light curve $f_p(t)$ as the time-series data. Depending on the feature to be mapped, other choices for the time-series data are possible.
For example, \cite{2011ApJ...739L..62K} used colors defined as the flux difference
between 0.85 and 0.45 microns, or 0.85 and 0.65 microns, to eliminate a flat component from the clouds and to retrieve the feature of the continents and ocean. In addition, in \cite{2019ApJ...882L...1F}, the second component of a principle component analysis is applied to extract the continental features from the multicolor light curve of DSCOVR. 


\subsection{Bayesian Formulation of the Static Spin-Orbit Tomography}\label{ss:bayesstatic}

Our dynamic mapping technique is based on Bayesian statistics. Before discussing the dynamic framework,
we first revisit the original static
spin-orbit tomography from a Bayesian perspective. 
The main difference between our formulations and those in \cite{2018AJ....156..146F} is that we take full advantage of analytic expressions for the posterior probability distribution of the map before resorting to a numerical sampling of the distribution.
This approach helps to significantly reduce the computational complexity of the modeling of the dynamic spin-orbit tomography, as described in Section \ref{ss:dysot}, which would otherwise be computationally almost intractable in practice.

The original spin-orbit tomography places the static assumption of the BRDF, $a(t,\Omega) = a(\Omega)$, in addition to the isotropic assumption. The static assumption converts Equation (\ref{eq:fredfirstjyanai}) into the form of a Fredholm integral equation of the first kind as follows:
\begin{eqnarray}
\label{eq:fredfirst}
f_p (t) = \int d \Omega  \, W (t,\Omega; \gvv) \, a(\Omega).
\end{eqnarray}
If we assume that the spin parameters are fixed, a discritization of Equation (\ref {eq:fredfirst}) yields a linear inverse problem for the geography $\av$ in the following manner:
\begin{eqnarray}
\label{eq:sot}
\dv = W \av, 
\end{eqnarray}
where $d_i = f_p (t_i)$  for $i=0,1,...,N_i-1$, $W_{ij} =  W (t_i,\Omega_j; \gvv)$, and $a_j = a(\Omega_j)$ for $j=0,1,...,N_j-1$ is the geography vector representing the surface map. 
 
The original spin-orbit tomography solves Equation (\ref{eq:sot}) in terms of the geography vector $\av$ given $\dv$. We call the retrieval technique of the form of Equation (\ref{eq:sot}) ``static spin-orbit tomography'' in contrast to the dynamic mapping technique developed in this study. When we regard the spin parameters $\gvv$ as free parameters, the problem becomes nonlinear. Note that the spin parameters can be optimized even using a brute force search \citep{2010ApJ...720.1333K}. Alternatively, 
the spin parameters can be inferred from the frequency modulation of the light curve before solving for the geography \citep{2016ApJ...822..112K,2020arXiv200611437N}.
In this study, we infer the spin parameters simultaneously with the geography using
a Bayesian approach \citep{2018AJ....156..146F}\footnote{Note that the previous studies take $\zeta$ and $\Theta_\mathrm{eq}$ as free parameters and fix the rotation period as the input value \citep[e.g.][]{2010ApJ...720.1333K,2018AJ....156..146F}. However, we regard the spin rotation period as a free parameter too in this study. }.
 
The prior of the geography $\av$ plays a key role in stabilizing the map. In this paper, we adopt a multivariate Gaussian for the prior:
\begin{eqnarray}
 p(\av|\thetaGP) = \Ng (\av| 0 ,\Sigma_{\av}),
\end{eqnarray}
where we define the multivariate normal distribution of the stochastic variable $\xv$ with the mean of $\muv$ and the covariance of $\Sigma$ by
\begin{eqnarray}
\Ng (\xv|\muv,\Sigma) = \frac{1}{(2 \pi)^{N/2} (\det{\Sigma})^{1/2}} e^{ - \frac{1}{2} (\xv - \muv)^T \Sigma^{-1} (\xv - \muv)  }.
\end{eqnarray}
The covariance of the model prior $\Sigma_{\av}$ is modeled using the spatial kernel as a function of the hyperparameter $\thetaGP$:
\begin{eqnarray}
 \Sigma_{\av} = K_S (\thetaGP).
\end{eqnarray}
In \cite{2018AJ....156..146F}, the authors proposed a multivariate Gaussian as the spatial kernel, and used a spherical harmonic expansion as the basis of the geography. 
Here, we continue to apply a pixel-based expression. The kernel for the Gaussian process (GP kernel) is then expressed as follows:
\begin{eqnarray}
\label{eq:Kkernels}
 (K_S)_{jj^\prime} = \alpha k(\Omega_j, \Omega_{j^\prime};\gamma),
\end{eqnarray}
where $\Omega_j$ is the spherical coordinate of the $j$-th pixel and $\gamma$ is the spatial correlation scale; in addition, $\alpha$ should be interpreted as the amplitude of the covariance the model prior. In this model, the hyperparameter of the spatial kernel is $\thetaGP = (\alpha, \gamma)^T$. 

In practice, we assume that $k(\Omega_j, \Omega_{j^\prime};\gamma)$ is a function of the angular separation between the $j$ and $j^\prime$-th pixels,
\begin{eqnarray}
 (K_S)_{jj^\prime} = \alpha k(\eta_{jj^\prime};\gamma),
\end{eqnarray}
where $k(\eta;\gamma)$ is the kernel function and $\eta_{jj^\prime}$ is the angular separation between the $j$-th and $j^\prime$ pixels. As the kernel function, for instance, one might use a radial-basis function (RBF) kernel,
\begin{eqnarray}
 k_\mathrm{RBF}(\eta;\gamma) = \exp{\left(- \frac{\eta^2}{2 \gamma^2} \right)},
\end{eqnarray}
or the Mat\'{e}rn -3/2 kernel,
\begin{eqnarray}
\label{eq:MaternA}
 k_\mathrm{M3/2}(\eta;\gamma) = \left( 1 + \frac{\sqrt{3} \eta}{\gamma} \right) e^{- \sqrt{3} \eta/\gamma}. 
\end{eqnarray}
Note that the L2 regularization corresponds to the extreme case of $\gamma \to 0$ in Equation (\ref{eq:Kkernels}). 

We assume that the observational noise is also described by a correlated Gaussian with the covariance matrix $\Sigma_{\dv}$.
The likelihood is therefore given by the following:
\begin{eqnarray}
 p(\dv|\av, \gvv) = \Ng (\dv| W \av ,\Sigma_{\dv} ).
\end{eqnarray}
In this section, we assume that we know the data covariance $\Sigma_{\dv}$ for simplicity although it is possible to infer the data covariance as well.

The principle of the Bayesian inference is summarized as follows. We consider the joint posterior distribution of all model (hyper)parameters $\av$, $\gvv$, and $\thetaGP$, which is expressed as $ p(\av, \gvv, \thetaGP |\dv)$. The Bayesian inference of the model parameter is achieved by marginalizing the joint posterior $ p(\av, \gvv, \thetaGP |\dv)$ for the target parameter. For instance, to infer the geography, we compute $p(\av|\dv)$ by marginalizing the joint probability of $p(\av, \gvv, \thetaGP |\dv)$ over $\gvv$ and $\thetaGP$.

As a notable feature of static spin-orbit tomography, Equation (\ref{eq:sot}) becomes a linear inverse problem when fixing the spin parameters $\gvv$. As descrived above, we assumed a multivariate normal distribution for both the likelihood and the prior distributions. 
In this case, the problem can be described as a Bayesian linear inverse problem with Gaussian priors (Appendix \ref{ap:bayes}). In this framework,
the posterior distribution of $\av$ given $\thetaGP$ and $\gvv$ is also a Gaussian and can be analytically expressed as follows:
\begin{eqnarray}
\label{eq:posteriorbayes}
p(\av|\dv,\thetaGP, {\gvv}) &=& \Ng(\av| \muv_{\av|\dv,\thetaGP, \gvv},\Sigma_{\av|\dv,\thetaGP, \gvv}) \\
\muv_{\av|\dv,\thetaGP,\gvv} &=& ( W^T \Pi_{\dv} W + \Pi_{\av} )^{-1} W^T \Pi_{\dv} \dv \nonumber \\
\label{eq:mean_static}
&=& ( W^T \Pi_{\dv} W + K_S^{-1} )^{-1} W^T \Pi_{\dv} \dv \\
\Sigma_{\av|\dv,\thetaGP, \gvv} &=& (W^T \Pi_{\dv} W + \Pi_{\av})^{-1} \nonumber \\
&=& (W^T \Pi_{\dv} W + K_S^{-1})^{-1},
\end{eqnarray}
where we define the precision matrices $\Pi_{\dv} = \Sigma_{\dv}^{-1}$ and $\Pi_{\av} = \Sigma_{\av}^{-1} = K_S^{-1}$. Note that $\muv_a$ is identical to the 
maximum a posteriori (MAP) 
solution given $\thetaGP$ and $\gvv$, which was 
adopted for the point estimate of the map by \cite{2011ApJ...739L..62K}. 

In addition, the marginal likelihood (known as ``evidence'') for 
$\gvv$ and $\thetaGP$ can be expressed in the following manner:
\begin{eqnarray}
p(\dv|\thetaGP, \gvv) &=&  \frac{p(\dv|\av,\thetaGP, {\gvv}) p(\av|\thetaGP, {\gvv})}{p(\av|\dv,\thetaGP, {\gvv})} \\
&=& \frac{p(\dv|\av, {\gvv}) p(\av|\thetaGP)}{p(\av|\dv,\thetaGP, {\gvv})} \\
\label{eq:anamarli0}
&=& \Ng(\dv|{\bf 0}, \Sigma_{\dv} + W \Sigma_{\av} W^T) \\
\label{eq:anamarli}
&=& \Ng(\dv|{\bf 0}, \Sigma_{\dv} + K_W ),
\end{eqnarray}
where we define the weighted spatial kernel by
\begin{eqnarray}
K_W = K_W (\thetaGP,\gvv) \equiv  W K_S W^T,
\end{eqnarray}
the derivation of which is provided in Appendix \ref{ap:bayes}. Once we have the analytic expression of Equation (\ref{eq:anamarli}), 
we can sample the sets of ($\thetaGP$, $\gvv$) from the marginal posterior distribution
\begin{eqnarray}
 p(\thetaGP, \gvv|\dv) \propto p(\dv|\thetaGP, \gvv) p(\thetaGP) p(\gvv)
\end{eqnarray}
using, for example, an MCMC algorithm as 
 \begin{eqnarray}
 \label{eq:sampleg_axialtilt}
\thetaGPdag, \gvvdag \sim p(\thetaGP,\gvv|\dv),
\end{eqnarray}
where $\thetaGPdag$ and $ \gvvdag$ are the $n$-th sample of the hyperparameter and the spin parameters, respectively.  
We emphasize that this sampling using the marginal likelihood is possible without inferring the geography, which significantly reduces the number of dimensions of the MCMC sampling.
We adopt the prior of the spin parameters as 
$p(\zeta) p(\ThetaM) \propto \sin{\zeta}$, 
the uniform prior for $\Pspin$, the log-uniform hyperprior for $\gamma$ and $\alpha$.

Using the sampling of Equation (\ref{eq:sampleg_axialtilt}), we can also infer the marginal posterior of the geography as
\begin{eqnarray}
p(\av|\dv) &=& \int d \thetaGP \int d \gvv \, p(\av, \thetaGP, \gvv|\dv) \\
&=& \int d \thetaGP \int d \gvv \, p(\av|\dv,\thetaGP,{\gvv}) p(\thetaGP, \gvv|\dv)  \\
\label{eq:papx}
&\approx& \frac{1}{N_s} \sum_{n=0}^{N_s-1} p(\av|\dv,\thetaGPdag, {\gvvdag}),
\end{eqnarray}
where $N_s$ is the number of the samples.

Using Equation (\ref{eq:papx}), we can compute the expectation of any statistical variable $f(\av)$ under the probability $p(\av|\dv)$ as follows:
\begin{eqnarray}
&\,&\langle f(\av) \rangle \approx \frac{1}{N_s} \sum_{n=0}^{N_s-1} \langle f(\av) \rangle_{ p(\av|\dv,\thetaGPdag, {\gvvdag}) } \\
&=& \frac{1}{N_s} \sum_{n=0}^{N_s-1} \int d \av f(\av) p(\av|\dv,\thetaGPdag, {\gvvdag}) \\
&=& \frac{1}{N_s} \sum_{n=0}^{N_s-1} \int d \av f(\av) \Ng(\av| \muv_{\av|\dv,\thetaGPdag,\gvvdag},\Sigma_{\av|\dv,\thetaGPdag, \gvvdag}) \nonumber \\
\end{eqnarray}
where $\langle f \rangle_P$ is an expectation of a statistical variable $f$ under the probability $P$. Herein, we omit the subscript when $P = p(\av|\dv)$, that is, $\langle \cdot \rangle \equiv \langle \cdot \rangle_{p(\av|\dv)}$. For instance, we can compute the mean of the marginal posterior for the geography using
\begin{eqnarray}
\muv_{\av|\dv} &=& \langle \av \rangle \approx \frac{1}{N_s} \sum_{n=0}^{N_s-1} \langle \av \rangle_{ p(\av|\dv,\thetaGPdag, {\gvvdag}) } \\
&=& \frac{1}{N_s} \sum_{n=0}^{N_s-1} \muv_{\av|\dv,\thetaGPdag, \gvvdag}.
\end{eqnarray}
 The computation of $\muv_{\av|\dv,\thetaGPdag, \gvvdag}$ from Equation (\ref{eq:mean_static}) requires solving the inverse matrices twice. Using the Woodbury matrix identity (Eq. \ref{eq:Woodbury}), we obtain
\begin{eqnarray}
\label{eq:kailath}
&\,& ({W}^T \Pi_{\dv} W + {\Pi_{\av}})^{-1}\nonumber \\ 
\label{eq:wood_ex}
&\,& = \Sigma_{\av} -  \Sigma_{\av} W^T ( \Sigma_{\dv} + W  \Sigma_{\av} W^T)^{-1}  W \Sigma_{\av} \\
\label{eq:woodburied_conversion}
&\,& = K_S -  K_S W^T ( \Sigma_{\dv} + K_W )^{-1}  W K_S.
\end{eqnarray}
We can then reduce Equation  (\ref{eq:mean_static}) to 
\begin{eqnarray}
\muv_{\av|\dv,\thetaGP,\gvv} &=& K_S W^T [ I -  (\Sigma_{\dv} +  K_W )^{-1} K_W  ] \Pi_{\dv} \dv \nonumber \\
\label{eq:analytic_static_}
 &=& K_S W^T (I + \Pi_{\dv} K_W)^{-1} \Pi_{\dv} \dv,
\end{eqnarray}
where we used the matrix identity of Equation (\ref{eq:IWoodbury}) with $U=I$, $V=K_W$, and $Y=\Sigma_{\dv}^{-1}=\Pi_{\dv}$ for the derivation of the second line. Equation (\ref{eq:analytic_static_}) can be computed by the solver of the linear equation 
\begin{eqnarray}
\Pi_{\dv} \dv = (I + \Pi_{\dv} K_W) \yv,
\end{eqnarray}
 for $\yv$ using the Cholesky decomposition, which is more stable than the direct computation of the inverse matrix. Finally, we obtain the mean map of the marginal posterior of $\av$ as follows:
\begin{align}
&\boxed{ \text{\bf Mean Map for Static Geography}  }\nonumber \\
&\langle \av \rangle = \frac{1}{N_s} \sum_{n=0}^{N_s-1}  K_S (\thetaGPdag) W(\gvvdag)^T \yv_n \\
&\yv_n = (I + \Pi_{\dv} K_W(\thetaGPdag,\gvvdag))^{-1} \Pi_{\dv} \dv \nonumber
\end{align}

In short, unlike the approach by \cite{2018AJ....156..146F}, our method avoids the direct sampling of the geography in an MCMC. This feature is even more powerful in a dynamic spin-orbit tomography where the number of parameters is much larger than in a static tomography.

\subsection{Dynamic Spin-Orbit Tomography  \label{ss:dysot}}

\begin{figure}[]
 \begin{center}
   \includegraphics[width=1.0\linewidth]{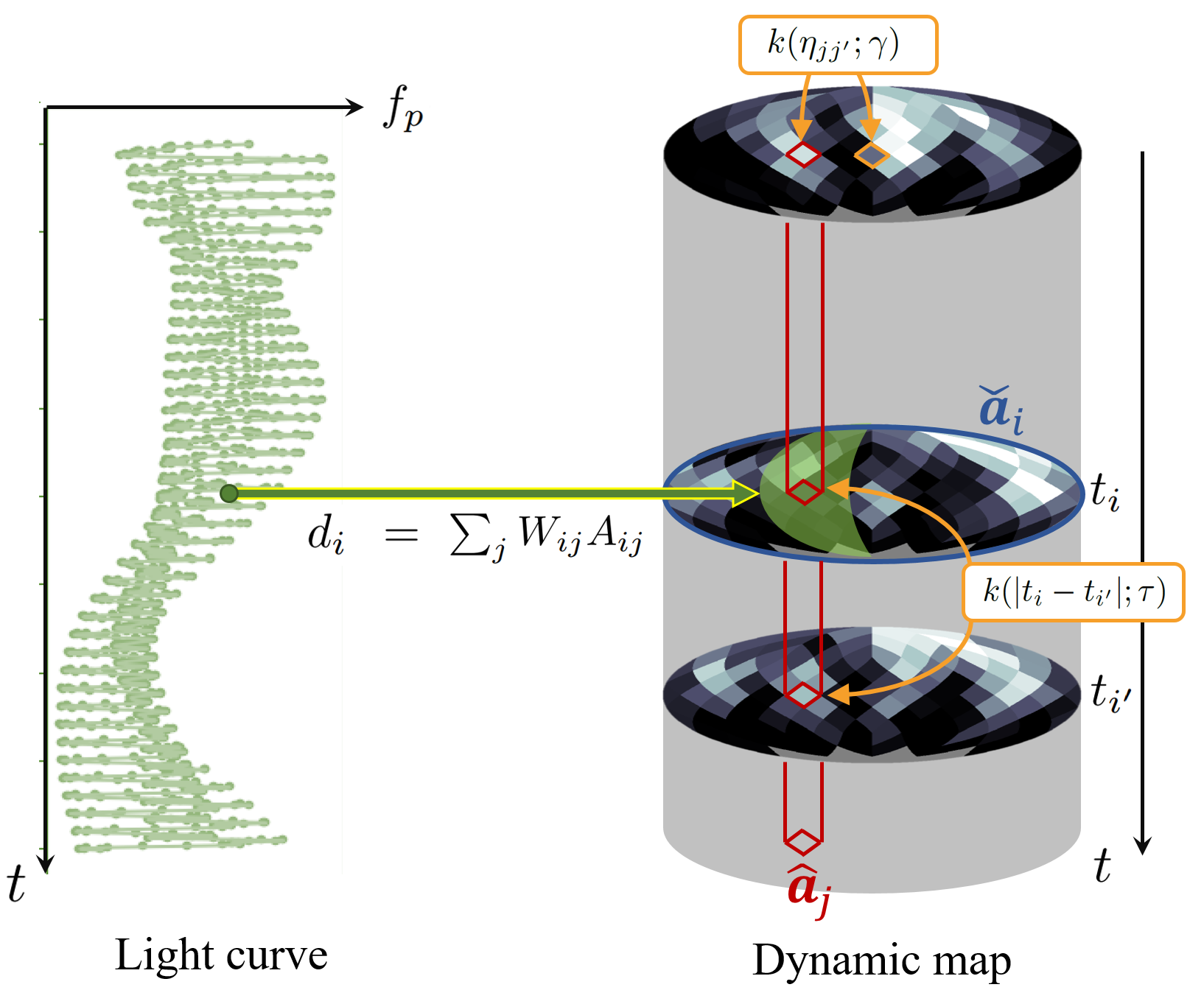}
 \end{center}
 \caption{Schematic of dynamic spin-orbit tomography. The data point at time $t_i$ conveys the information in the IV region (colored in green) of the dynamic map at the epoch of $t_i$. The different epochs in the $j$-th pixel are stochastically connected by $k(|t_i-t_{i^\prime}|;\tau)$ in the temporal GP kernel with a correlation timescale of $\tau$. The spatial distribution of the map is stabilized by the spatial GP kernel $k(\eta_{jj^\prime};\gamma)$ with an angular correlation scale of $\gamma$. \label{fig:scheme}}
\end{figure}

Extending the Bayesian framework for static mapping to time-varying mapping, we construct dynamic spin-orbit tomography. The structure of the dynamic spin-orbit tomography is schematically summarized in Figure \ref{fig:scheme}. We infer the time-varying geography, called a ``dynamic map'' (right), from the light curve (left). The data point $d_i = d(t_i)$ at time $t_i$ conveys the geographic information in the IV area at time $t_i$. This information is transmitted pixel by pixel along the time axis through the Gaussian process. In addition, the geography is stabilized by the spatial kernel against an overfitting. Below, we formulate the concept shown in Figure \ref{fig:scheme} in a step-by-step manner.

%
%
Dynamic spin-orbit tomography does not assume a static geography. Therefore, we start from the discretization of Equation (\ref{eq:fredfirstjyanai}) using the time-varying geography matrix
\begin{eqnarray}
A = (A_{ij}),
\end{eqnarray}
where $A_{ij}$ is the map value of the $j$-th pixel at time $t=t_i$. We then obtain the discretized version of Equation (\ref{eq:fredfirstjyanai}) as
\begin{eqnarray}
\label{eq:modelgplip}
\dv = \psiv (W,A)
\end{eqnarray}
where $\psiv = \psiv (W,A)$ is an operator indicating $\psi_i = \sum_j W_{ij} A_{ij}$. In other words, $\psiv$ extracts the diagonal components of a matrix $W A^T$ as a vector. 

Equation (\ref{eq:modelgplip}) is not a regular form of a linear inverse problem. However, by expanding the geometric kernel from $ \mathbb{R}^{N_i \times N_j} $ to $ \mathbb{R}^{N_i \times N_i N_j}$ and by using the isomorphism of $A$, one can convert it into a linear inverse problem. First, we define the expanded form of the geometric kernel as
\begin{eqnarray}
\tilde{W} &=& \left(
    \begin{array}{cccc}
      \mathcal{D}{(\wvj_0)} & \mathcal{D}{(\wvj_1)} & \cdots & \mathcal{D}{(\wvj_{N_j-1})} \\
    \end{array}
  \right) 
  \in \mathbb{R}^{N_i \times N_i N_j},  \nonumber \\
\end{eqnarray} 
where $\wvj_j$ is the column vector of the column of $W$\footnote{In this paper, we identify the column vector of the {\it column} of a matrix by a hat symbol ($\hat{\cdot}$) and the column vector of the {\it row} of a matrix by a check ($\check{\cdot}$). }, and $\mathcal{D}{(\wvj_j)}$ is an operator that creates a diagonal matrix from a vector $\wvj_j$, that is, $\acute{W} = \mathcal{D}{(\wvj_j)}$ indicates that $\acute{W}_{ij} = \delta_{ij} \wvj_j $ ($\delta_{ij}$ is the Kronecker delta).  We also define the vector isomorphic to $A$ using the following:
\begin{eqnarray}
{\av} &=& \mathrm{vec}(A) \equiv \left(
    \begin{array}{c}
      \avj_0 \\
      \avj_1 \\
      \vdots \\
      \avj_{N_j-1} 
    \end{array}
  \right) \in \mathbb{R}^{N_i N_j},
\end{eqnarray}
 where $\mathrm{vec}(A) $ indicates the vectorization of $A$ and $\avj_j$ is the time-series vector of the $j$-th pixel (the $j$-th column of $A$) as
\begin{eqnarray}
\avj_j \equiv (A_{0j},A_{1j},\hdots,A_{(N_i-1)j})^T.
\end{eqnarray}

Using $\tilde{W}$ and $\av$, one can rewrite Equation (\ref{eq:modelgplip}) by the linear inverse problem,
\begin{eqnarray}
\label{eq:linproblem}
\dv = \tilde{W} \av.
\end{eqnarray}

The dynamic spin-orbit tomography introduces both time and spatial correlations of the geography in a pixel-by-pixel manner. We assume a multivariate normal distribution
as the model prior of the geography as follows:
\begin{eqnarray}
\label{eq:prior_dsot}
p(\av|\thetaGP) = \Ng(\av|{\bf 0}, \Sigma_{\av}),
\end{eqnarray}
where $ \Sigma_{\av}$ is the model covariance of $\av$. We model $ \Sigma_{\av}$ using the grand kernel whose element between the $j$-th pixel at time $t_i$ and the $j^\prime$-th pixel at $t_i^\prime$ is given by
\begin{eqnarray}
\label{eq:grank}
K_{i j i^\prime j^\prime} = \alpha k(\eta_{jj^\prime}; \gamma) k(|t_i - t_{i^\prime}|; \tau), 
\end{eqnarray}
where $\alpha$ is the amplitude of the grand kernel, $\gamma$ is the spatial correlation scale and $\tau$ is the temporal correlation scale. 
 From Equation (\ref{eq:grank}), the grand kernel can be expressed as follows:
\begin{eqnarray}
\label{eq:Kkroneck}
K = \alpha K_S \otimes K_T, 
\end{eqnarray}
where $\otimes$ is the Kronecker product and $K_S$ and $K_T$ are the spatial and temporal kernels defined as follows:
 \begin{eqnarray}
\label{eq:grank2}
(K_S)_{j j^\prime} &=& k(\eta_{jj^\prime}; \gamma)  \\
(K_T)_{i i^\prime} &=& k(|t_i - t_{i^\prime}|; \tau).
\end{eqnarray}
In this model, the hyperparameter vector of the grand kernel becomes $\thetaGP = (\gamma,\alpha,\tau)^T$. 

Given the model of Equation (\ref{eq:linproblem}), the likelihood function is expressed as
\begin{eqnarray}
\label{eq:likelihood_dsot}
p(\dv|\av, \gvv) = \Ng (\dv|\tilde{W} \av ,\Sigma_{\dv}),
\end{eqnarray}
where $\Sigma_{\dv}$ is the data covariance. We still assume that we know the data covariance, although one can also model the data covariance using the GP kernel and so on.

From the framework of the Bayesian inverse problem (Appendix \ref{ap:bayes}), the 
posterior distribution conditioned on
$\dv, \thetaGP, \gvv$ is 
\begin{eqnarray}
\label{eq:posteriorgiven}
p(\av|\dv,\thetaGP,\gvv) &=& \Ng(\av|\muv_{\av|\dv,\thetaGP,\gvv},\Sigma_{\av|\dv,\thetaGP,\gvv}) \\
\label{eq:analytic_atami}
\muv_{\av|\dv,\thetaGP,\gvv} &=& (\tilde{W}^T \Pi_{\dv} \tilde{W} + {K^{-1}})^{-1} \tilde{W}^T \Pi_{\dv} \dv \\
\label{eq:cov}
\Sigma_{\av|\dv,\thetaGP,\gvv} &=&  (\tilde{W}^T \Pi_{\dv} \tilde{W} + {K^{-1}})^{-1}.
\end{eqnarray}

There remain two computational difficulties to implementing Equation (\ref{eq:analytic_atami}). One is the computational complexity because the operation of the inverse matrix in Equation (\ref{eq:analytic_atami}), 
\begin{eqnarray}
\label{eq:analytic_atami_inv}
 (\tilde{W}^T \Pi_{\dv} \tilde{W} + K^{-1})^{-1} \in \mathbb{R}^{N_i N_j \times N_i N_j},
\end{eqnarray}
requires a cost of $\mathcal{O} (N_i^3 N_j^3)$. The other is the memory size. The matrices in Equation (\ref{eq:analytic_atami}) requires the memory allocation of $\mathcal{O} (N_i^2 N_j^2)$, which corresponds to 
tens of terabytes for $N_i=10^3$ and $N_j=10^3$. To mitigate these difficulties, 
we reduce the dimensions of the inverse matrix and obtain compact forms of Equations (\ref{eq:analytic_atami}) and (\ref{eq:cov}) by reshaping
the vectors and recontracting the dimension of the matrices.  

First, from the same derivation as Equation (\ref{eq:woodburied_conversion}) using the Woodbury matrix identity (Eq. \ref{eq:Woodbury}), we obtain
\begin{eqnarray}
\label{eq:kailath}
&\,& \Sigma_{\av|\dv,\thetaGP,\gvv} = (\tilde{W}^T \Pi_{\dv} \tilde{W} + {K^{-1}})^{-1}\nonumber \\ 
\label{eq:kailath_expand}
&\,& = K -  K \tilde{W}^T ( \Sigma_{\dv} + K_W )^{-1}  \tilde{W} K
\end{eqnarray}
 The weighted kernel $K_W \in \mathbb{R}^{N_i \times N_i}$ in the inverse matrix is defined by
\begin{eqnarray}
\label{eq:Kred}
K_W &\equiv& \tilde{W} K \tilde{W}^T =  \alpha \tilde{W} (K_S \otimes K_T) \tilde{W}^T \\
\label{eq:wkerneldy}
&=&  \alpha K_T \odot (W K_S W^T),
\end{eqnarray}
where $\odot$ is the element-wise product (the Hadamard product). The derivation is given in Appendx \ref{ap:recont1}.
The computational cost of the last term of Equation (\ref{eq:kailath}) is $\mathcal{O} (N_i^3)$ because it only involves an inversion of $( \Sigma_{\dv} + K_W) \in \mathbb{R}^{N_i \times N_i}$.\footnote{Note that $K_W$ is not a Toeplitz matrix even for a equal grid spacing of time. Therefore, the Toeplitz method, which is often used to reduce the computational complexity of an inverse matrix of a Gaussian process regression, cannot be applied.}

Second, we reshape Equation (\ref{eq:analytic_atami}) into a compact form by re-contracting the dimensions of the matrices. 
In the same way as derived in Equation (\ref{eq:analytic_static_}), Equation (\ref{eq:analytic_atami}) reduces to
\begin{eqnarray}
\label{eq:analytic_Kailath_fast}
\muv_{\av|\dv,\thetaGP,\gvv} &=& 
 \alpha (K_S \otimes K_T) \tilde{W}^T (I + \Pi_{\dv} K_W)^{-1} \Pi_{\dv} \dv.
\end{eqnarray}
The reshaping of Equation (\ref{eq:analytic_Kailath_fast}) with the re-contraction formula in Appendix \ref{ap:recont2} provides us with a compact form of the mean of the posterior distribution of $A$ given $\gvv$ and $\thetaGP$ (i.e. $\muv_{\av|\dv,\thetaGP,\gvv} = \mathrm{vec}(A^\ast)$), which is summarized below.
\begin{align}
&A^{\ast} = \alpha K_T \mathcal{D} (\yv) W K_S \label{eq:analytic_dysot_map}  \\
&\yv \equiv  (I + \Pi_{\dv} K_W)^{-1} \Pi_{\dv} \dv \nonumber\\
&K_W \equiv  \alpha K_T \odot (W K_S W^T) \nonumber, 
\end{align}
Note that the element of $\mathcal{D} (\yv) W$ is expressed as follows:
\begin{eqnarray}
\label{eq:proje}
(\mathcal{D} (\yv) W )_{ij} = W_{ij} y_i.
\end{eqnarray}
 
The implementation of Equation (\ref{eq:analytic_dysot_map}) requires the computational complexity of $\mathcal{O} (N_i^3)$ to solve $\yv$ and the memory size of $\mathcal{O} (N_i^2)$ or $\mathcal{O} (N_i N_j)$ for the allocation of $K$ or $W$. For $N_i=1024$ and $N_j=3072$, i.e.,  $\sim 3 \times 10^6$ parameters in total, it only took $\sim$ 0.3 s to compute Equation (\ref{eq:analytic_dysot_map}) using a 2.6-GHz Intel Core i7-9750H CPU.

Third, we obtain the compact and tractable form of the posterior distribution of Equation (\ref{eq:posteriorgiven}).
To do so, we consider the \rev{marginal} posterior of each snapshot of the geography. Defining the geography vector at time $t_i$ (the $i$-th row of $A$) by
\begin{eqnarray}
\avi_i \equiv (A_{i0},A_{i1},\hdots,A_{i(N_j-1)})^T,
\end{eqnarray}
we can \rev{derive the marginal posterior of $\avi_i$ by extracting a submatrix of the covariance matrix\footnote{ \rev{ Suppose that the distribution of $\av$ is given by the multivariate Gaussian $\Ng(\av|\muv,\Sigma)$. The marginal distribution of a subvector $\av^{(L)} = ( a_k )_{k \in L} $ is given by $\Ng(\av^{(L)}|\muv^{(L)},\Sigma^{(L)})$, where  $L$ is a subset of indices $L \subset (0,1,\cdots,N-1)$, $\muv^{(L)} = (\mu_k)_{k \in L} $, and $\Sigma^{(L)} = ( \Sigma_{kl} )_{k \in L, l \in L}$  \citep[\S 2.3 in ][]{bishop2006pattern}. }}} from Equation (\ref{eq:kailath_expand}) as
\begin{align}
&\boxed{ \text{\bf  Snapshot Given $\gvv$ and $\thetaGP$}  }\nonumber \\
\label{eq:dysot_posterior_map} 
&p(\avi_i|\dv, \thetaGP, \gvv) = \Ng(\avi_i|\avi_i^\ast,\Sigma_{\avi_i|\dv, \thetaGP, \gvv}) 
\end{align}
where
\begin{eqnarray}
\avi_i^\ast \equiv (A_{i0}^\ast,A_{i1}^\ast,\hdots,A_{i(N_j-1)}^\ast)^T,
\end{eqnarray}
 is the snapshot of $A^\ast$ at time of $t_i$, and \rev{the submatrix of the covariance (the snapshot covariance)}
 \begin{align}
\label{eq:dysot_posterior_map_cov} 
&{\Sigma}_{\avi_i|\dv, \thetaGP, \gvv}= \alpha (K_T)_{ii} K_S - B_i^T (\Sigma_{\dv} + K_W)^{-1} B_i, \\
&{B}_i = \alpha \mathcal{D}(\tvi_i) W K_S \\ 
&{\tvi}_i \equiv ((K_T)_{i0},\hdots, (K_T)_{i (N_i-1)})^T 
\end{align}
is \rev{constructed by applying the $\mathcal{S}$ extractor defined in Appendix \ref{ap:tsext} into Equation (\ref{eq:kailath_expand}).}
In the above form of a posterior snapshot, we require a memory size of $\mathcal{O} (N_i^2)$ or $\mathcal{O} (N_j^2)$ for each snapshot. 

Likewise, we can also consider the \rev{marginal} posterior for the time-series of the $j$-th pixel using the $\mathcal{T}$ extractor defined in Appendix \ref{ap:tsext} as follows:
\begin{align}
&\boxed{ \text{\bf  Pixel-wise Evolution Given $\gvv$ and $\thetaGP$} }\nonumber \\
 \label{eq:dysot_posterior_pix}
&p(\avj_j|\dv, \thetaGP, \gvv) = \Ng(\avj_j|\avj_j^\ast,\Sigma_{\avj_j|\dv, \thetaGP, \gvv})
\end{align}
where
\begin{eqnarray}
\avj_j^\ast \equiv (A_{0j}^\ast,A_{1j}^\ast,\hdots,A_{(N_i-1)j}^\ast)^T,
\end{eqnarray}
 is the pixel-wise evolution of $A^\ast$ at the $j$-th pixel, and
\begin{align}
 \label{eq:dysot_posterior_pix_cov}
&{\Sigma}_{\avj_j|\dv, \thetaGP, \gvv}= \alpha (K_S)_{jj} K_T - C_i^T (\Sigma_{\dv} + K_W)^{-1} C_i, 
\end{align}
is \rev{the submatrix of the covariance matrix} (the pixel-wise covariance) derived by adopting $\mathcal{T}$ extractor in Appendix \ref{ap:tsext} into Equation (\ref{eq:kailath_expand}) with 
\begin{eqnarray}
C_i &=& \alpha \mathcal{D}(\uvi_j) K_T \\
\uvi_j &\equiv& ((W K_S)_{0j},\hdots, (W K_S)_{(N_i-1)j})^T.
\end{eqnarray}
Equation (\ref{eq:dysot_posterior_pix}) also requires a memory size of $\mathcal{O} (N_i^2)$ or $\mathcal{O} (N_j^2)$ for each pixel. 
Either Equation (\ref{eq:dysot_posterior_map}) or (\ref{eq:dysot_posterior_pix}) can be used
to compute the posterior of Equation (\ref{eq:posteriorgiven}) depending on the specific purpose.  In the following discussion, we use the snapshot posterior without a loss of generality.  

The evidence of dynamic mapping is derived through the same procedure for deriving Equations (\ref{eq:anamarli0}) and (\ref{eq:anamarli}), namely, 
\begin{eqnarray}
\label{eq:posteaixal0dy}
p(\dv|\thetaGP,\gvv) &=& \Ng(\dv|{\bf 0}, \Sigma_{\dv} + \tilde{W} \Sigma_{\av} \tilde{W}^T) \\
\label{eq:marginallikelidy}
&=& \Ng(\dv|{\bf 0}, \Sigma_{\dv} + K_W ).
\end{eqnarray}
Interestingly, the computational cost of Equation (\ref{eq:marginallikelidy}) is almost the same as that for the static mapping because $\Sigma_{\dv} + K_W \in \mathbb{R}^{N_i \times N_i}$. 
The analytic form of $p(\dv|\thetaGP,\gvv)$ allows us to efficiently sample
 \begin{eqnarray}
 \label{eq:sampleg_axialtiltagain}
\thetaGPdag, \gvvdag \sim p(\thetaGP,\gvv|\dv)
\propto p(\dv|\thetaGP, \gvv) p(\thetaGP) p(\gvv),
\end{eqnarray}
using e.g., an MCMC algorithm.

Using the sample of $\thetaGPdag$ and $\gvvdag$, the marginal posterior of the dynamic map can be approximated as follows:
\begin{eqnarray}
 p(\avi_i|\dv) \approx \frac{1}{N_s} \sum_{n=0}^{N_s-1} p(\avi_i|\dv, \thetaGPdag, \gvvdag).
\end{eqnarray}
In addition, the summary statistics are
\begin{eqnarray}
&\,&\langle f(\avi_i) \rangle \approx \frac{1}{N_s} \sum_{n=0}^{N_s-1} \int d \av f(\avi_i) \Ng(\avi_i| (\avi_i^{\ast})_n^\dagger,\Sigma_{\avi_i|\dv,\thetaGPdag, \gvvdag}), \nonumber \\
\end{eqnarray}
where $(\avi_i^{\ast})_n^\dagger$ is the snapshot of $A^\ast$ given $\thetaGP = \thetaGPdag$ and $\gvv = \gvvdag$. The mean of the marginal snapshot for a dynamic geography is given by the following:
\begin{align}
\label{eq:snapfin}
&\langle \avi_i \rangle  \approx \frac{1}{N_s} \sum_{n=0}^{N_s-1} (\avi_i^{\ast})_n^\dagger.
\end{align}
Reshaping $\avi_i$ and $\avi_i^{\ast}$ in Equation (\ref{eq:snapfin}) to $A$ and $A^\ast$, we find the mean geography matrix as
\begin{align}
&\boxed{ \text{\bf Mean Map Matrix for Dynamic Geography}  }\nonumber \\
\label{eq:snapfinA}
&\langle A \rangle  \approx \frac{1}{N_s} \sum_{n=0}^{N_s-1} \alpha_n^\dagger  K_T(\tau_n^\dagger) \mathcal{D} (\yv_n) W(\gvvdag) K_S(\gamma_n^\dagger),
\end{align}
where
\begin{align}
&\yv_n \equiv  [I + \Pi_{\dv} (K_W)_n]^{-1} \Pi_{\dv} \dv, \nonumber\\
&(K_W)_n \equiv  \alpha_n^\dagger K_T(\tau_n^\dagger) \odot [W(\gvvdag) K_S(\gamma_n^\dagger) W(\gvvdag)^T] \nonumber,
\end{align}
and
\{$\thetaGPdag = (\gamma_n^\dagger, \alpha_n^\dagger, \tau_n^\dagger)^T$, $\gvv^\dagger_n$\}
is 
the $n$-th set of hyperparameters sampled from $p(\thetaGP, \gvv|\dv)$ (Equation \ref{eq:sampleg_axialtiltagain}). 

\section{Test using a Toy Model}\label{sec:toy}

\begin{figure*}[]
 \begin{center}
   \includegraphics[width=0.65\linewidth]{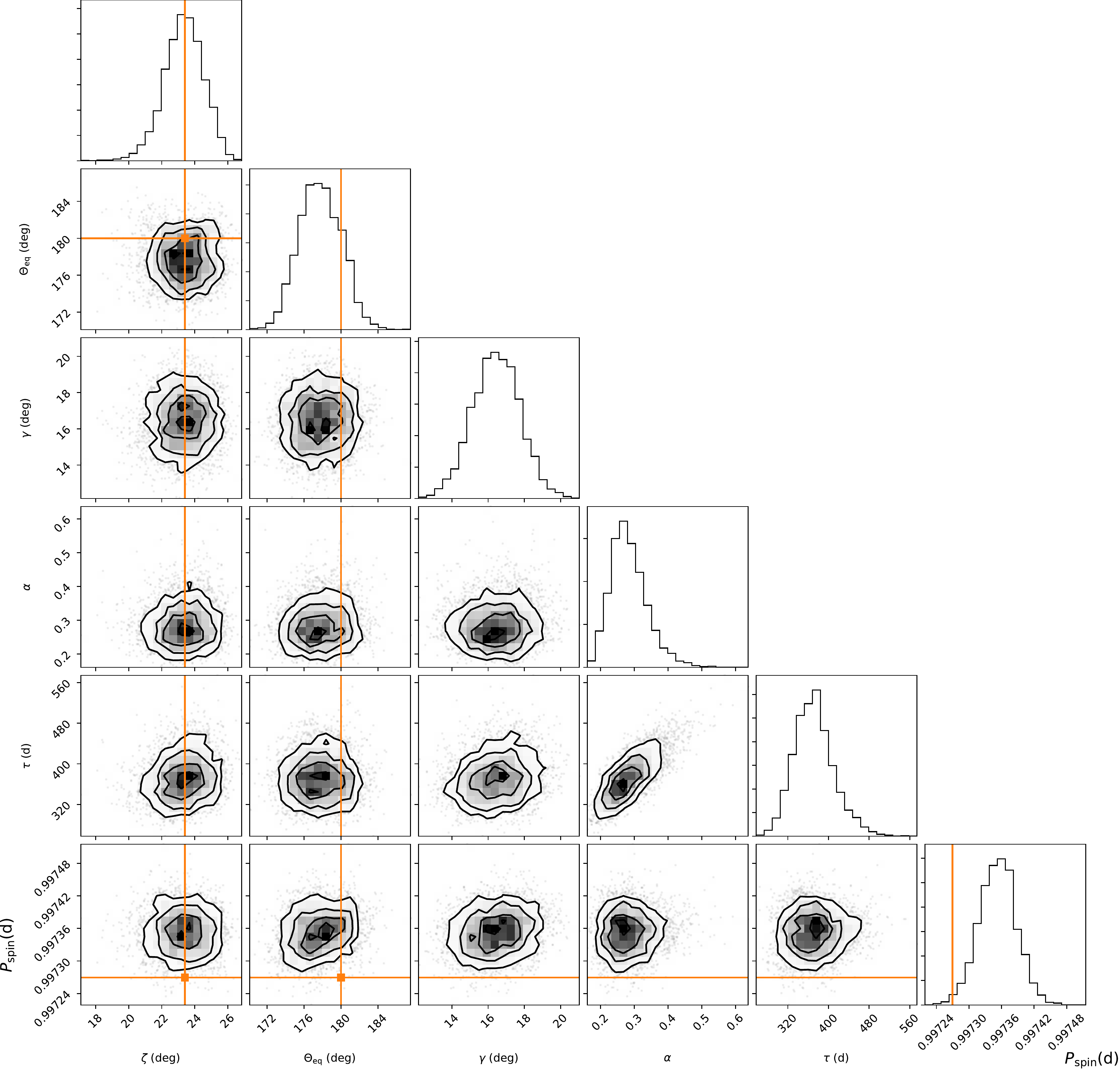}
 \end{center}
 \caption{Marginal posterior $p(\thetaGP,\gvv|\dv)$ of the dynamic spin-orbit tomography. The input values of the spin parameters are indicated by the orange solid lines. We used the sidereal day of Earth $\Pspin = 0.99727$ d as the spin rotation period.  \label{fig:corner_dyRBF}}
\end{figure*}

\begin{figure*}[]
 \begin{center}
   \includegraphics[width=0.32\linewidth]{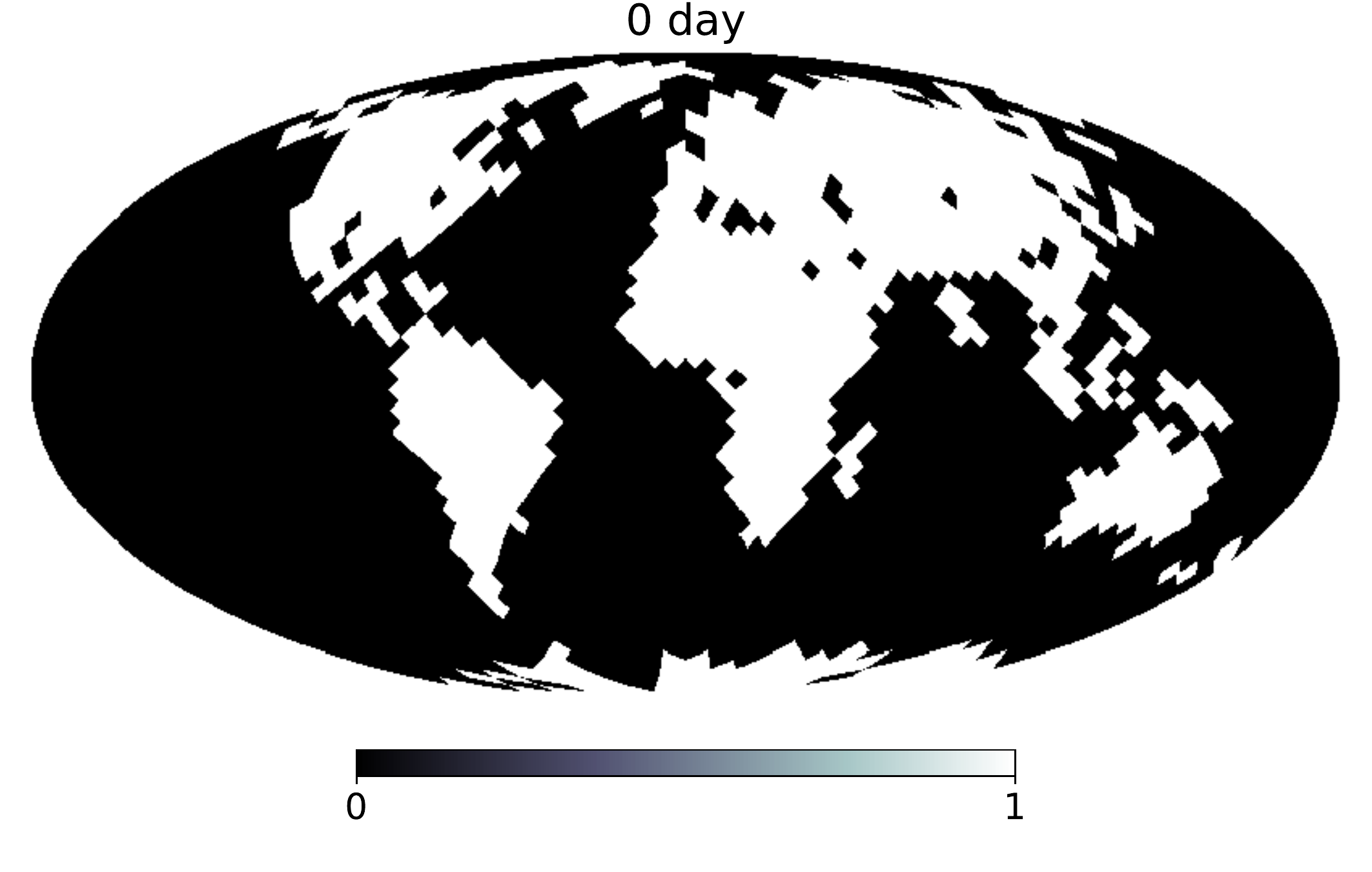}
    \includegraphics[width=0.32\linewidth]{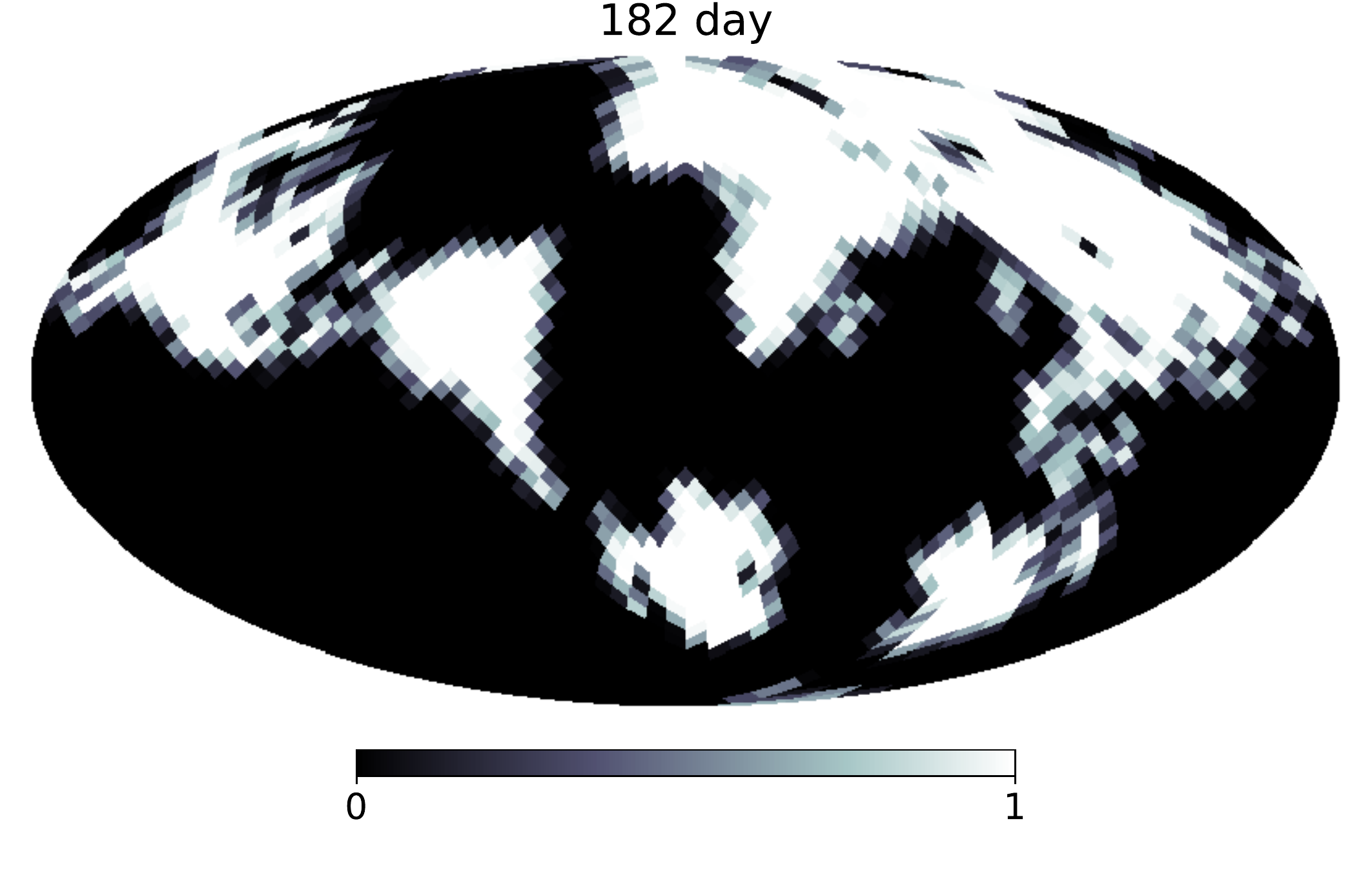}  
   \includegraphics[width=0.32\linewidth]{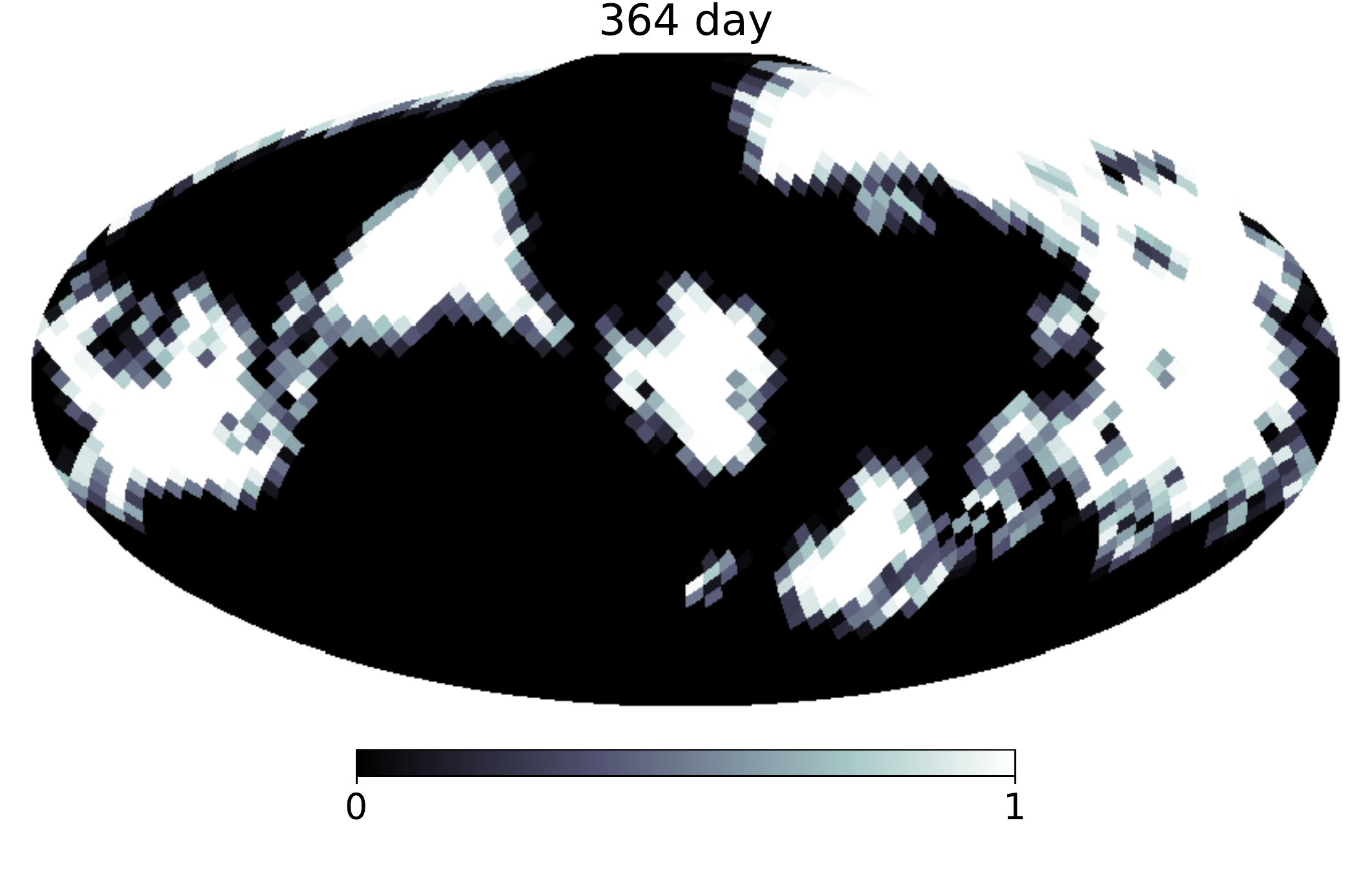}
         
   \includegraphics[width=0.32\linewidth]{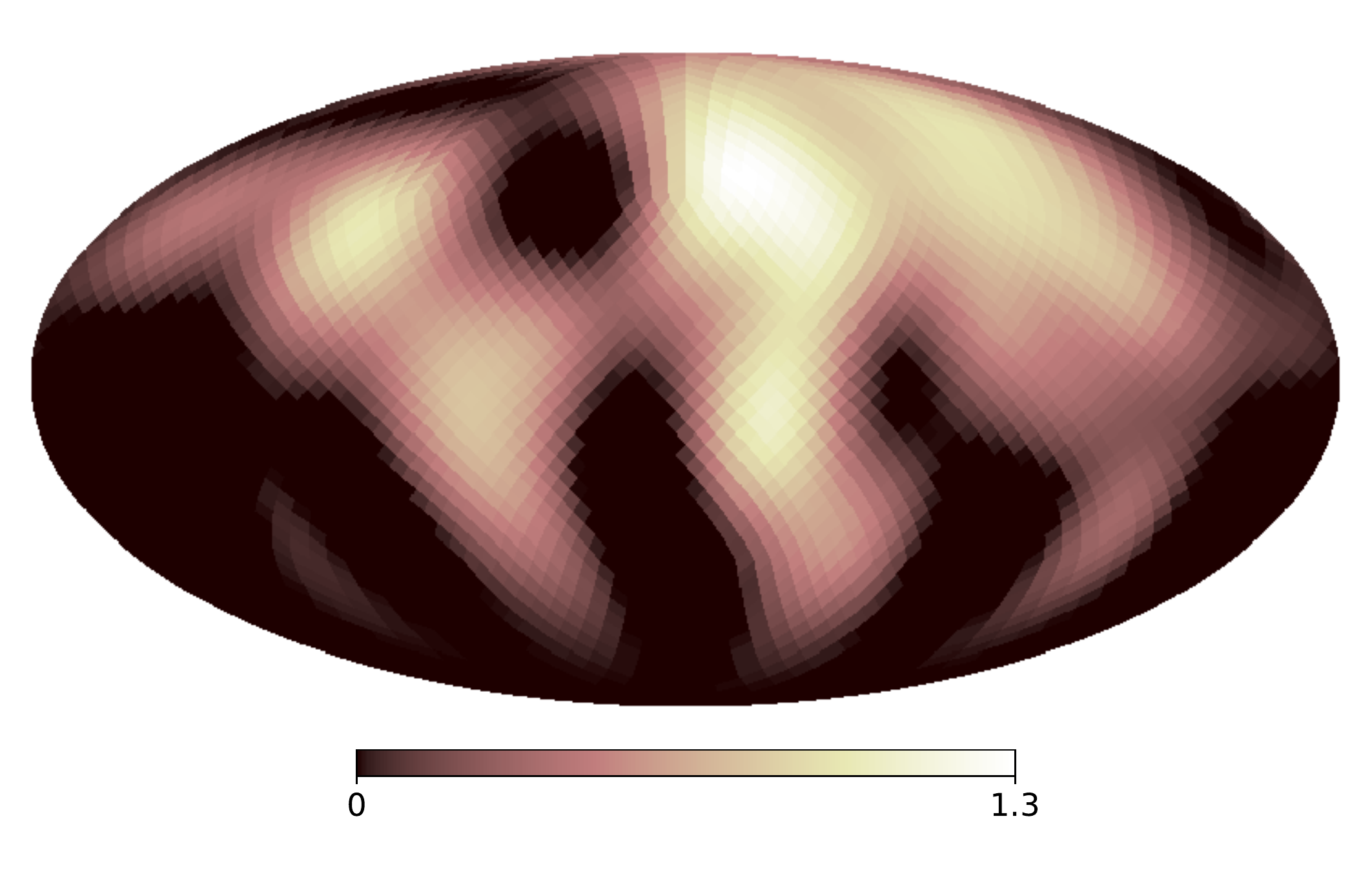}
   \includegraphics[width=0.32\linewidth]{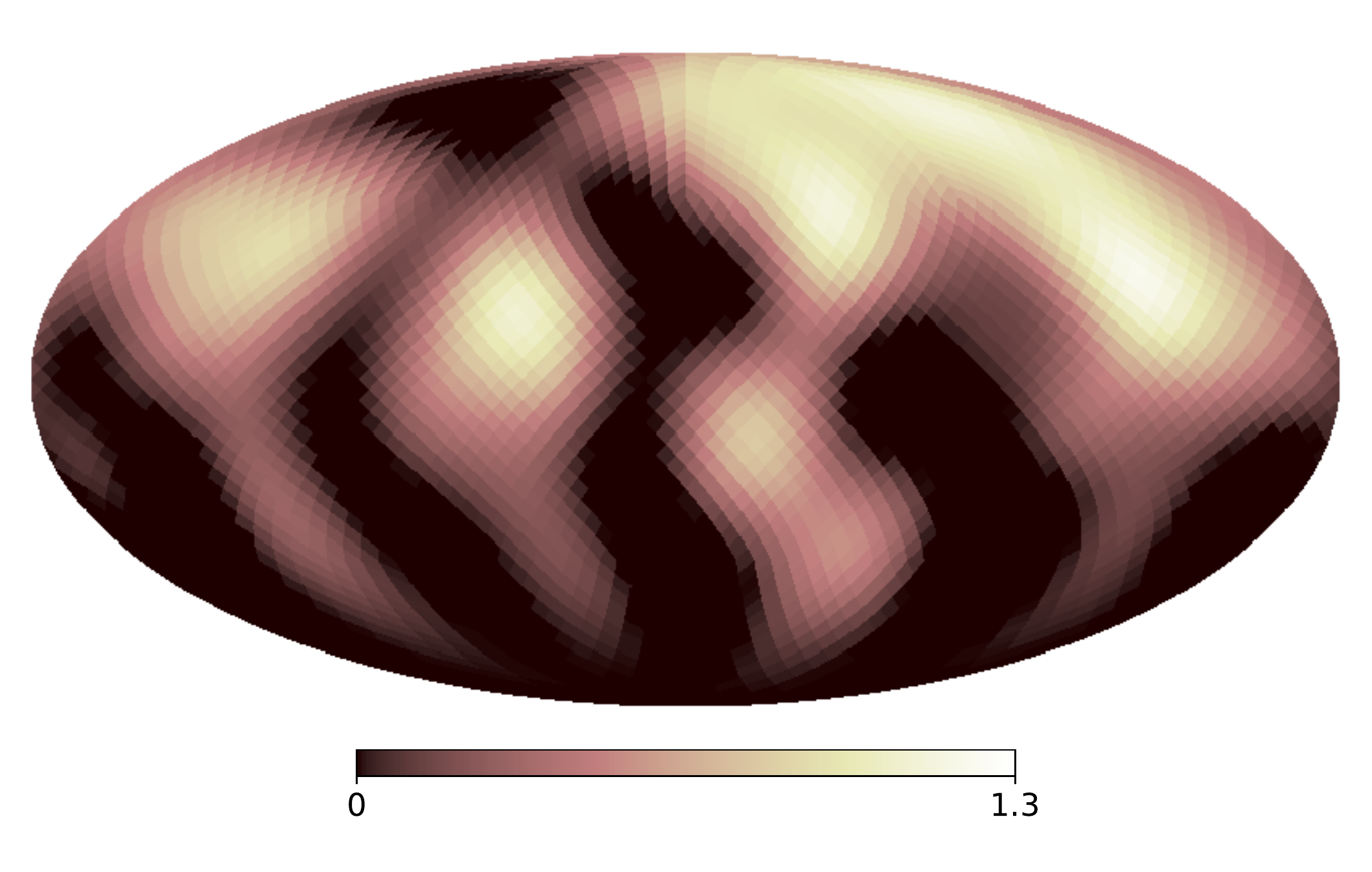}
   \includegraphics[width=0.32\linewidth]{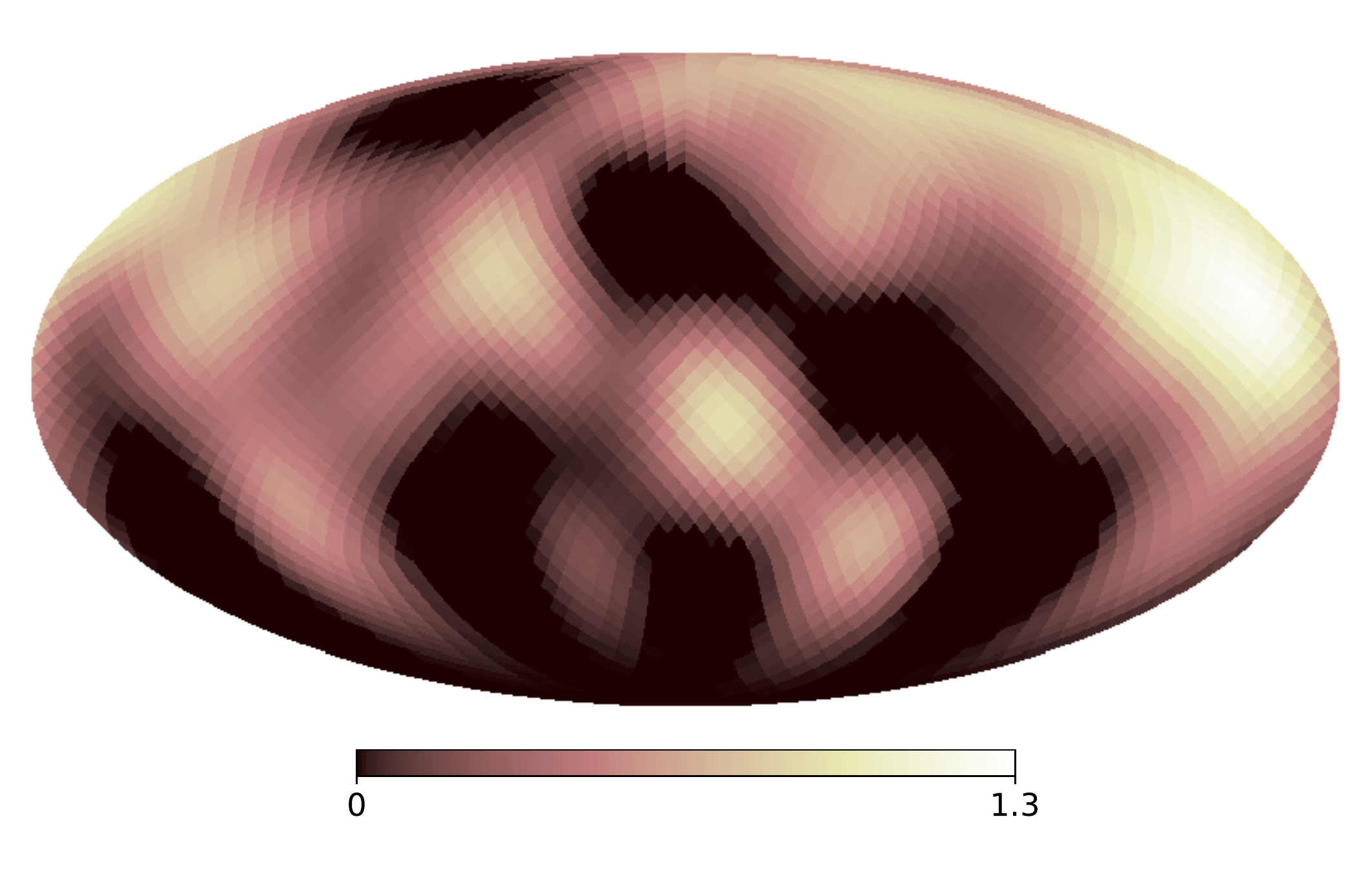}
 \end{center}
 \caption{Input (top) and the retrieved mean snapshot (bottom) for three different times, $t=0, 182,$ and 364 days from left to right. \rev{We restrict the pixel value to the range of 0--1.3 for comparison with the input maps although some pixels are above or under the range.}
 \label{fig:dysot}}
\end{figure*}

We test the dynamic spin-orbit tomography using a toy model. The toy model assumes an extreme case in which a cloudless Earth has a rapid continental drift. 
The shape and congifuration of the continents are identical to those on the Earth, but they rotate by $90^\circ$ per planet year around the axis perpendicular to the spin axis (top panels in Figure \ref{fig:dysot}).
We use the geometric settings of \cite{Kawahara2020}, an orbital inclination of 45$^\circ$, an axial tilt of $23.4^\circ$, and an orbital phase at an equinox of $\ThetaM=90^\circ$, the spin rotation period of $\Pspin=$ 23.9345/24.0 = 0.99727 d, and the orbital revolution period of 365 d. We computed the integrated light curve and injected observational noise assuming an independent Gaussian with a standard deviation of 1 \% of the mean of the light curve. We took $N_i=$1024 frames of a light curve evenly distributed during a 1-year period.

For this test, we used the Mat\'{e}rn-3/2 kernel for the temporal regularization and the RBF kernel for the spatial regularization. The number of pixels is $N_j=3072$. 
To sample from the marginal posterior distribution $p(\gvv, \thetaGP|\dv)$,
we used a Python-based MCMC package {\it emcee} \citep{2013PASP..125..306F} and assumed independent, log-flat priors for the hyperparameters $\thetaGP$ ($0.01 \le \gamma \le \pi/2$, $10^{-4} \le \alpha \le 10^4$, $10^{-4} \le \tau \le 10^{4}$) and 
$p(\gvv)d\gvv \propto \sin{\zeta}d\zeta d\ThetaM d \Pspin$ ($0 \le \zeta \le \pi$,$0 \le \ThetaM \le 2 \pi$,
and $0.5 \le \Pspin \le 1.5$ d). 

Figure \ref{fig:corner_dyRBF} shows a corner plot for the MCMC samples from $p(\gvv, \thetaGP|\dv)$.
We find that a dynamic spin-orbit tomography can infer an obliquity and the orbital phase at an equinox as well as in a static mapping.
The inferred spatial correlation scale $\gamma \sim 16$ deg ($1800$ km) is reasonable because it roughly corresponds to the scale of the continents (approximately half the size of Australia). The time scale of $\tau \sim 1 $yr is also reasonable because our toy continents rotate by 90 degrees per year.
\rev{
On the other hand, the marginal posterior of the spin-rotation is slightly biased from the input value. We do not fully understand the origin of this bias. However, the bias is comparable to the period error that causes longitudinal shift of only one pixel per year, $\Delta \Pspin = 4 \times 10^{-5}$ d, and we suspect that this is associated with a finite (but still high) resolution of our map. In this sense, the rotational period is reasonably well recovered within the model uncertainty. Alternatively, the bias is potentially owing to the fact that the input map, which consists of 0 or 1 only, is not well described by a Gaussian process prior.}

Note that many of the previous studies on an obliquity estimate have only considered the prograde rotation ($0^\circ \le \zeta \le 90^\circ$) \citep{2010ApJ...720.1333K,2011ApJ...739L..62K,2012ApJ...755..101F,2018AJ....156..146F,2019AJ....158..246B}. However, it is in fact possible to break the degeneracy between the prograde and retrograde solutions, as pointed out by \cite{2016MNRAS.457..926S} using the theoretical analysis of the motion of the geometric kernel and also a specfic example of a light curve. This is the case in both static and dynamic mapping, and we discuss the issue further in Appendix \ref{ap:proretro}.

\begin{figure}[]
 \begin{center}
   \includegraphics[width=0.7\linewidth]{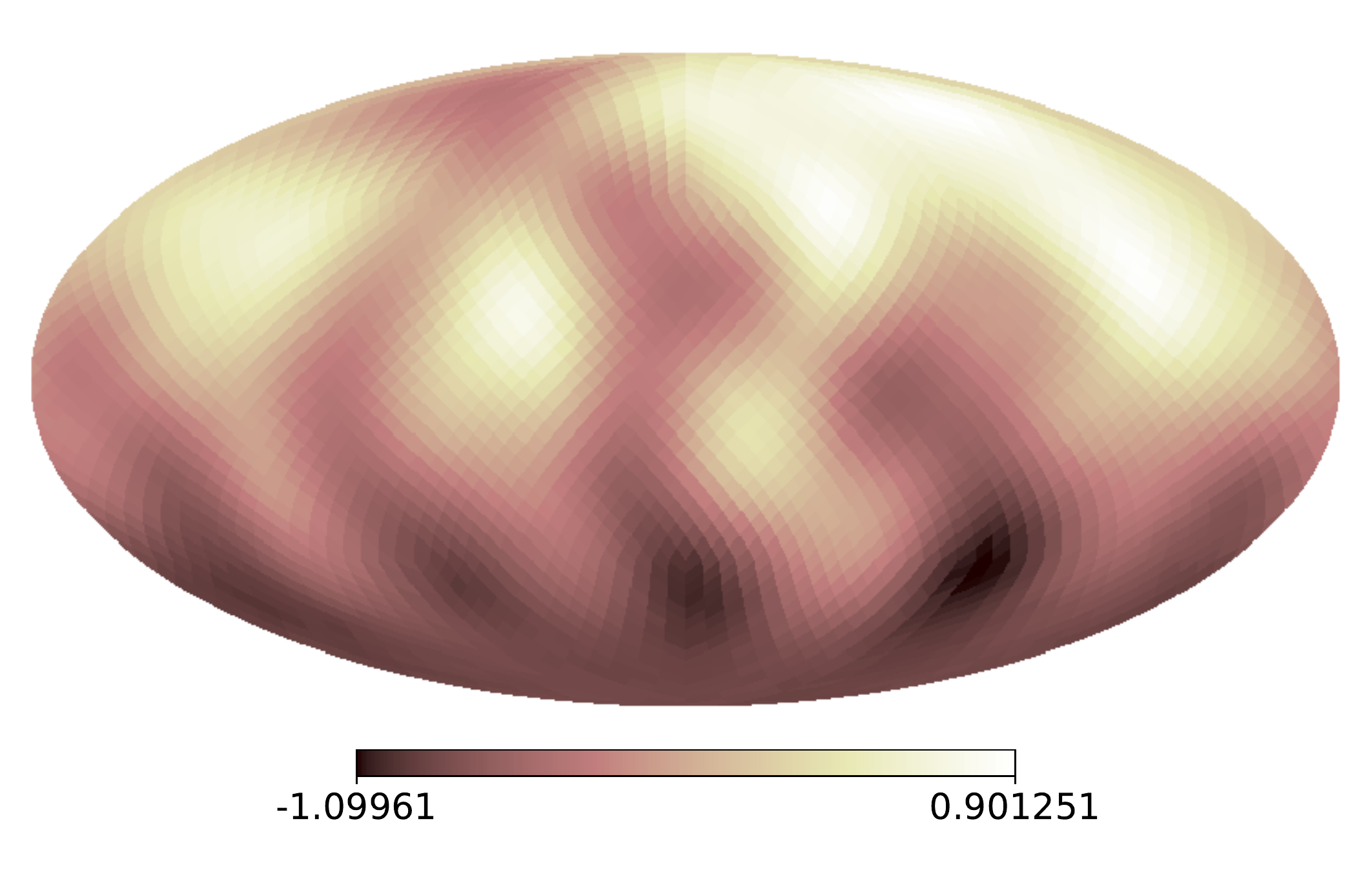}
      \includegraphics[width=0.7\linewidth]{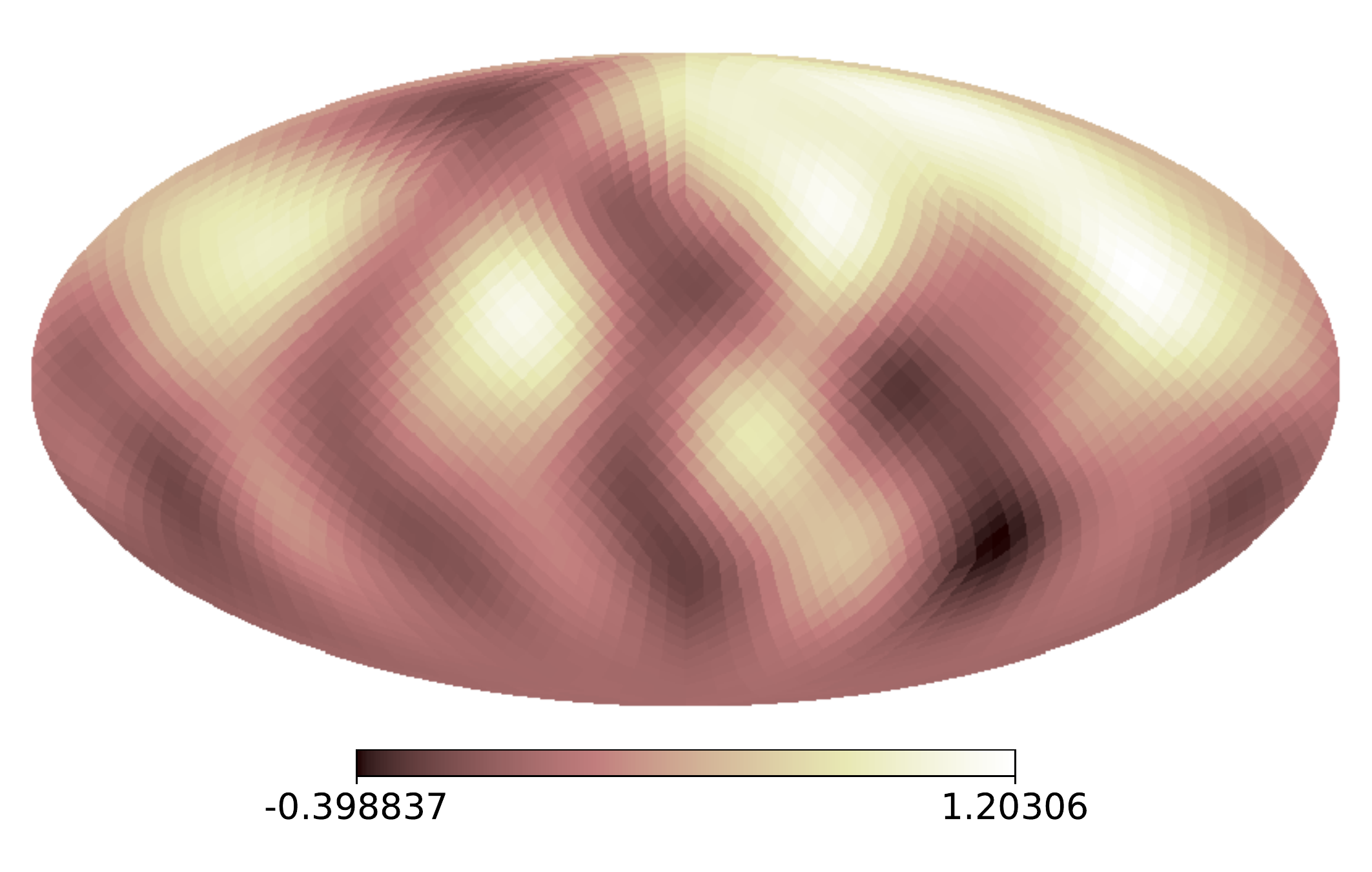}
    \includegraphics[width=0.7\linewidth]{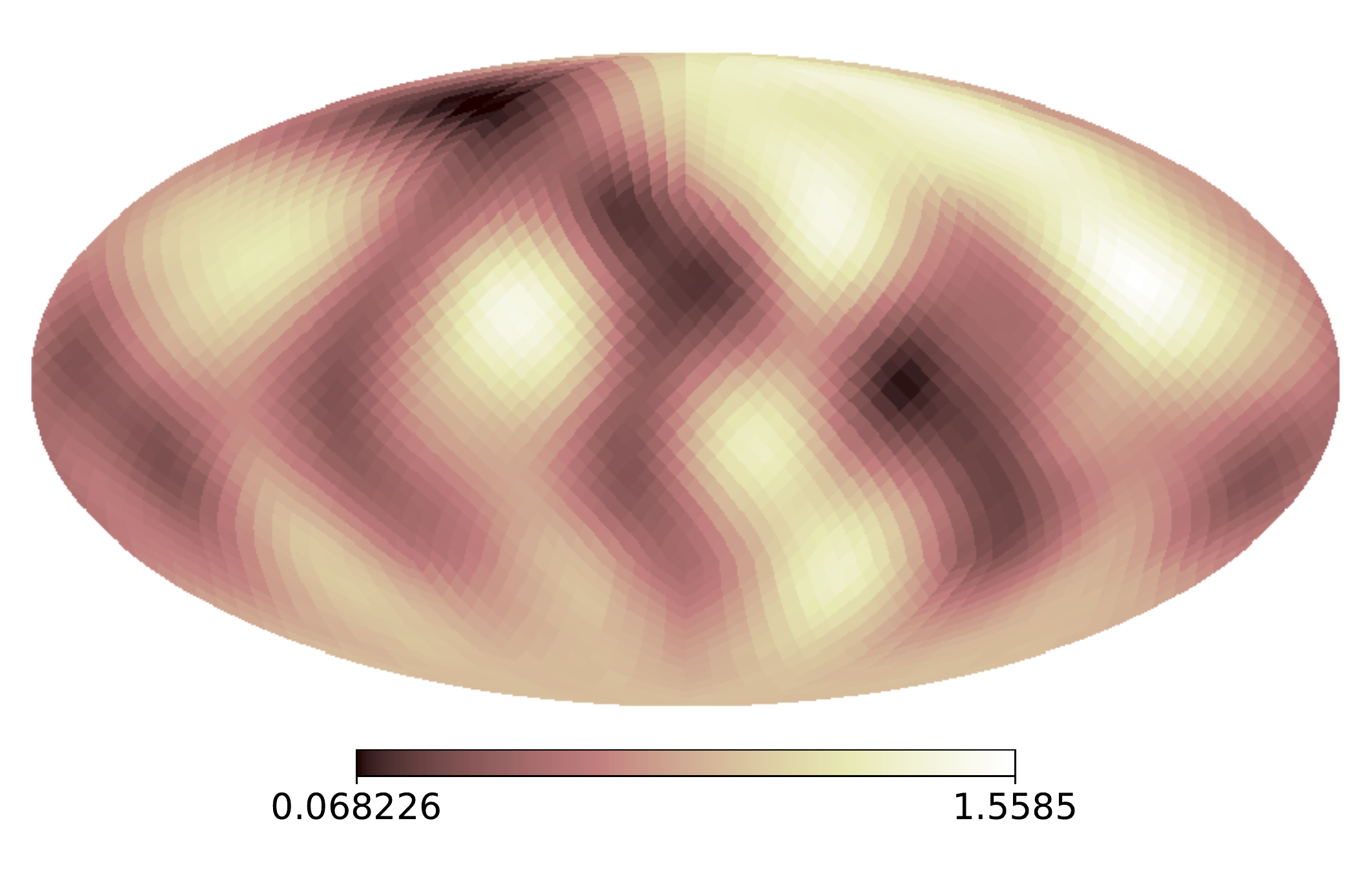}
 \end{center}
 \caption{\rev{The 5, 50, and 95 \% percentile maps of each pixel from top to bottom. These maps indicate the geography uncertainity of the snapshot of t=$182$ day, corresponding to the middle column in Figure \ref{fig:dysot}.}
 \label{fig:credible}}
\end{figure}

Figure \ref{fig:dysot} shows the results for the geography\footnote{ \rev{We note that we do not have any constraint of the model range such as nonnegativity although the pixel value should be regarded as an albedo in the case of the toy model.}}.
The snapshots in the middle row computed using Equation (\ref{eq:snapfin}) with $N_s=4032$ capture well the characteristics of the input maps in the top row at each of three different times (0, 182, and 364~days from left to right). \rev{In a one-dimensional problem, this information can be shown in the form of a corner plot. The visualization is more difficult in the current problem, however, because the covariance matrix has $N_j^2$ elements, whereas the map has only $N_j$ pixels. \cite{2018AJ....156..146F} used the credible boundaries for each pixel to visulaize the geography uncertainties. Following their approach, we compute the 5, 50 (median), and 95 \% percentile maps for each pixel for $t=$ 182 day, as shown in Figure \ref{fig:credible}. All of the major structures in the nothern hemisphere appear in all the percentile maps\footnote{\rev{We also test another visualization, called the randomized map in Appendix \ref{ap:ranmap}, which captures the geography uncertainty from the other perspective.}}. This implies that these structures are robustly inferred.}

\section{Demonstration using Real data by DSCOVR}
 
Finally, we demonstrate the dynamic spin-orbit tomography by applying it to a real multiband light curve of the Earth as observed by DSCOVR \citep{2018AJ....156...26J}. 
Although the geometry of the DSCOVR observations is not the same as the one relevant for the direct imaging of an exo-Earth,
the continuous monitoring of the Earth from the L1 point conveys both latitudinal and longitudinal information on the geography. This enables a 2-D mapping to be conducted \citep{2019ApJ...882L...1F,Kawahara2020,2020arXiv200403941A}. 

A principle component analysis (PCA) is a traditional scheme used to extract the spectral components of the planet surface \citep{2009ApJ...700..915C}.
In \cite{2019ApJ...882L...1F}, the authors demonstrated that the L2 regularization of the second principle component (PC2) of the DSCOVR
data recovers structures that resemble the continents of
the Earth. They also found that the land/ocean fraction is independent of the first principle component (PC1). 
These results are naturally explained if the largest variation (PC1) is mainly due to cloud components.
This motivated us to apply the dynamic spin-orbit tomography to PC1 of the DSCOVR data to test if the time-varying structures of clouds can be recovered.

\begin{figure}[]
 \begin{center}
   \includegraphics[width=0.95\linewidth]{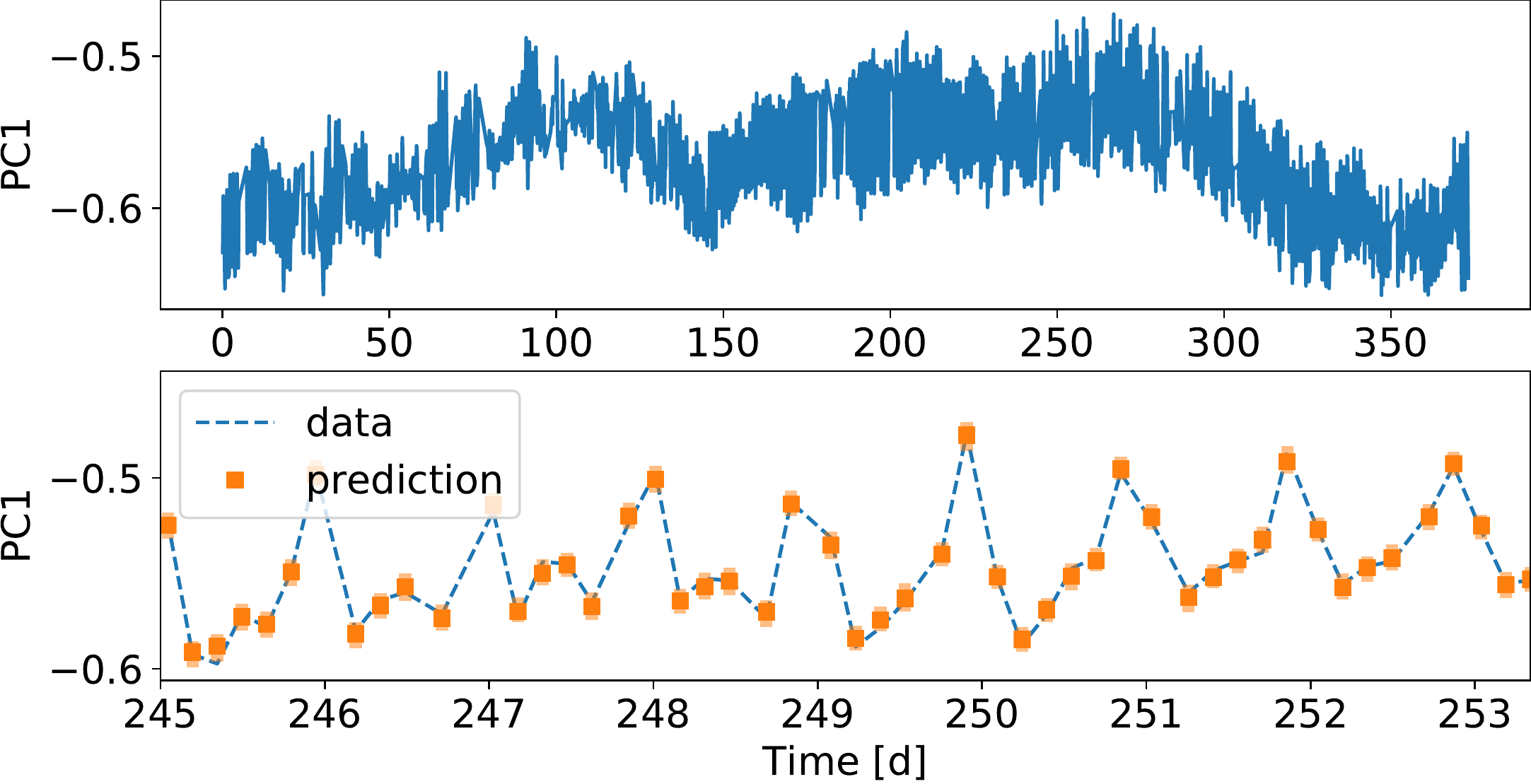}
 \end{center}
 \caption{One-year light curve of PC1 (top). The bottom panel enlarges the top panel in early Semptember. The orange dots and error bars indicate the median of the prediction and its 10 -- 90 \% percentile.  \label{fig:predata}}
\end{figure}

\begin{figure*}[]
 \begin{center}
   \includegraphics[width=0.65\linewidth]{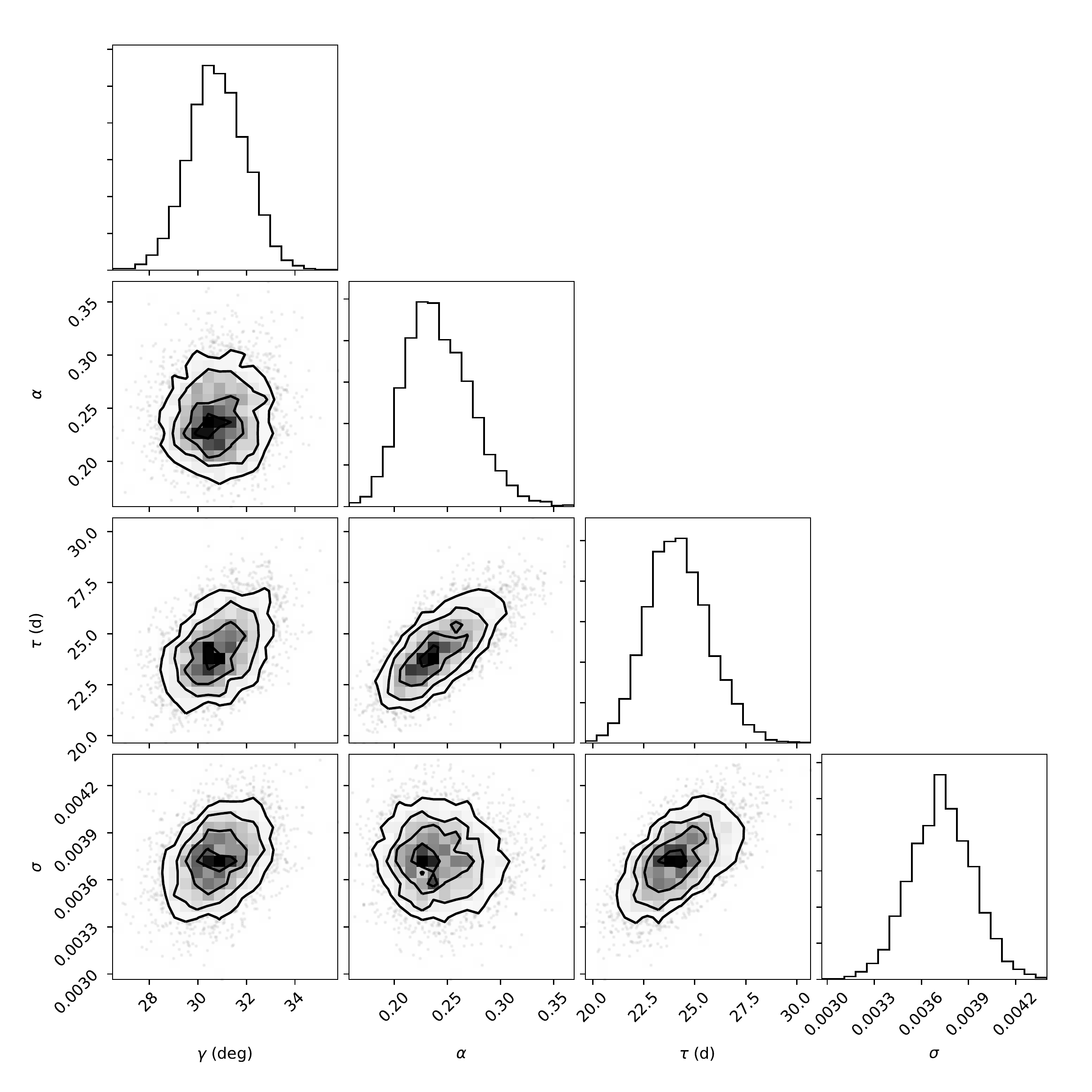}
 \end{center}
 \caption{The marginal posterior $p(\thetaGP,\sigma|\dv)$ of the dynamic spin-orbit tomography for the DSCOVR data. \label{fig:corner_dyRBF_dscovr}}
\end{figure*}

Excluding three UV bands (0.3175, 0.325, and 0.340 $\mu$m) in ten DSCOVR filters, seven optical bands (0.388, 0.443, 0.552, 0.680, 0.688, 0.764, and 0.779 $\mu$m) are used for PCA\footnote{Note that strong oxygen B and A absorptions exist in the filters of 0.688, 0.764 $\mu$m.}.  
To reduce the computational cost, we 
used one of two bins of 1-year data of 2016 used in \cite{2019ApJ...882L...1F}. Figure \ref{fig:predata} (top) shows the 1-year data of PC1 (the number of frames is $N_i=2435$). We model the noise of the PC1 using an independent Gaussian with $\Sigma_{\dv} = \sigma^2 I$ and 
infer $\sigma$ simultaneously with the other parameters.
 In addition, we use the geometric kernel $W$ in \cite{2019ApJ...882L...1F}. Thus, we have four parameters ($\gamma$, $\tau$, $\alpha$, and $\sigma$) in addition to the geography. 
 
\begin{figure*}[]
 \begin{center}
   \includegraphics[width=0.32\linewidth]{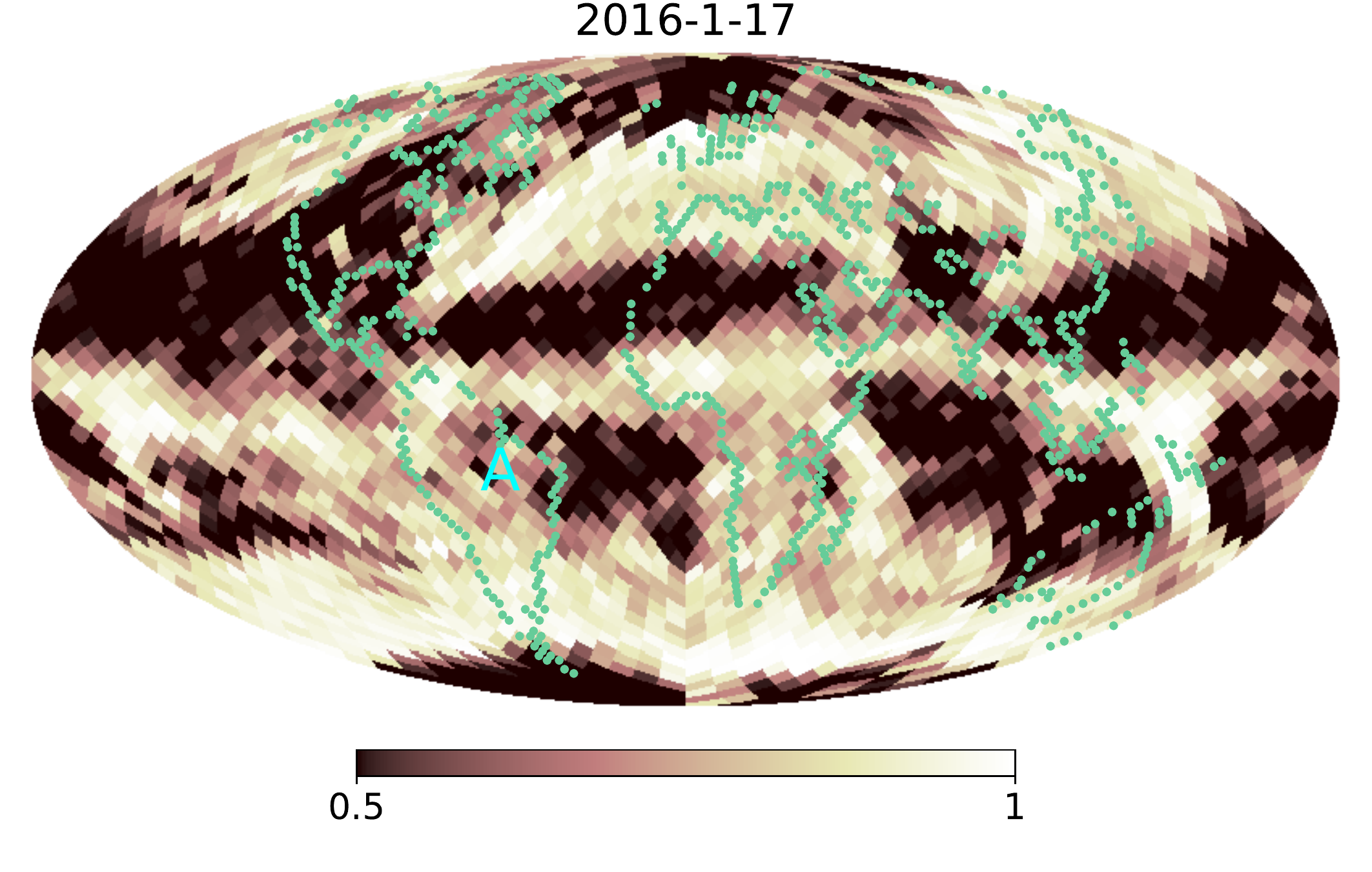}
   \includegraphics[width=0.32\linewidth]{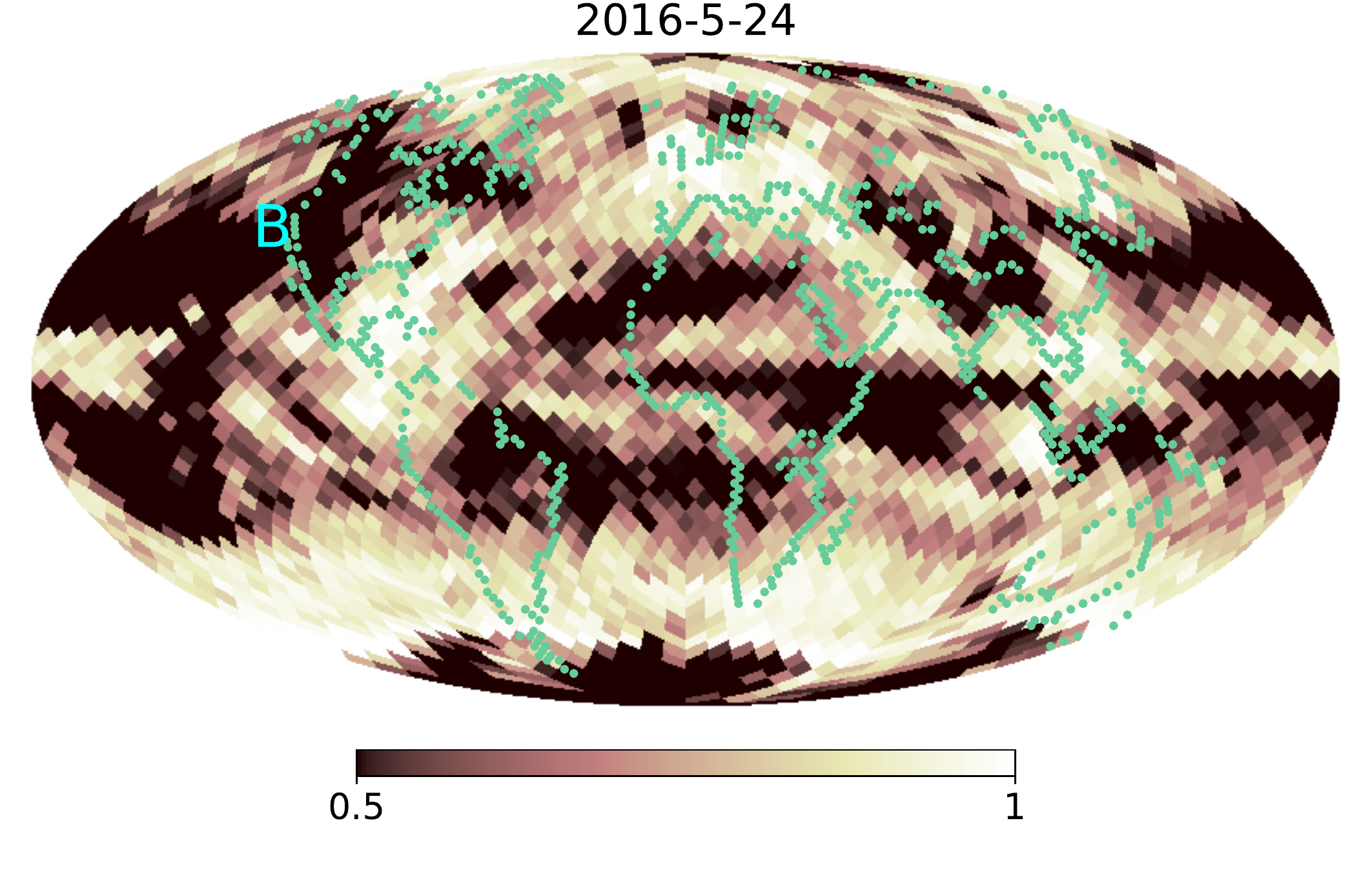}
   \includegraphics[width=0.32\linewidth]{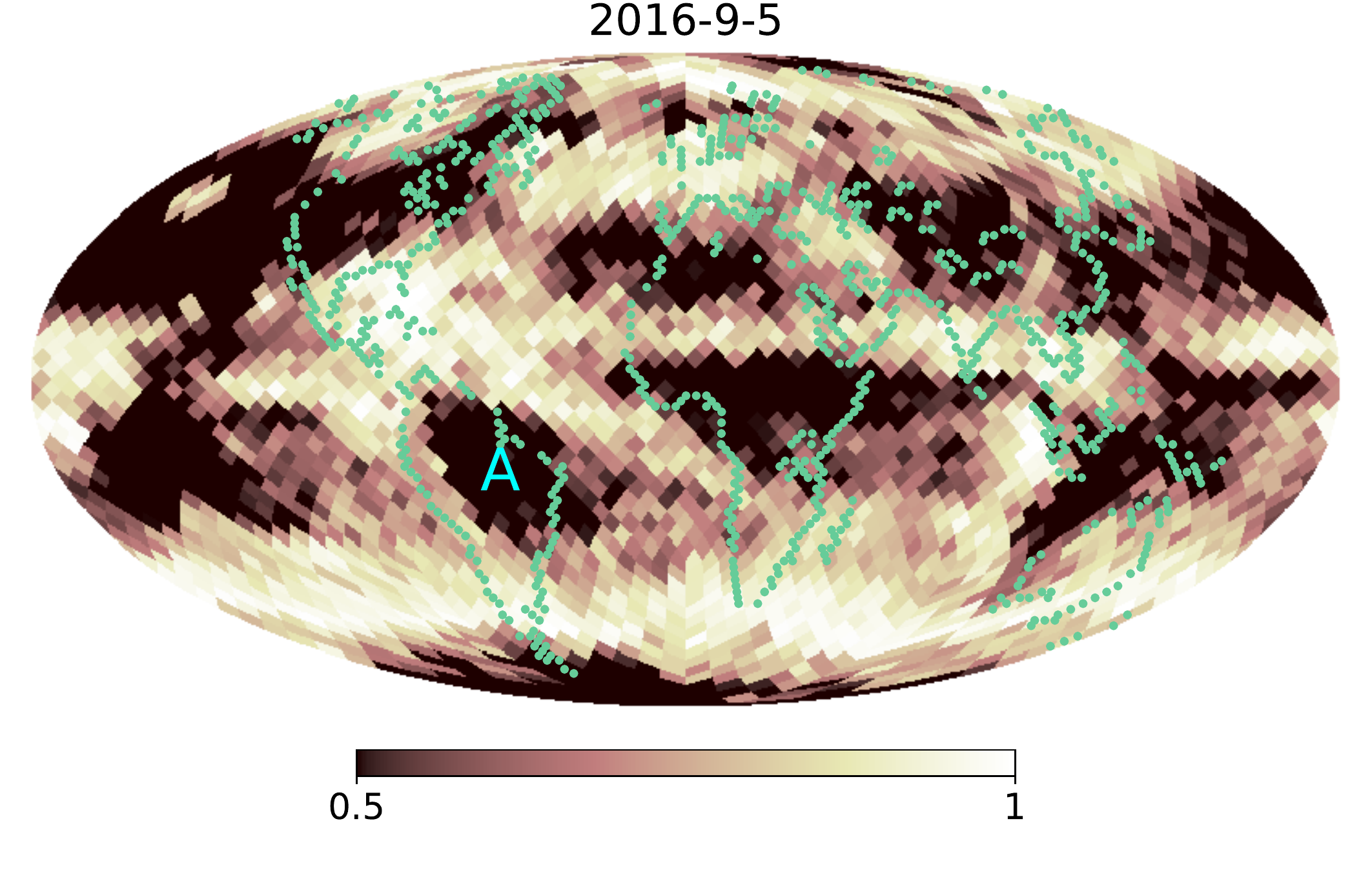}           \includegraphics[width=0.32\linewidth]{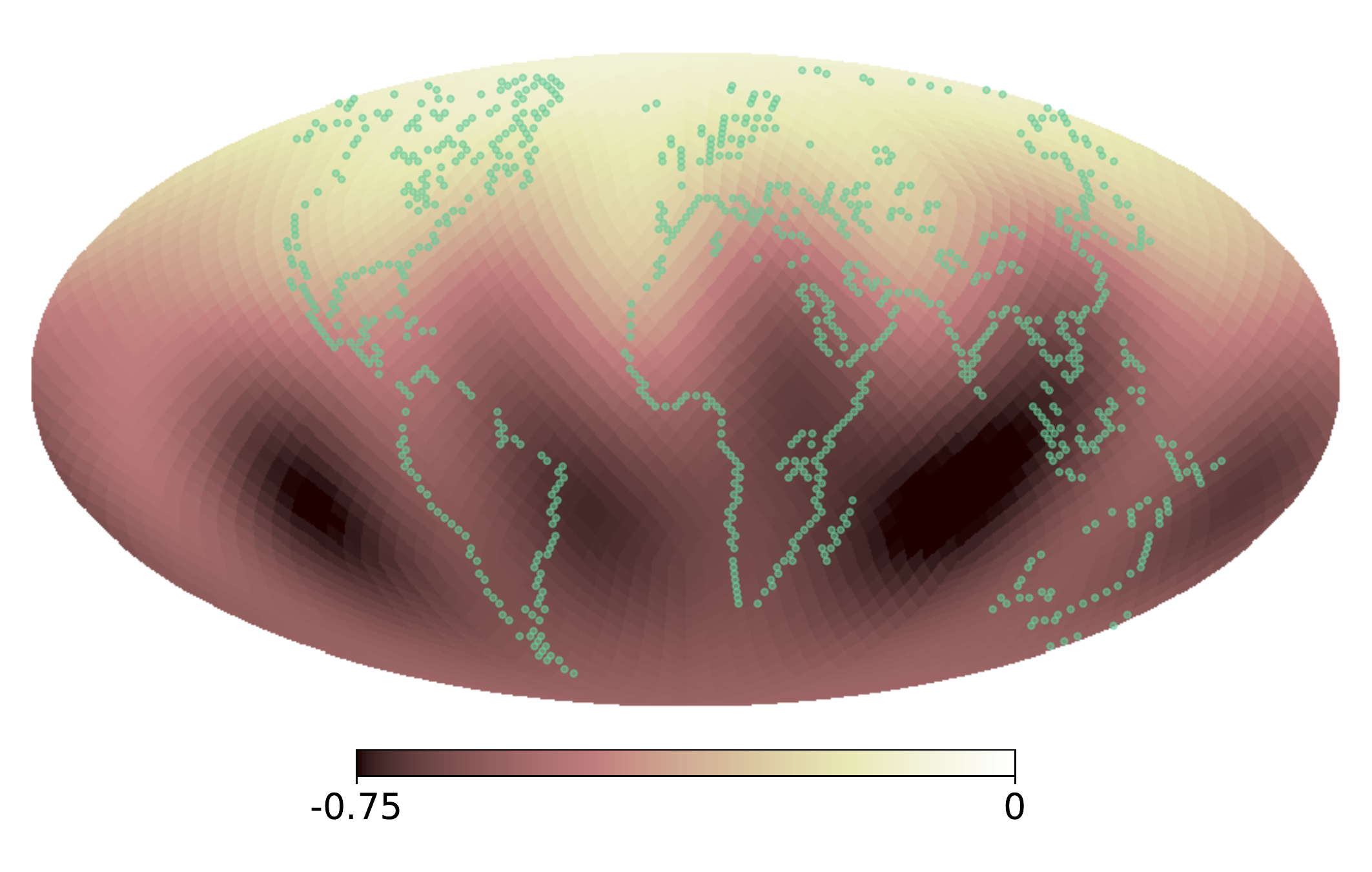}
   \includegraphics[width=0.32\linewidth]{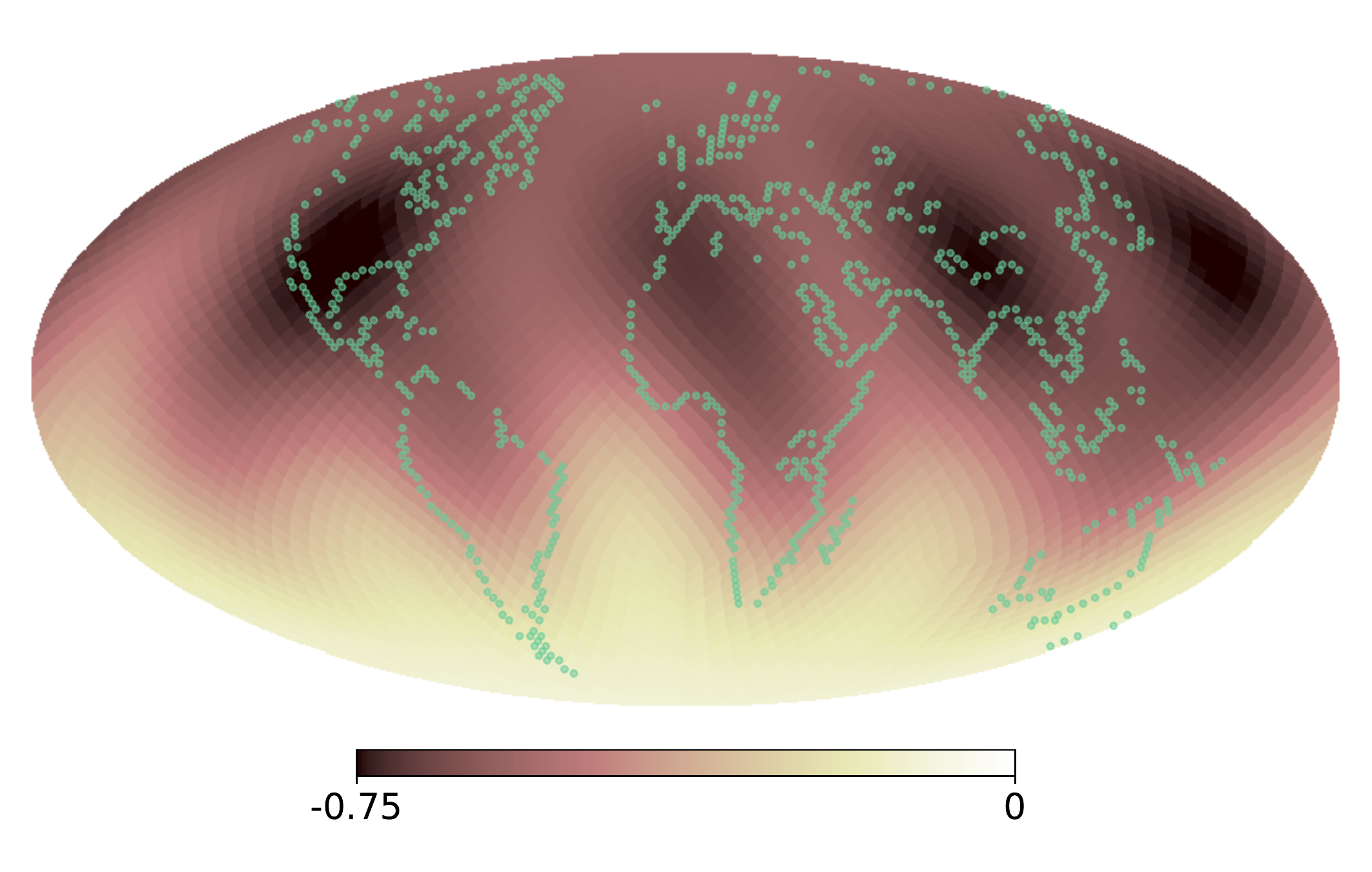}
   \includegraphics[width=0.32\linewidth]{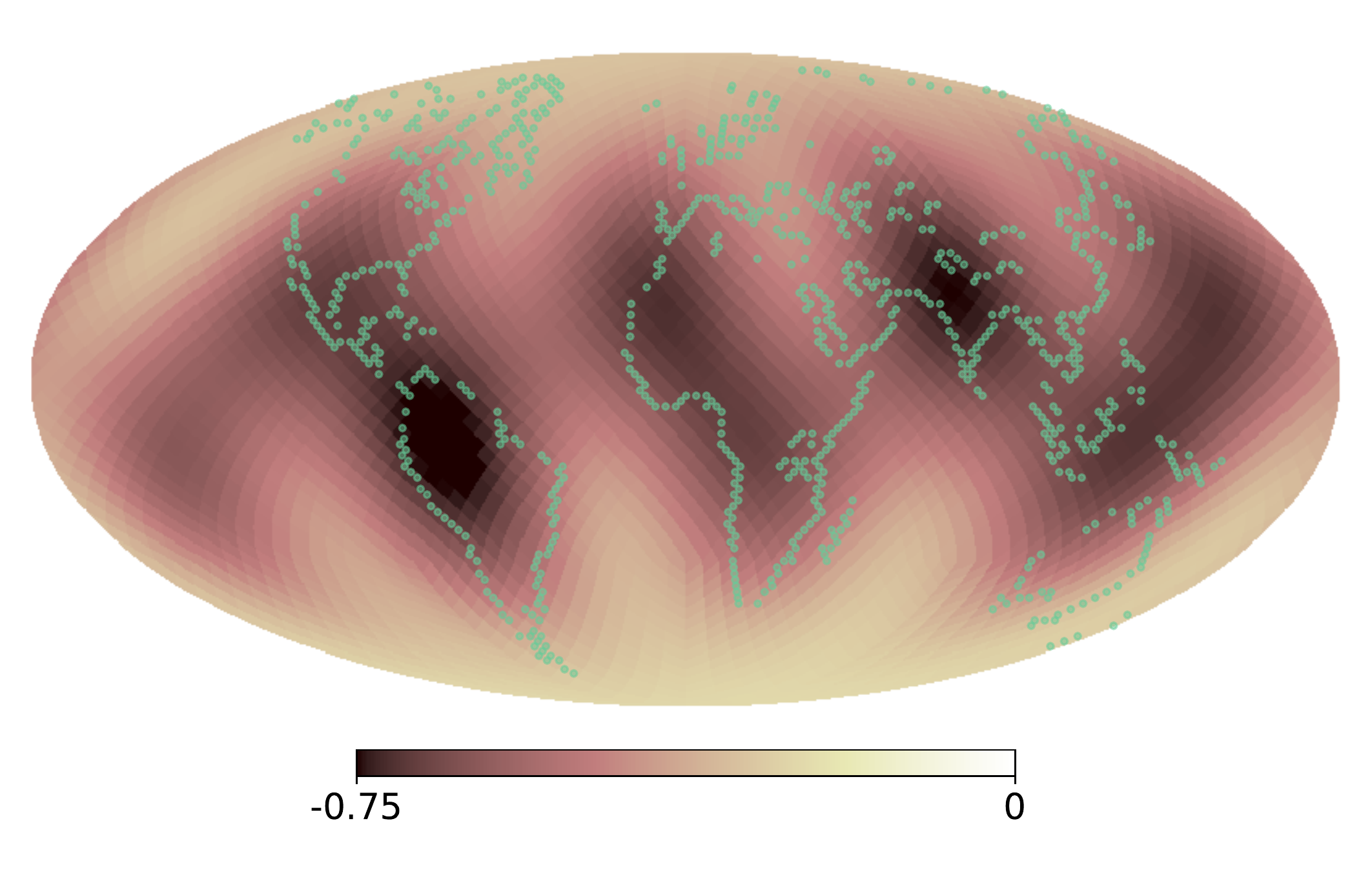}   
 \end{center}
 \caption{Observed 8-day mean cloud fraction (top),
 the retrieved mean snapshot (bottom), for three different dates in January, May, and September in 2016 (from left to right). \rev{The range of the color bar is commonly fixed to -0.75 -- 0 to compare three snapshot with each other. We note that the pixel value can be negative because these are the mapping of PC1, not albedo.}
 \label{fig:dysotdscovr}}
\end{figure*}

\begin{figure}[]
 \begin{center}
   \includegraphics[width=0.7\linewidth]{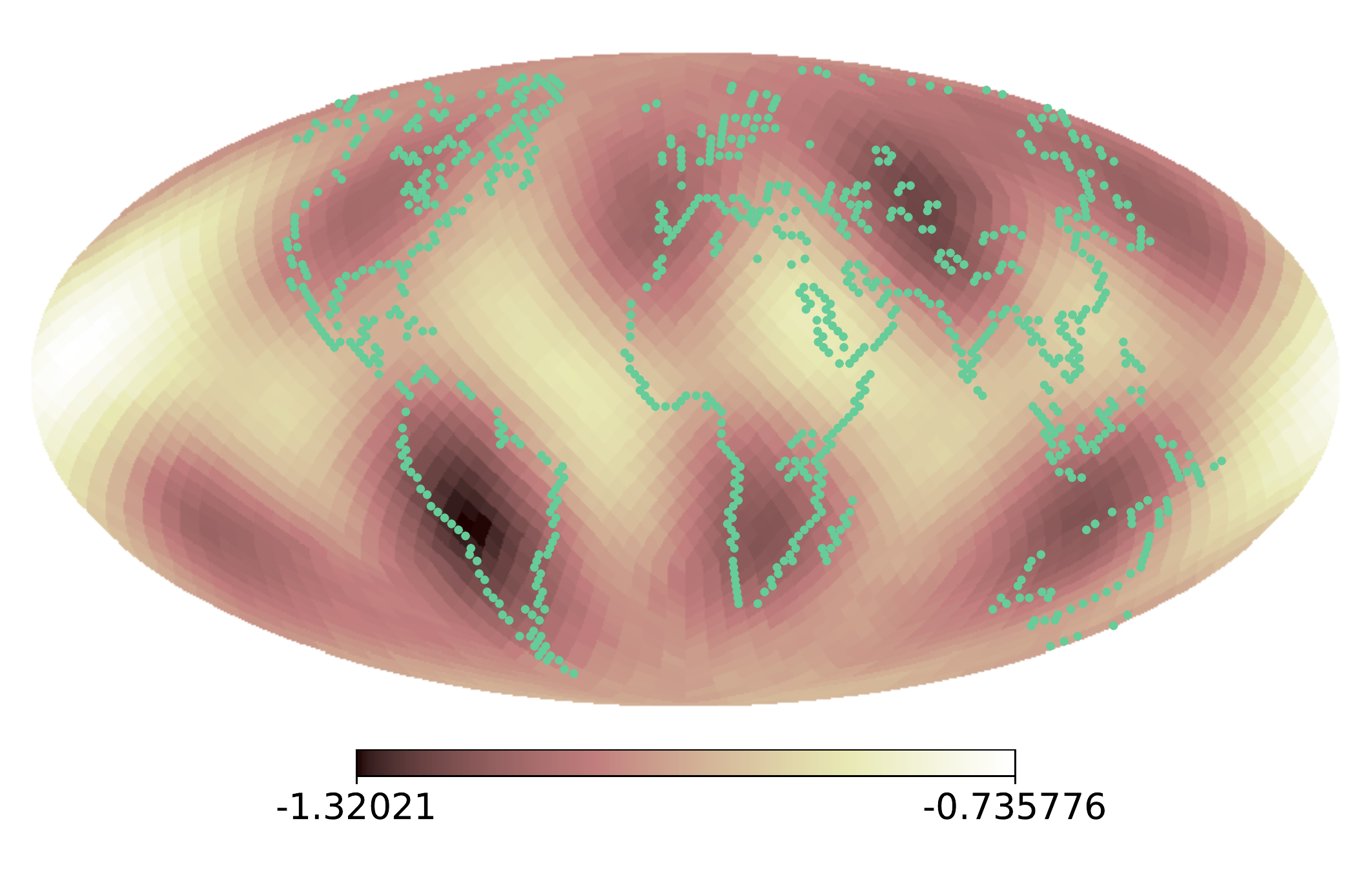}
      \includegraphics[width=0.7\linewidth]{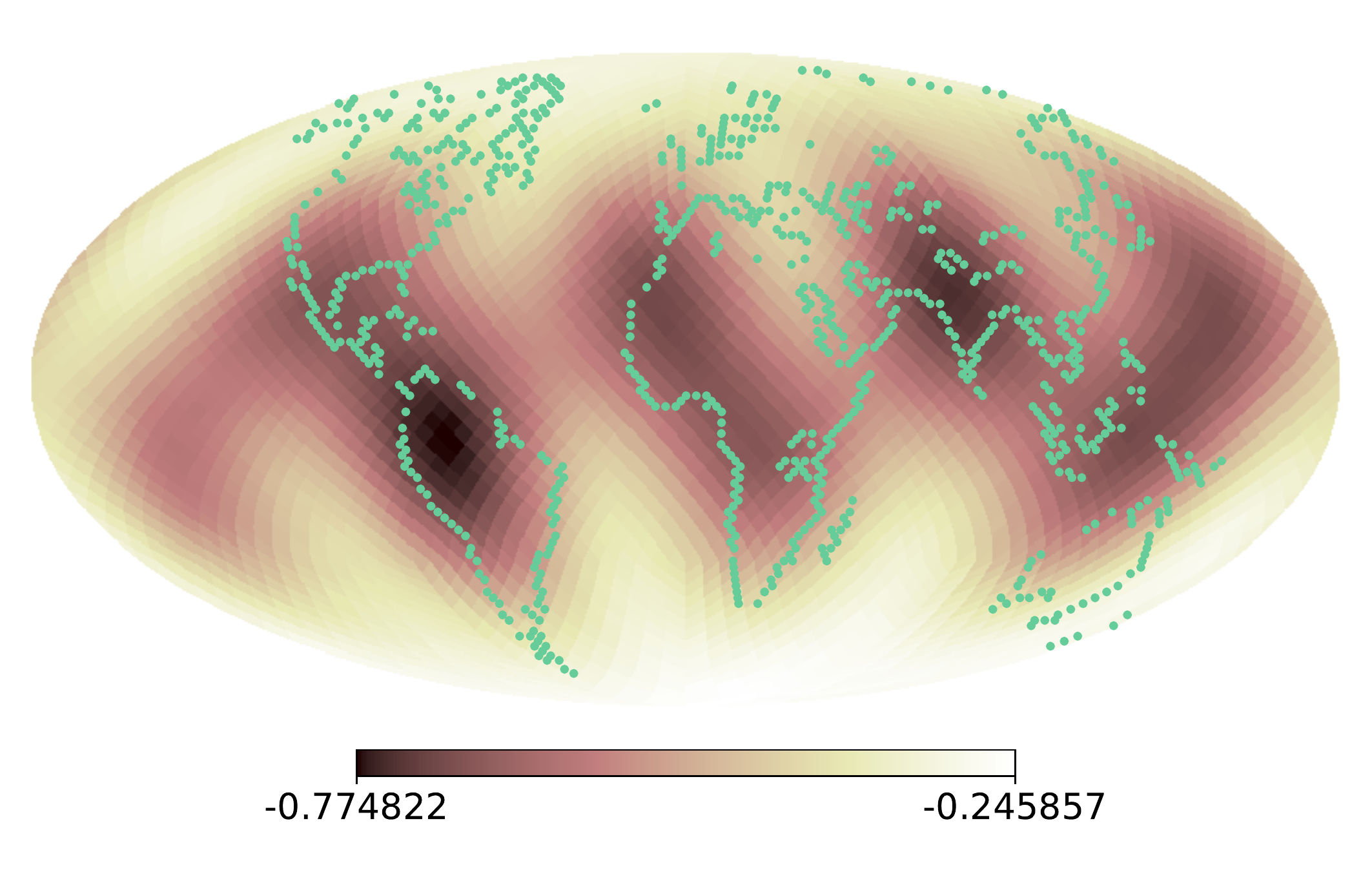}
    \includegraphics[width=0.7\linewidth]{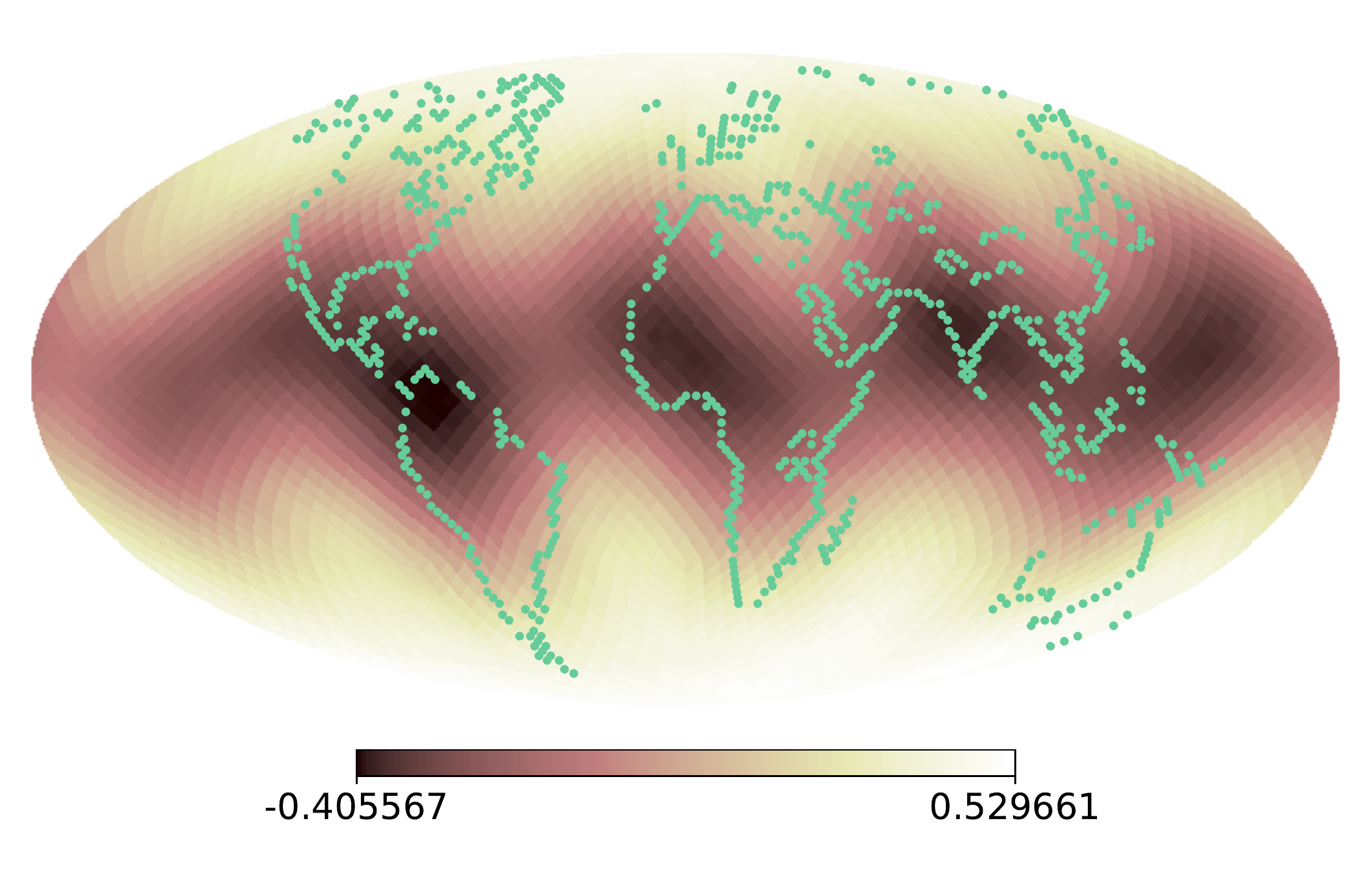}
 \end{center}
 \caption{\rev{The 5, 50, and 95 \% percentile maps (from top to bottom) for September in 2019, corresponding to the right column in Figure \ref{fig:dysotdscovr}.  }
 \label{fig:credible_dscovr}}
\end{figure}

The bottom panel in Figure \ref{fig:predata} shows the posterior prediction from our model, which matches well with the data.
Figure \ref{fig:corner_dyRBF_dscovr} shows a corner plot for the parameters aside from the geography. The time scale of the dynamic map ($\tau$) is found to be 3--4 weeks. 
Figure \ref{fig:dysotdscovr} shows the inferred snapshots (middle and bottom rows) at three differnet dates, along with the observed 8-day mean cloud fraction (top) provided by Moderate Resolution Imaging Spectroradiometer \citep[MODIS;][]{platnick2003modis}. These snapshots capture some of the temporal features despite the limited spatial resolution: For instance, the rainy (January; left) and \rev{clear-sky} (September; right) seasons in the Amazon (indicated by ``A'' in the top panels ) and the \rev{clear-sky} area in north America (indicated by ``B'' in the top panels) in May (middle), are captured. In particular, the overall cloud pattern is reproduced best in September (right) likely because the best viewing geometry is achieved at the equinox. Because the temporal resolution is largely from the spin rotataion,
patterns in the snapshot are elongated along the latitudial direction. This elongation makes it impossible to retrieve the narrow cloudy band at the equator known as the Intertropical Convergence Zone. 
\rev{Figure \ref{fig:credible_dscovr} shows the 5, 50, 95 \% percentile maps for the snapshot in September 2016, corresponding to the right panel of Figure \ref{fig:dysotdscovr}. 
These maps assure that some of the global features in the reconstructed maps are not statistical fluctuations but reflect the actual cloud distributions. For instance, the clear-sky region indicated by ``A'' in the right panel of Figure \ref{fig:dysotdscovr} appears in all the percentile maps.
}

\section{Summary and Discussion}

In this study, we developed a retrieval method of the time-varying geography from a single component reflected light curve of a directly imaged exoplanet.
We tested this method, which we call dynamic spin-orbit tomography,
using a toy model of an Earth-like planet with drifting continents
and found that the time-varying geogpraphy is accurately retrieved from dynamic mapping
and that the axial-tilt and the spin rotation period are well recovered.
We also demonstrated the dynamic spin-orbit tomography by applying it to the real light curve of the PC1 observed by DSCOVR and found that the dynamic map 
roughly captured
the largerst structures of time-varying clear-sky and cloudy areas.

The other findings and idea presented are summarized as follows.
\begin{enumerate}
    \item The marginal distribution of the spin parameters and the hyperparameter $p(\thetaGP, \gvv|\dv)$ were analytically derived for both static and dynamic mapping.
    \item The analytic solution of the Bayesian inverse problem allows us to sample from the marginal posterior distribution of the geography $p(\av|\dv)$ without directly sampling from $\av$.
    \item We numerically showed that spin-orbit tomography is able to distinguish the prograde rotation from the retrograde rotation, which has mostly been overlooked since \cite{2010ApJ...720.1333K}, except for in the theoretical analysis by \cite{2016MNRAS.457..926S}.
\end{enumerate}

In this paper, we considered a single component light curve as the data vector. The next step is to develop a multicolor version of the dynamic mapping. For instance, \cite{Kawahara2020} proposed spin-orbit unmixing, which disentangles spectral and geometric information from the multi-color light curve using the non-negative matrix factorization. The author derived an algorithm for the point estimate of the geography with L2 regularization and spectral components with volume regularization. A dynamic version of the spin-orbit unmixing is one of the potential solutions for multicolor dynamic mapping.

\acknowledgements

The authors would like to thank to the DSCOVR team for making the data publicly available. We are also grateful to Siteng Fan and Yuk L. Yung for providing the processed light curves and their geometric kernel from the DSCOVR dataset. The authors thank Masataka Aizawa for the fruitful discussions on the Bayesian statistics. We also appreciate the discussion with Joel Schwartz for regarding the prograde and retrograde rotation. \rev{We would also like to thank an anonymous reviewer for insightful comments. } This study was supported by JSPS KAKENHI grant nos. JP18H04577, JP18H01247, and JP20H00170 (H.K.). In addition, this study was supported by the JSPS Core-to-Core Program Planet2 and SATELLITE Research from the Astrobiology center (AB022006). H.K. would like to thank Niiharu Yokohama Citizen Forest for providing the time and space under social distancing to allow content of this study to be reflected upon outside the university owing to the lockdown in Tokyo against the Covid-19 crisis.

\vspace{5mm}

\software{scikit-learn \citep{scikit-learn}, matplotlib \citep{Hunter:2007},  numpy \citep{2011CSE....13b..22V}, scipy \citep{scipy}, python3,  emcee \citep{2013PASP..125..306F}, HealPix \citep{2005ApJ...622..759G}, corner \citep{corner}}
\appendix

\section{Bayesian Inverse Problem with a Gaussian Prior}\label{ap:bayes}

In this Appendix, we first consider the Bayesian linear inverse problem
\begin{eqnarray}
\dv = W \av
\end{eqnarray}
with Gaussian covariances described through a multivariate normal distribution,
\begin{eqnarray}
\label{eq:mvgap}
\Ng (\xv|\muv,\Sigma) = \frac{1}{(2 \pi)^{N/2} (\det{\Sigma})^{1/2}} e^{ - \frac{1}{2} (\xv - \muv)^T \Sigma^{-1} (\xv - \muv)}.
\end{eqnarray}
In other words, we assume the likelihood given by
\begin{eqnarray}
\label{eq:LIlikeli}
p(\dv|\av) &=& \Ng(\dv|W \av, \Sigma_{\dv})
\end{eqnarray}
and the model prior expressedhttps as  
\begin{eqnarray}
\label{eq:LIprior}
p(\av) &=& \Ng(\av|{\bf 0}, \Sigma_{\av}),
\end{eqnarray}
where $\dv$ is the data vector, $\av$ is the model vector, and $\Sigma_{\dv}$ and $\Sigma_{\av}$ are the covariance matrices of the data and model, respectively. We also define the precision matrices by
\begin{eqnarray}
\label{eq:precmat}
\Pi_{\dv} &\equiv& \Sigma_{\dv}^{-1} \\
\Pi_{\av} &\equiv& \Sigma_{\av}^{-1}.
\end{eqnarray}

\subsection{Maximum a Posteriori}\label{apss:dervatami}

Bayes' Theorem provides a posterior distribution of the model as 
\begin{eqnarray}
\label{eq:Byaespr}
p(\av|\dv) = \frac{p(\dv|\av) p(\av)}{p(\dv)} \propto e^{-\frac{1}{2} Q(\av)}
\end{eqnarray}
where, $Q(p)$ is the cost function
\begin{eqnarray}
\label{eq:minim_tildeap}
Q(\av) &=& (\dv - {W} {\av})^T \Pi_{\dv} (\dv - W \av) + {\av}^T {\Pi_{\av}} {\av} \\
&=& \av^T (W^T \Pi_{\dv} W + \Pi_{\av}) \av -2 \av^T W^T \Pi_{\dv} \dv + c,
\end{eqnarray} 
where $c$ is a constant term for $\av$.

A maximum a posteriori (MAP) is defined by the model that maximizes the posterior distribution. This is equivalent to the maximization of $Q$.
Equating the derivative of $Q$ by $\av$ to be zero, we obtain the MAP solution as
\begin{eqnarray}
\label{eq:normalposidef}
\av^\mathrm{MAP}  =  ( W^T \Pi_{\dv} W + \Pi_{\av} )^{-1} W^T \Pi_{\dv} \dv.
\end{eqnarray}

\subsection{Posterior Distribution} \label{apss:bayes}

We start from the multivariate Gaussian distribution of Equation (\ref{eq:mvgap}). Expanding the negative logarithm of Equation (\ref{eq:mvgap}) for $\av$, we obtain the following:
\begin{eqnarray}
\label{eq:gp}
- 2 \log{ \Ng(\av|\muv, \Sigma) } = (\av - \muv)^T \Sigma^{-1} (\av - \muv) = \av^T \Sigma^{-1} \av - 2 \av^T \Sigma^{-1} \muv + \mathrm{const}. 
\end{eqnarray}
If we can express the negative logarithm of the Gaussian distribution as
\begin{eqnarray}
\label{eq:gpqu}
- 2 \log{ p(\av) } = \av^T P \av - 2 \av^T \qv + \mathrm{const},
\end{eqnarray}
then, compared with Equation (\ref{eq:gp}), we obtain 
\begin{eqnarray}
\label{eq:gpqusol}
p(\av) = \Ng(\av| P^{-1} \qv ,P^{-1}).
\end{eqnarray}

We consider the posterior distribution $p(\av|\dv) \propto p(\dv|\av) p(\av) $, defined by the linear model with the Gaussian process of Equations (\ref{eq:LIlikeli}) and (\ref{eq:LIprior}). The negative logarithm of the posterior is proportional to the cost function of Equation (\ref{eq:minim_tildeap}).
Using Equation (\ref{eq:gpqusol}), we obtain the posterior distribution as 
\begin{eqnarray}
\label{eq:pposte}
p(\av|\dv) &=& \Ng(\av| \muv, \Sigma_{a|d}) \\
\muv &=& ( W^T \Pi_{\dv} W + \Pi_{\av} )^{-1} W^T \Pi_{\dv} \dv \\
\Sigma_{a|d} &=& (W^T \Pi_{\dv} W + \Pi_{\av})^{-1}.
\end{eqnarray}
It can be seen that the MAP solution is identical to the mean of the posterior, i.e. $\av^\mathrm{MAP}  = \muv$ \citep{tarantola}.


\subsection{\rev{Posteriors with Nonlinear Parameters}}\label{ss:evidence}

We then introduce nonlinear parameters $\thetaGP$ and $\gvv$ and compute the marginal likelihood (evidence) for the nonlinear parameters. The evidence $p(\dv|\thetaGP,\gvv)$ obeys a multivariate normal distribution for the Gaussian process with a linear transformation. To derive the explicit expression of the evidence, we start from Bayes' theorem 
\begin{eqnarray}
p(\av|\dv,\thetaGP,\gvv) &=& \frac{p(\dv|\av,\thetaGP,\gvv) p(\av|\thetaGP,\gvv)}{p(\dv|\thetaGP,\gvv) } = \frac{p(\dv|\av,\gvv) p(\av|\thetaGP)}{p(\dv|\thetaGP,\gvv) }.
\end{eqnarray}
The negative logarithm of the evidence is expressed as
\begin{eqnarray}
&\,&-2 \log{p(\dv|\thetaGP, \gvv)} = -2 \log{p(\dv|\av,\gvv)} + 2 \log{p(\av|\dv, \thetaGP, \gvv)} + c \\
&\,& = (\dv - W \av)^T \Pi_{\dv} (\dv - W \av) \nonumber \\ 
&\,&- [\av - (W^T \Pi_{\dv}  W + \Pi_{\av})^{-1} W^T \Pi_{\dv} \dv]^T \left( W^T \Pi_{\dv} W + \Pi_{\av} \right) [\av - (W^T \Pi_{\dv}  W + \Pi_{\av})^{-1} W^T \Pi_{\dv} \dv] + c \\
&\,& = \dv^T \left[ \Pi_{\dv} - \Pi_{\dv} W (W^T \Pi_{\dv} W + \Pi_{\av})^{-1} W^T \Pi_{\dv} \right] \dv  + c\\
\label{eq:minimK}
&\,& = \dv^T (\Sigma_{\dv} + W \Sigma_{\av} W^T)^{-1} \dv + c 
\end{eqnarray}
where we used the Woodbury matrix identity of Equation (\ref{eq:Woodbury}) for the last transformation. Comparing with Equations (\ref{eq:minimK}) and (\ref{eq:pposte}), we obtain the following:
\begin{eqnarray}
\label{eq:anaevigen}
p(\dv|\thetaGP,\gvv) = \Ng(\dv|{\bf 0}, \Sigma_{\dv} + W \Sigma_{\av} W^T)
\end{eqnarray}

\rev{Using Equation (\ref{eq:anaevigen}), 
 an MCMC algorithm can sample the sets of ($\thetaGP$, $\gvv$) based on Bayes' Theorum, 
\begin{eqnarray}
 p(\thetaGP, \gvv|\dv) \propto p(\dv|\thetaGP, \gvv) p(\thetaGP, \gvv).
\end{eqnarray}
Denoting the $n$-th sample of the hyperparameter and the spin parameters by $\thetaGPdag$ and $ \gvvdag$ from $ p(\thetaGP, \gvv|\dv) $, one can also infer the marginal posterior of the geography as
\begin{eqnarray}
p(\av|\dv) &=& \int d \thetaGP \int d \gvv \, p(\av, \thetaGP, \gvv|\dv) 
= \int d \thetaGP \int d \gvv \, p(\av|\dv,\thetaGP,{\gvv}) p(\thetaGP, \gvv|\dv)  \\
&\approx& \frac{1}{N_s} \sum_{n=0}^{N_s-1} p(\av|\dv,\thetaGPdag, {\gvvdag}),
\end{eqnarray}
where $N_s$ is the number of the samples.
}

\subsection{Optimization of Evidence for dynamic spin-orbit tomography}

An optimization of the evidence is useful before applying the time-consuming MCMC.  We show the derivative of the negative logarithm of the evidence of Equation (\ref{eq:marginallikelidy}) for a dynamic spin-orbit tomography,  
\begin{eqnarray}
- 2 \log{p(\dv|\thetaGP,\gvv,\phiGP)} = \log{\det{(\Sigma_{\dv} + K_W)}} + \dv^T (\Sigma_{\dv} + K_W)^{-1} \dv + c.
\end{eqnarray}
Here we also consider the hyperparameter for the data covariance $\phiGP$ as well as $\thetaGP, \gvv$. Instead of the direct use of such evidence, we use the cost function,
\begin{eqnarray}
\mathcal{L} \equiv \log{\det{(\Sigma_{\dv} + K_W)}} +  \dv^T (\Sigma_{\dv} + K_W)^{-1} \dv.
\end{eqnarray}
to optimize the evidence. Unfortunately, $\partial \mathcal{L} /\partial \gvv$ is intractable. We only consider the derivative of $\mathcal{L}$ using $\thetaGP$ and $\phiGP$.

We minimize $\mathcal{L}$ to achieve the maximum evidence.  Using the relations 
\begin{eqnarray}
\frac{\partial}{\partial \thetaGP} \log{\det{(\Sigma_{\dv} + K_W)}} = \mathrm{tr} \left[ (\Sigma_{\dv} + K_W)^{-1} \frac{\partial K_W}{\partial \thetaGP} \right] \\
\frac{\partial}{\partial \thetaGP} (\Sigma_{\dv} + K_W)^{-1} = - (\Sigma_{\dv} + K_W)^{-1} \frac{\partial K_W}{\partial \thetaGP} (\Sigma_{\dv} + K_W)^{-1},
\end{eqnarray}
we obtain the derivative of $\mathcal{L}$ by $\thetaGP$ as
\begin{eqnarray}
\frac{\partial  \mathcal{L}}{\partial \thetaGP} &=& \mathrm{tr} \left[ (\Sigma_{\dv} + K_W)^{-1} \frac{\partial K_W}{\partial \thetaGP} \right]  - \yv^T \frac{\partial K_W}{\partial \thetaGP} \yv \\
\yv &\equiv& (\Sigma_{\dv} + K_W)^{-1} \dv.
\end{eqnarray}
The term of $\partial K_W/\partial \thetaGP$ depends on the GP kernel, the derivative of which is usually tractable. 

We take the hyperparameters as a form of $\acute{p} = \log{p}$ to ensure the nonnegativity. The derivative of the Mat\'{e}rn -3/2 kernel by $\acute{\tau}$ is given in the following:
\begin{eqnarray}
\label{eq:dMaternA}
 \frac{\partial}{\partial \acute{\tau}} k_\mathrm{M3/2}(t_i,t_{i^\prime};\acute{\tau}) =  3 |t_i - t_{i^\prime}|^2  e^{- \sqrt{3} |t_i - t_{i^\prime}| \exp{(-\acute{\tau})} - 2\acute{\tau} }.
\end{eqnarray}
In addition, the derivative of the RBF kernel by $\acute{\gamma}$ is given by
\begin{eqnarray}
\label{eq:dMaternA}
 \frac{\partial}{\partial \acute{\gamma}} k_\mathrm{RBF}(\eta_{jj^\prime};\acute{\gamma}) = \eta_{jj^\prime}^{2} e^{- 2\acute{\gamma} - \eta_{jj^\prime}^{2} \exp{(- 2 \acute{\gamma})}/2}
\end{eqnarray}
 The derivative of $K_W$ by $\thetaGP = ({\acute{\gamma},\acute{\alpha}},\acute{\tau})^T$ is given by the following:
\begin{eqnarray}
\label{eq:partKw}
\frac{\partial K_W}{\partial \thetaGP} = \left(
    \begin{array}{c}
      \frac{\partial }{\partial \acute{\gamma}} \\
           \frac{\partial}{\partial \acute{\alpha}} \\
           \frac{\partial}{\partial \acute{\tau}} 
   \end{array} \right) K_W
 = \left(
    \begin{array}{c}
     K_T \odot \left(W \left[ \alpha \frac{\partial}{\partial \acute{\gamma}}  k(\eta_{jj^\prime};\acute{\gamma}) \right] W^T \right) \\
     K_W \\ 
  \alpha \frac{\partial}{\partial \acute{\tau}} k(|t_i-t_{i^\prime}|;\acute{\tau})  \odot W K_S W^T 
  \end{array} \right) 
\end{eqnarray}


Likewise, we obtain the derivative of $\mathcal{L}$ using $\phiGP$ as follows:
\begin{eqnarray}
\frac{\partial  \mathcal{L}}{\partial \phiGP} &=& \mathrm{tr} \left[ (\Sigma_{\dv} + K_W)^{-1} \frac{\partial \Sigma_{\dv}}{\partial \phiGP} \right]  - \yv^T \frac{\partial \Sigma_{\dv}}{\partial \phiGP} \yv 
\end{eqnarray}
In the independent Gaussian case of $\Sigma_d = \sigma^2 I$ and $\phiGP=\acute{\sigma}$, where $\acute{\sigma}= 2 \log{\sigma}$, we obtain $\partial \Sigma_d/\partial \phiGP = \Sigma_d$.

\section{Matrix Identities}\label{ss:matriidentity}

We next derive the matrix identities we use in this study. We start from the famous Woodbury matrix identity,
\begin{eqnarray}
\label{eq:Woodbury}
(Z + UYV)^{-1} = Z^{-1} - Z^{-1} U ( Y^{-1} + V Z^{-1} U )^{-1} V Z^{-1}. 
\end{eqnarray}
Adopting $Z=I$, we obtain the following relation:
\begin{eqnarray}
\label{eq:IWoodbury}
(I + UYV)^{-1} = I -  U ( Y^{-1} + V U )^{-1} V. 
\end{eqnarray}
Alternatively, adopting $Y=I$ into the Woodbury matrix identity of Equation (\ref{eq:Woodbury}), we obtain the Kailath variant \citep{kailath1980linear}, which is expressed as follows:
\begin{eqnarray}
\label{eq:Kailathap}
(Z + UV)^{-1} = Z^{-1} - Z^{-1} U ( I + V Z^{-1} U )^{-1} V Z^{-1}. 
\end{eqnarray}
In addition, $Z=I$ yields the following relation
\begin{eqnarray}
\label{eq:UV}
(I + UV)^{-1} = I - U (I + VU)^{-1} V.
\end{eqnarray}
Adopting $U=I$ or $V=I$, we have the following two identities.
\begin{eqnarray}
\label{eq:UVleft}
(I + V)^{-1} &=& I - (I + V)^{-1} V \\
\label{eq:UVright}
(I + U)^{-1} &=& I - U (I + U)^{-1}.
\end{eqnarray}

\section{Isomorphism of Kernel Expansion and Re-contraction}\label{eq:isom}

We consider the isomorphism of $\mathbb{R}^{N_i \times N_j} \cong \mathbb{R}^{N_i N_j}$ for a matrix and its vectorization. In general, we denote the tensor reshaping operator of $A$ as $\reshape^{(p \to q)} (A)$, where $p$ and $q$ are the shapes of the tensors before and after reshaping, respectively. We can express the vectorization of a matrix $A$ and its inverse as follows:
\begin{eqnarray}
\label{eq:reshape}
\mathrm{vec} (A) &=& \reshape^{(N_i \times N_j \to N_i N_j)} (A) = {\av} \in \mathbb{R}^{N_i N_j}\\
\mathrm{mat} (\av) &=& \reshape^{(N_i N_j \to N_i \times N_j )} (\av) = A \in \mathbb{R}^{N_i \times N_j}
\end{eqnarray}
We define the corresponding linear operator $\tilde{W}$ for $\av$ to the operator $\psiv$ for $A$ as
\begin{eqnarray}
\label{eq:equality}
\psiv(W,A) = \tilde{W} \av
\end{eqnarray}
where $\psiv = \psiv (W,A)$ is the operator that indicates $\psi_i = \sum_j W_{ij} A_{ij}$. 
In the matrix form, we can express $\tilde{W}$ as
\begin{eqnarray}
\tilde{W} &=& \left(
    \begin{array}{cccc}
      \mathcal{D}{(\wvj_0)} & \mathcal{D}{(\wvj_1)} & \cdots & \mathcal{D}{(\wvj_{N_j-1})} \\
    \end{array}
  \right) 
  \in \mathbb{R}^{N_i \times N_i N_j}  \nonumber \\
\end{eqnarray} 
where $\wvj_j$ is the column vector of the column of $W$, and $\mathcal{D}{(\wvj_j)}$ is the operator that makes a diagonal matrix from a vector $\wvj_j$, that is, $\acute{W} = \mathcal{D}{(\wvj_j)}$ indicates $\acute{W}_{ij} = \delta_{ij} \wvj_j $ ($\delta_{ij}$ is the Kronecker delta). 

The equality of Equation (\ref{eq:equality}) can also be expressed as
\begin{eqnarray}
\label{eq:equality_a}
\psi_i = \sum_j W_{ij} A_{ij} = \sum_{J} \tilde{W}_{iJ}  (\mathrm{vec} (A))_J
\end{eqnarray}
where $J$ is an unfold index of $(i,j)$. 

From Equation (\ref{eq:equality_a}), the linear equality for the tensor $\mathcal{A}$ is given by 
\begin{eqnarray}
\label{eq:equality_atens}
 \sum_j W_{ij} \mathcal{A}_{ijk} = \sum_{J} \tilde{W}_{iJ} A_{Jk} 
\end{eqnarray}
where 
\begin{eqnarray}
\label{eq:atensor}
 A = \reshape^{(N_i \times N_j \times N_k \to N_i N_j \times N_k)} (\mathcal{A}).
\end{eqnarray}

\subsection{Re-contraction Formula 1: $ \tilde{W} (S \otimes T) \tilde{W}^T = T \odot (W S W^T)$}\label{ap:recont1}
We prove the re-contraction formula 1 used in Equation (\ref{eq:wkerneldy}) as follows:
\begin{eqnarray}
\label{eq:wstw}
   \tilde{W} (S \otimes T) \tilde{W}^T = T \odot (W S W^T),
\end{eqnarray}
where, we define the matrices $S \in \mathbb{R}^{N_j \times N_j}$ and $T \in \mathbb{R}^{N_i \times N_i}$, and where $\otimes$ is the Kronecker product and $\odot$ is the Hadamard product. The  $ii'$ element of the righthand side denoted by $Y_{i\ip}$ is written using  a tensor of $\mathcal{P} \in \mathbb{R}^{N_i \times N_j \times N_{\ip} \times N_{\jp}}$, the element of which is given by $\mathcal{P}_{ij\ip\jp} \equiv S_{j\jp} T_{i\ip}$ as follows:
\begin{eqnarray}
\label{eq:Yiip}
 Y_{i\ip} = T_{i\ip} \sum_{j,\jp} W_{ij} S_{j\jp} W_{\ip\jp} &=& \sum_{j} W_{ij} \sum_{\jp} \mathcal{P}_{ij\ip\jp} W_{\ip\jp} = \sum_{j} {W}_{ij} \mathcal{Q}_{ij\ip} 
\end{eqnarray}
where 
\begin{eqnarray}
\mathcal{Q}_{ij\ip} &\equiv& \sum_{\jp} \mathcal{P}_{ij\ip\jp} W_{\ip\jp}.
\end{eqnarray}

Adopting the relation of Equation (\ref{eq:equality_atens}) to Equation (\ref{eq:Yiip}) with the reshaped matrix, we have
\begin{eqnarray}
Q &=& \reshape^{(N_i \times N_j \times N_{\ip} \to N_i N_j \times N_{\ip})} (\mathcal{Q}) = \sum_{\jp} \mathcal{P}^\ast_{J\ip\jp} W_{\ip\jp} = \sum_{\jp} \mathcal{P}^\ast_{J\ip\jp} W^T_{\jp\ip} = \sum_{\Jp} \mathcal{P}^{\ast\ast}_{J\Jp} \tilde{W}^T_{\Jp \ip}
\end{eqnarray}
where 
\begin{eqnarray}
\mathcal{P}^\ast &=& \reshape^{(N_i \times N_j \times N_{\ip} \times N_{\jp} \to N_i N_j \times N_{\ip} \times N_{\jp})} (\mathcal{P})\\
\mathcal{P}^{\ast\ast} &=&  \reshape^{(N_i N_j \times N_{\ip} \times N_{\jp} \to N_i N_j \times N_{\ip} N_{\jp})} (\mathcal{P}^\ast) \\
&=& \reshape^{(N_i \times N_j \times N_{\ip} \times N_{\jp} \to N_i N_j \times N_{\ip} N_{\jp})} (\mathcal{P}) = S \otimes T,
\end{eqnarray}
and thus we obtain 
\begin{eqnarray}
 Y_{i\ip} = \sum_{j} {W}_{ij} \mathcal{Q}_{ij\ip} = \sum_{J} \tilde{W}_{iJ} Q_{J\ip} =
\sum_{J} \tilde{W}_{iJ} \sum_{\Jp} P^{\ast\ast}_{J\Jp} \tilde{W}^T_{\Jp\ip}  = \sum_{J,\Jp} \tilde{W}_{iJ} (S \otimes T)_{J\Jp} \tilde{W}^T_{\Jp\ip}. 
\end{eqnarray}
In the matrix form, we can find the relation of Equation (\ref{eq:wstw}).

\subsection{Recontraction Formula 2: $(S \otimes T) \tilde{W}^T \xv = \mathrm{vec} ( T \mathcal{D}(\xv) W S^T ) $ }\label{ap:recont2}

Adopting $V = S^T$, $U = T$, and $X = \mathrm{mat} (\tilde{W}^T \xv) $ into the well-known relation of 
\begin{eqnarray}
 (V^T \otimes U) \mathrm{vec} (X) = \mathrm{vec} (UXV),
\end{eqnarray}
we obtain 
\begin{eqnarray}
\label{eq:idenetikr}
(S \otimes T) \tilde{W}^T \xv = \mathrm{vec} (T \, \mathrm{mat} (\tilde{W}^T \xv) \, S^T).
\end{eqnarray}
Considering the expression of 
\begin{eqnarray}
\tilde{W}^T \xv =\left(
    \begin{array}{c}
       \wvj_0 \odot \xv \\
       \wvj_1 \odot \xv  \\
      \vdots \\
       \wvj _{N_j-1} \odot \xv  
    \end{array}
  \right) \in \mathbb{R}^{N_i N_j},
\end{eqnarray}
we find that the matrix form of $X$ is given by the following:
\begin{eqnarray}
\mathrm{mat} (\tilde{W}^T \xv) &=& \left(
    \begin{array}{cccc}
       \wvj_0 \odot \xv, &  \wvj_1 \odot \xv,  & \cdots , &  \wvj _{N_j-1} \odot \xv  \\
    \end{array}
  \right) 
=  \mathcal{D}(\xv) W 
\end{eqnarray} 
\rev{Equation (\ref{eq:idenetikr}) then yields
\begin{eqnarray}
(S \otimes T) \tilde{W}^T \xv = \mathrm{vec} ( T \mathcal{D}(\xv) W  S^T ).
\end{eqnarray}
}
\subsection{$\mathcal{S}$  and $\mathcal{T}$  extractors}\label{ap:tsext}
The $\mathcal{S}$ extractor is used to extract the covariance for the $i$-th snapshot from the full covariance matrix. The $\mathcal{S}$ extractor generates a matrix $\in \mathbb{R}^{N_j \times N_j}$ consisting of the elements of the product of $T_{ii}$ from $S \otimes T \in \mathbb{R}^{N_i N_j \times N_i N_j}$ so that
\begin{eqnarray}
   \mathcal{S}_i(S \otimes T) &= T_{ii} S.
\end{eqnarray} 
This indicates that $\mathcal{S}_i(X)$ extracts the elements whose indices are $\boldsymbol{J}_i \otimes \boldsymbol{J}_i^T \in \mathbb{R}^{N_j \times N_j}$ from $X \in \mathbb{R}^{N_i N_j \times N_i N_j}$, where $\boldsymbol{J}_i = (i, i + N_i, i + 2 N_i, \cdots, i + (N_j - 1) N_i )^T$.  Likewise, the $\mathcal{T}$  extractor generates a matrix consisting of the elements of the product of $S_{jj}$ from $S \otimes T$ as
\begin{eqnarray}
   \mathcal{T}_j(S \otimes T) &= S_{jj} T,
\end{eqnarray} 
 indicating that $\mathcal{T}_j(X)$ extracts the elements whose indices are $\boldsymbol{I}_j \otimes \boldsymbol{I}_j^T \in \mathbb{R}^{N_i \times N_i}$ from $X$, where $\boldsymbol{I}_j = (j N_i, 1 + j N_i, \cdots, (N_i-1) + j N_i)^T$.

Then, let us consider $\mathcal{S}_i (Y)$ and $\mathcal{T}_j (Y)$ for 
\begin{eqnarray}
   Y &=& Z^T P Z \\
   Z &\equiv& \tilde{W} (S \otimes T) \in \mathbb{R}^{N_i \times N_i N_j}
\end{eqnarray}
where $P$ is a square matrix. The element of $Y$ is given by
\begin{eqnarray}
   Y_{J \Jp} = \hat{\zv}_J^T P \hat{\zv}_{\Jp},
\end{eqnarray}
where $\hat{\zv}_J$ is the column vector of the column of $Z$. Because $Z$ can be expressed as
\begin{eqnarray}
   Z = \left(
    \begin{array}{cccc}
       Z^\prime[0]  Z^\prime[1] \cdots  Z^\prime[N_j-1]\\
    \end{array}
  \right) 
\end{eqnarray}
where
\begin{eqnarray}
Z^\prime[j] = \sum_k S_{kj} \mathcal{D} (\wvj_k) T 
\end{eqnarray} 
we obtain 
\begin{eqnarray}
   \hat{\zv}_{i + j N_j}^T P  \hat{\zv}_{\ip + \jp N_j} &=& \sum_{l l^\prime} \left[ \left(\sum_k S_{kj} W_{lk} T_{li} \right) P_{l l^\prime}  \left(\sum_k S_{k\jp} W_{l^\prime k} T_{l^\prime \ip} \right) \right] \\
   &=& \sum_{l l^\prime} \left\{ \left[ \left(\sum_k W_{lk} S_{kj}\right) T_{li} \right] P_{l l^\prime}  \left[ \left(\sum_k  W_{l^\prime k} S_{k\jp}\right) T_{l^\prime \ip} \right] \right\}
\end{eqnarray}
Then, extracting ($\boldsymbol{I}_i \times \boldsymbol{I}_i$) from $Y$, the S extractor is expressed as follows:
\begin{eqnarray}
\mathcal{S}_i(Y) = [ \mathcal{D}(\tvi_i) W S ]^T P [ \mathcal{D}(\tvi_i) W S ],
\end{eqnarray}
where $\tvi_i$ is the column vector of the column of $T$. Likewise, extracting ($\boldsymbol{J}_j \times \boldsymbol{J}_j$) from $Y$, we also obtain
\begin{eqnarray}
 \mathcal{T}_j(Y) = (\mathcal{D} (\uvi_j) T )^T P (\mathcal{D} (\uvi_j) T),
\end{eqnarray}
where $\uvi_j$ is the column vector of the column of $W S$. 

\rev{
Note that $\mathcal{S}$  and $\mathcal{T}$  extractors are a linear operator, that is, $\mathcal{S}_i(X+Y) = \mathcal{S}_i(X) + \mathcal{S}_i(Y)$, $\mathcal{S}_i(\alpha X) = \alpha \mathcal{S}_i(X)$,  $\mathcal{T}_i(X+Y) = \mathcal{T}_i(X) + \mathcal{T}_i(Y)$, $\mathcal{T}_i(\alpha X) = \alpha \mathcal{T}_i(X)$ for matrices $X$ and $Y$ and a scalar value $\alpha$. Equation (\ref{eq:dysot_posterior_map_cov}) can be derived by applying $\mathcal{S}$ extractor to Equation (\ref{eq:kailath_expand}) as
\begin{eqnarray}
\mathcal{S}_i (\Sigma_{\av|\dv,\thetaGP,\gvv}) &=&  \alpha \mathcal{S}_i(K_S \otimes K_T) -  \mathcal{S}_i [K \tilde{W}^T ( \Sigma_{\dv} + K_W )^{-1}  \tilde{W} K] \\
&=& \alpha (K_T)_{ii} K_S -  [ \alpha \mathcal{D}(\tvi_i) W K_S ]^T ( \Sigma_{\dv} + K_W )^{-1} [ \alpha  \mathcal{D}(\tvi_i) W K_S ].
\end{eqnarray}
Likewise, Equation (\ref{eq:dysot_posterior_pix_cov}) is derived as 
\begin{eqnarray}
\mathcal{T}_j (\Sigma_{\av|\dv,\thetaGP,\gvv}) &=&  \alpha \mathcal{T}_j(K_S \otimes K_T) -  \mathcal{T}_j [K \tilde{W}^T ( \Sigma_{\dv} + K_W )^{-1}  \tilde{W} K] \\
&=& \alpha (K_S)_{jj} K_T -   [ \alpha  \mathcal{D} (\uvi_j) K_T ]^T ( \Sigma_{\dv} + K_W )^{-1}  [\alpha \mathcal{D} (\uvi_j) K_T].
\end{eqnarray}
}

\section{No degeneracy between prograde and retrograde rotations}\label{ap:proretro}

\begin{figure}[]
 \begin{center}
    \includegraphics[width=0.60\linewidth]{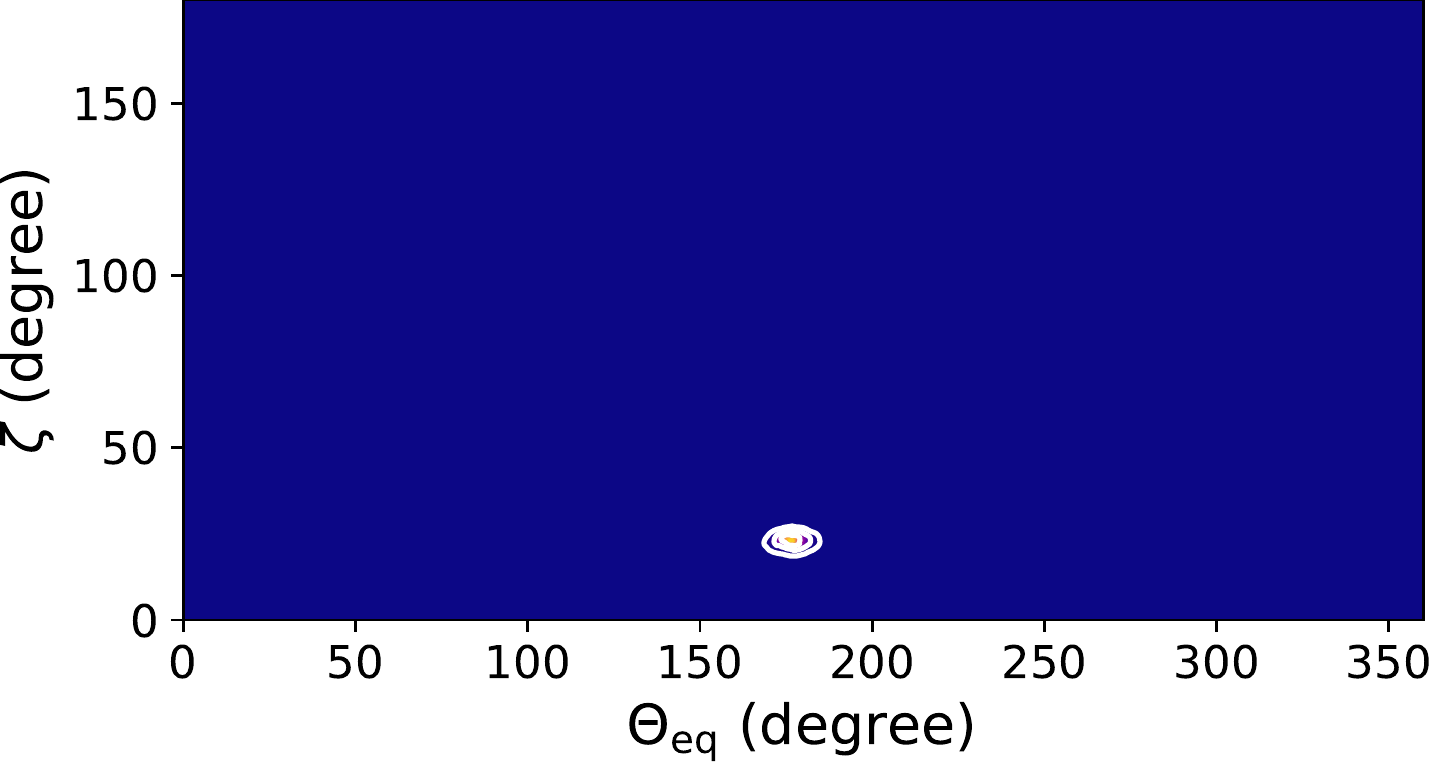}
    \includegraphics[width=0.35\linewidth]{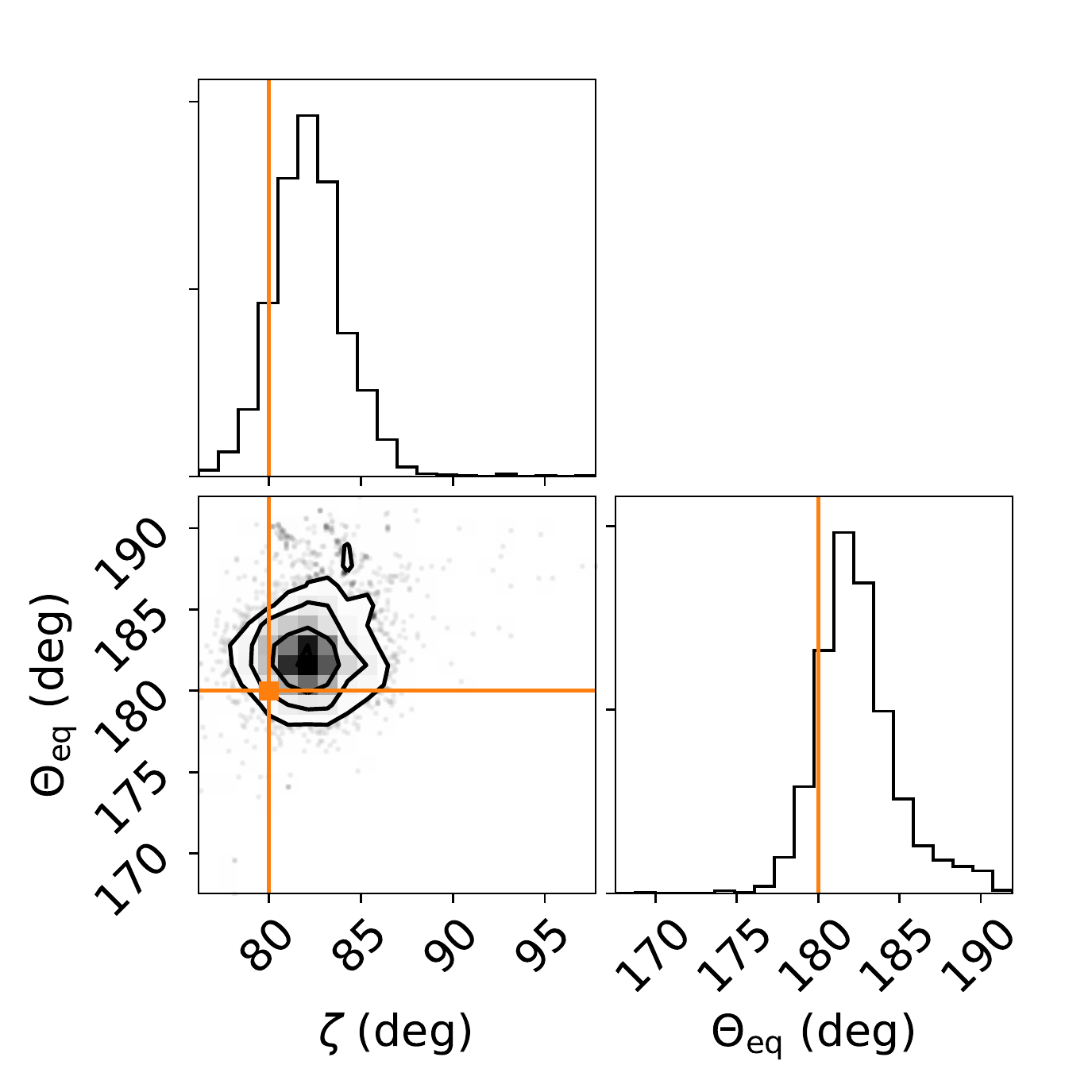}
 \end{center}
 \caption{The marginal distribution $p(\zeta,\Theta_\mathrm{eq}|\Pspin,\thetaGP)$ for the toy model (left). We adopt the mean values of $\thetaGP$ in Figure \ref{fig:corner_dyRBF} and the input value for $\Pspin$ in this panel. The three contours indicate 50\%, 95\%, and 99.5 \% encircled areas of the probability. The right panel shows the marginal distribution for another test, adopting $\zeta=80 ^\circ$ as the input value and 10 \% statistical noise. Both tests shows that there are no degeneracy between the prograde and the retrograde rotation. \label{fig:proret}}
 \end{figure}

In this section, we show that there is no degeneracy between the prograde ($\zeta \le 90^\circ$) and the retrograde ($\zeta \ge 90^\circ$) rotations in spin-orbit tomography. The reason why we discuss this issue in particular is that \cite{2010ApJ...720.1333K} restricted to the prograde case based on the misunderstanding by one of the authors (H.K.).
The related papers kept not considering the retrograde case \citep{2011ApJ...739L..62K,2012ApJ...755..101F,2018AJ....156..146F,2019AJ....158..246B} even after \cite{2016MNRAS.457..926S} correctly pointed the fact of the nondegeneracy from the theoretical argument of the geometric kernel. 

Because the discussion by \cite{2016MNRAS.457..926S} is the thoeretical argument, we here investigate the nondegeneracy of the prograde and retrograde rotations numerically. In general, it is insufficient to see the MCMC results to prove the nondegeneracy because the MCMC does not cover the overall parameter space. The full marginal distribution $p(\zeta,\Theta_\mathrm{eq})$ on the $\zeta$--$\Theta_\mathrm{eq}$ plane completely proves the nondenegenacy. However, it is quite time consuming to compute it. Instead, Figure \ref{fig:proret} shows the posterior distribution of $p(\zeta,\Theta_\mathrm{eq}|\Pspin,\thetaGP)$  on the $\zeta$--$\Theta_\mathrm{eq}$ plane, for the toy model in Section \ref{sec:toy}. This panel shows that there is almost no measure of the probability in the prograde area of $\zeta > 90^\circ$ at least for the fixed values of $\thetaGP$. Another test is given in the right panel of Figure \ref{fig:proret}. In this case, we used $\zeta=80^\circ$ and a larger statistical noise of the data (10 \% of the light curve). This is the resluts by MCMC, however, the retrograde solution ($\zeta = 100^\circ$) is close to the input one. So, the MCMC easily reaches the parameter space around the retrograde point. Despite, we find that there is almost no sampling points around the retrograde point.

\begin{figure}[]
 \begin{center}
    \includegraphics[width=0.5\linewidth]{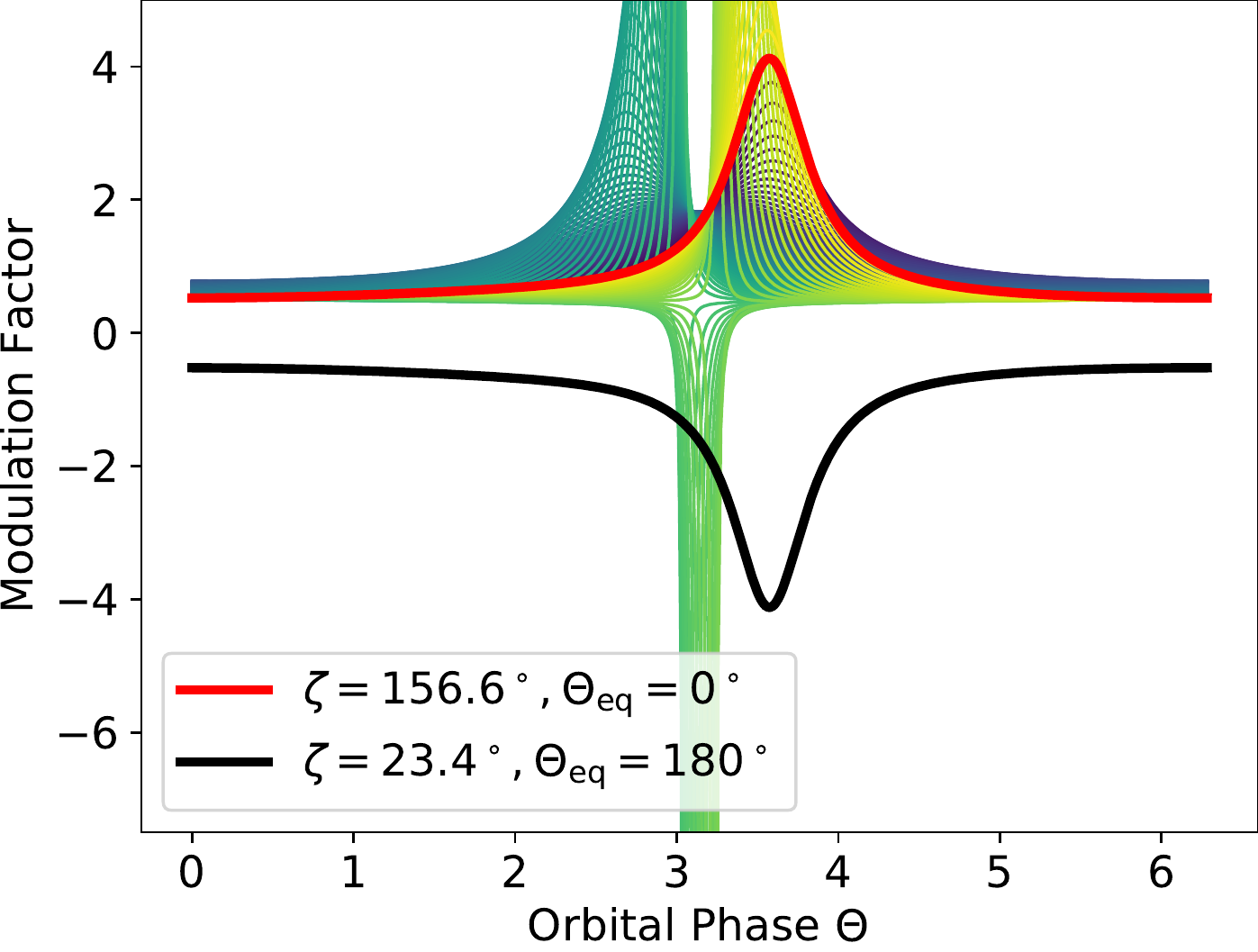}
 \end{center}
 \caption{ Modulation factor for a prograde case $\zeta=23.4^\circ$ and $\Theta_\mathrm{eq}=180^\circ$ (black) and retrograde cases ($\zeta=156.6^\circ$) for various $\Theta_\mathrm{eq}$ (thin lines). The red curve is the conjugate retrograde case for the prograde one ($\zeta=156.6^\circ$ and $\Theta_\mathrm{eq}=0^\circ$ ). The maximum weighted longitude approximation is applied to compute this panel \citep{2016ApJ...822..112K}. \label{fig:modfac}}
 \end{figure}

It might be intuitively a bit difficult to understand the nondeneracy from the perspective of the amplitude modulation. Here we try to explain the nondegeneracy from the point of the frequency modulation of the periodicity of the light curve.  \cite{2016ApJ...822..112K} showed that the instantaneous frequency curves of the prograde and the retrograde rotations are quite different for non-zero obliquity. The apparent instantaneous frequency is expressed as $f_\mathrm{obs} = f_\mathrm{spin} + \epsilon(\Theta) f_\mathrm{orb}$, where  $f_\mathrm{spin}$ and $f_\mathrm{orb}$ are the spin and orbit  rotation frequency, and $\epsilon(\Theta)$ is a modulation factor as a function of $\zeta$ and $\Theta_\mathrm{eq}$. In Figure \ref{fig:modfac}, we show the modulation factor for the prograde case $\zeta=23.4^\circ$ and $\Theta_\mathrm{eq}=180^\circ$ (black) and the retrograde cases ($\zeta=156.6^\circ$, thin lines) for different $\Theta_\mathrm{eq}$. Even for the conjugate retrograde case ($\zeta=156.6^\circ$ and $\Theta_\mathrm{eq}=0^\circ$ ) to the prograde one, one can see the different behavior of the modulation. This difference of the periodicity variation clearly shows the nondegeneracy between the prograde and retrograde rotations in the spin-orbit tomography.

\section{Randomized map}\label{ap:ranmap}

\begin{figure*}[]
 \begin{center}
  \includegraphics[width=0.32\linewidth]{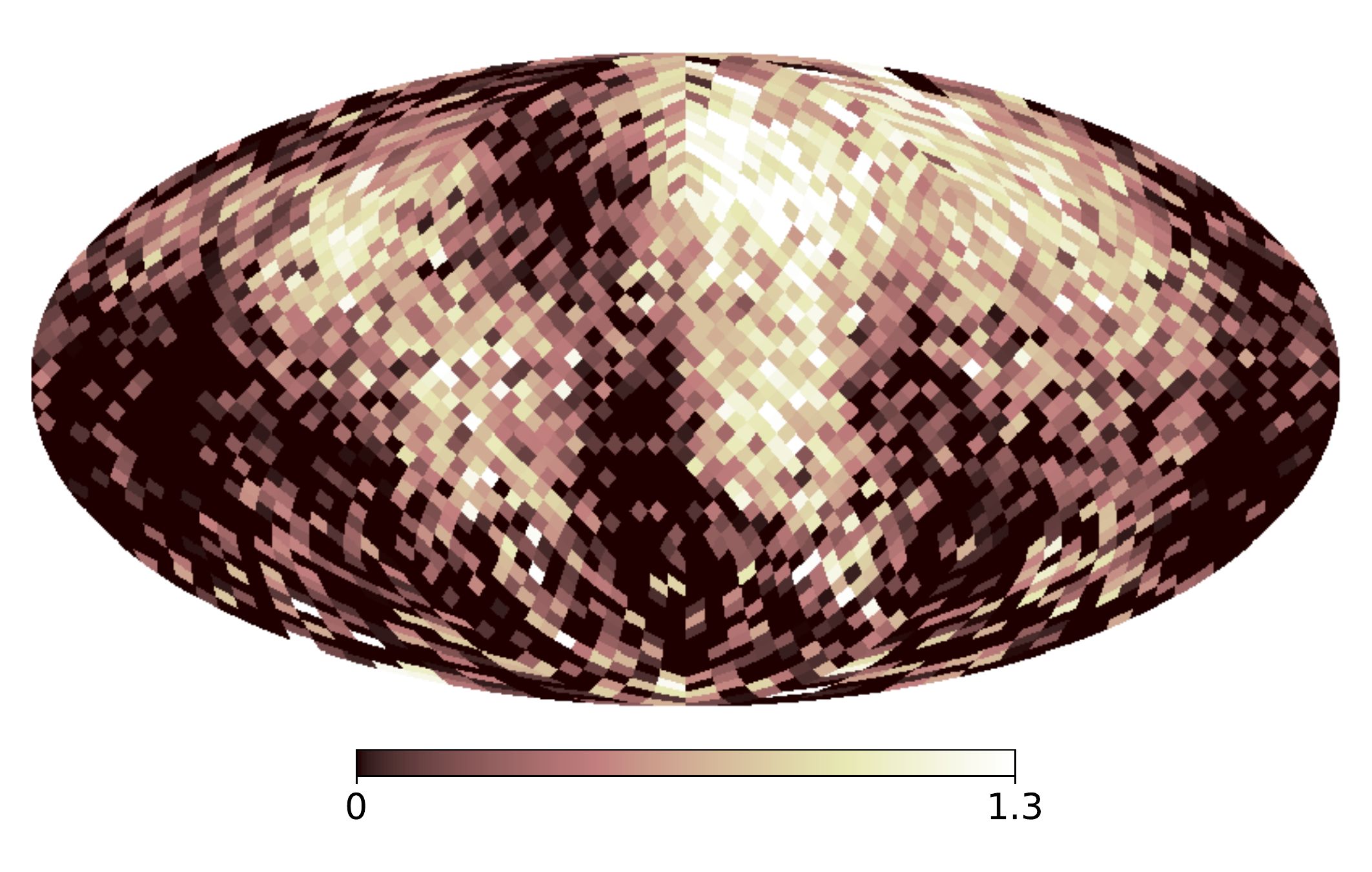}
      \includegraphics[width=0.32\linewidth]{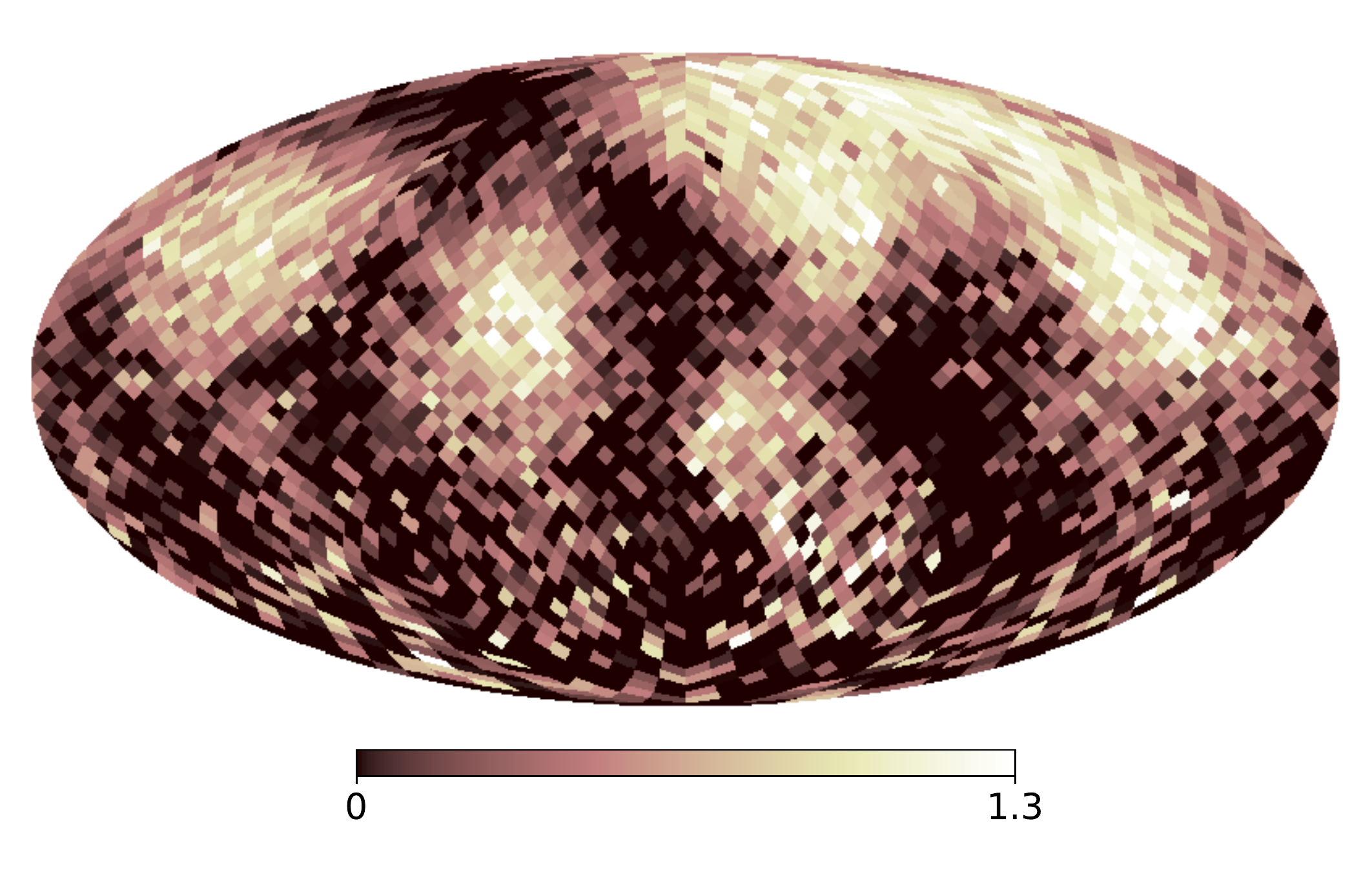}
        \includegraphics[width=0.32\linewidth]{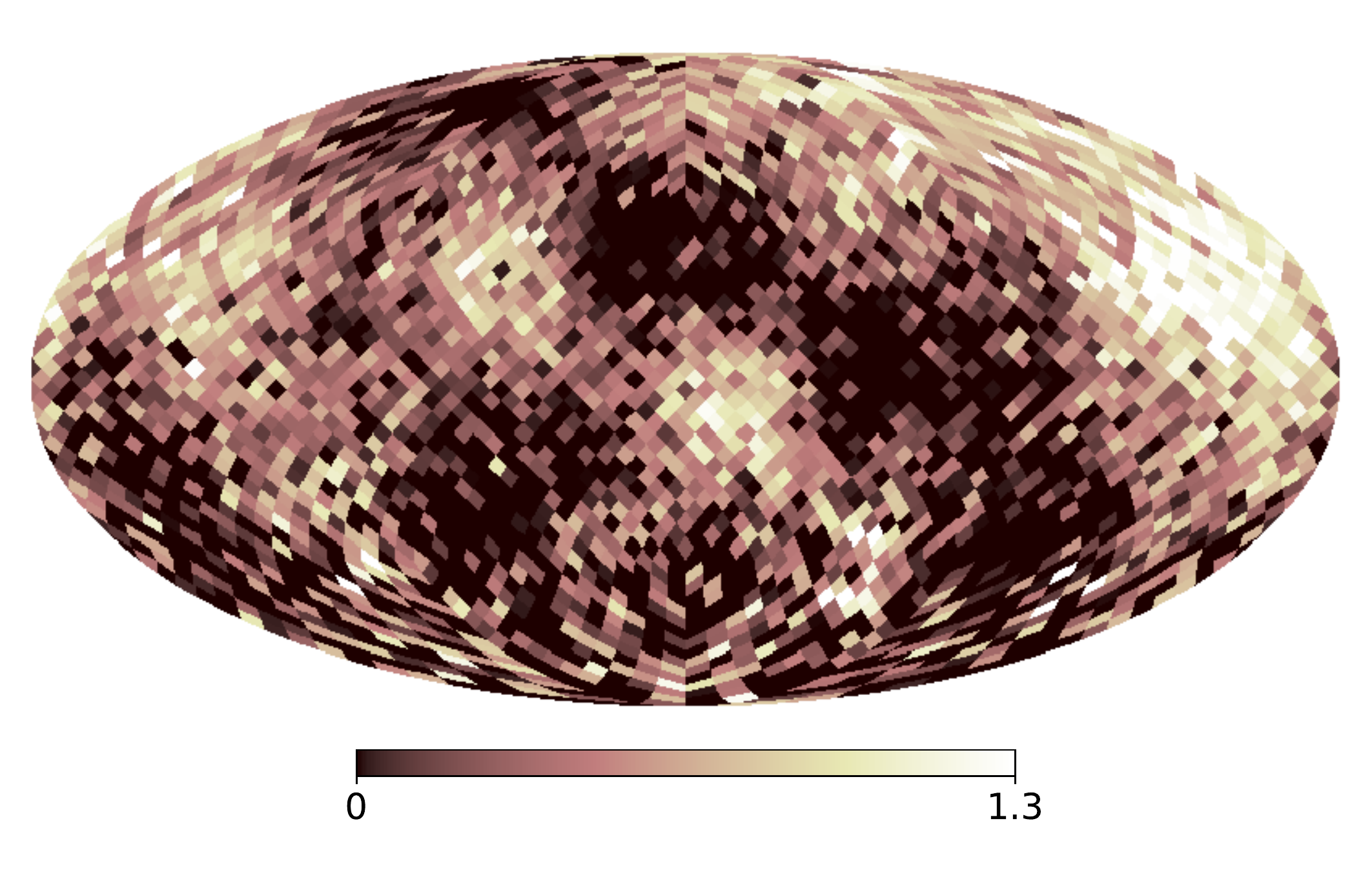}
 \end{center}
 \caption{The randomized maps of the toy model (see Figure \ref{fig:rand_dysot}) for three different times, $t=0, 182,$ and 364 days from left to right. \rev{We restrict the pixel value to the range of 0--1.3 for comparison with the input maps although some pixels are above or under the range.}
 \label{fig:rand_dysot}}
\end{figure*}

\rev{As another visulization of the geography uncertainty, we propose } ``randomized maps", as shown in Figure \ref{fig:rand_dysot} for the toy model corresponding to Figure \ref{fig:dysot}. The randomized map includes the information of the covariance between the pixels. The randomized maps here are thus generated by sampling from $p(\avi_i|\dv)$ {\it independently for each pixel}. In other words, the $j$-th pixel of the randomized map $\check{a}_i(\Omega_j)$ is sampled from its marginal posterior $p(\check{a}_i(\Omega_j)|\dv)$.\footnote{In practice, this random sampling is achieved as follows: (1) We generate $N_j$ sets of ($\thetaGPdag, \gvvdag$) using an MCMC with Equation (\ref{eq:sampleg_axialtiltagain}). (2) The corresponding $N_j$ maps are generated by sampling one map from the multivariate normal distribution given by Equation (\ref{eq:dysot_posterior_map}) for each set of ($\thetaGPdag, \gvvdag$), that is,  $(\avi_i)^\dagger_n  \sim p(\avi_i|\dv, \thetaGPdag, \gvvdag)$. (3) The value of the $n$-th pixel is chosen from the $n$-th map. This is equivalent to sampling each $\check{a}_i(\Omega_j)$ from $p(\check{a}_i(\Omega_j)|\dv) = \int d \gvv \int d \thetaGP \int d \avi_i^{\backslash j} \, p (\avi_i,\thetaGP,\gvv|\dv) $, where $\avi_i^{\backslash j} $ is $\avi_i$ minus the element of the $j$-th pixel.} 

In the maps generated in this way, structures appear more blurred than in the mean maps in the middle row when the inferred pixel value has a large uncertainty and/or the structures are maintained by a strong covariance between neighboring pixels. Therefore, the randomized maps are meant to visualize the robustness of the coherent structures shown in the retrieved map. For instance, the mean snapshots (middle row in Figure \ref{fig:rand_dysot}) exhibit coherent structures in the southern hemisphere, which become unclear or even disappear in the randomized maps (Figure \ref{fig:rand_dysot}). This indicates that the data are not sensitive to the southern hemisphere and the retrieved maps are less reliable in this region. By contrast, the structures in the northern hemisphere are clear in both the mean snapshots and the randomized maps because the majority of the realizations of the posterior has the northern structures. Again, this implies that these structures are robustly inferred as well as shown by the percentile maps. 

\begin{figure*}[]
 \begin{center}
   \includegraphics[width=0.32\linewidth]{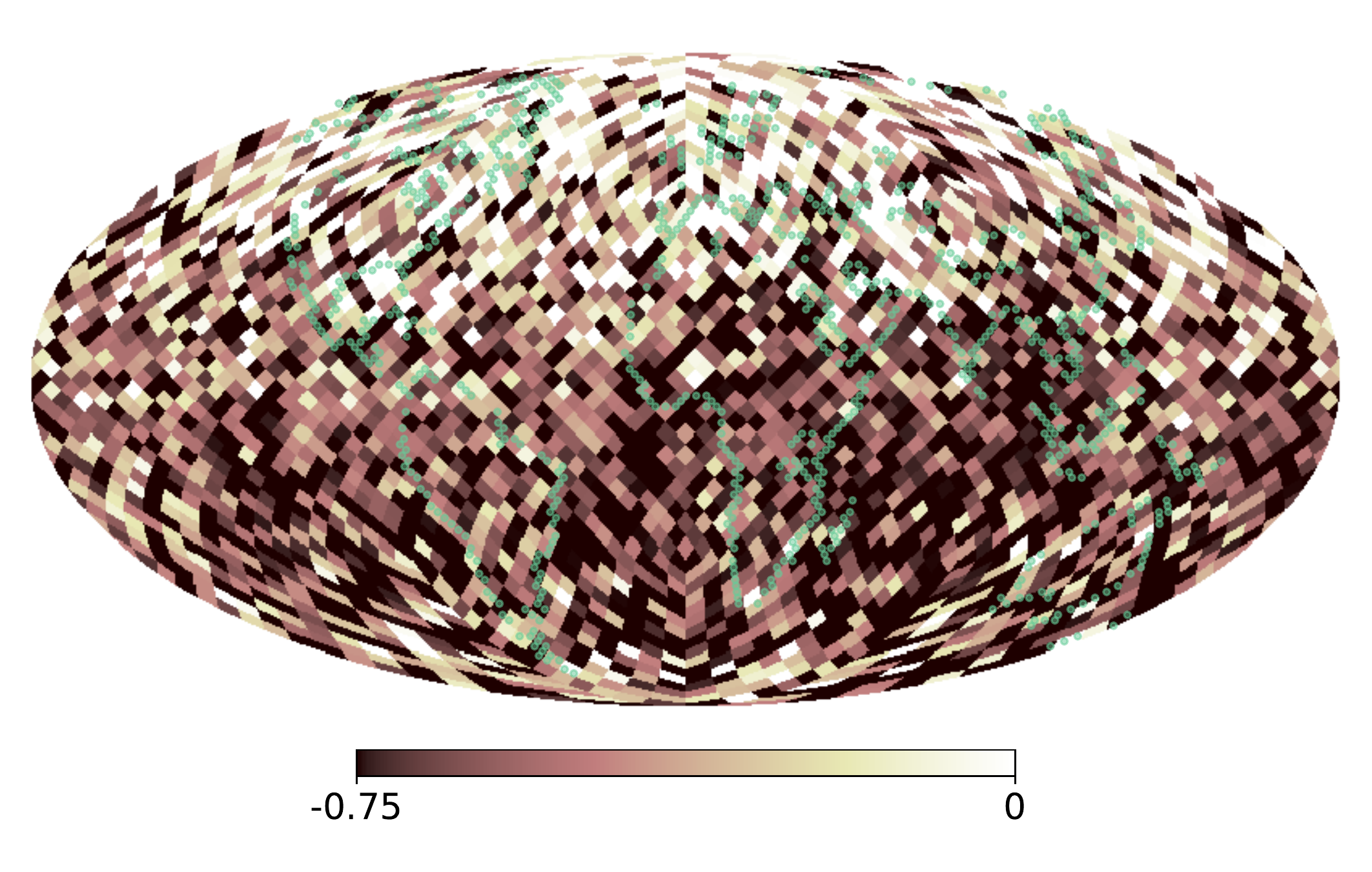}
   \includegraphics[width=0.32\linewidth]{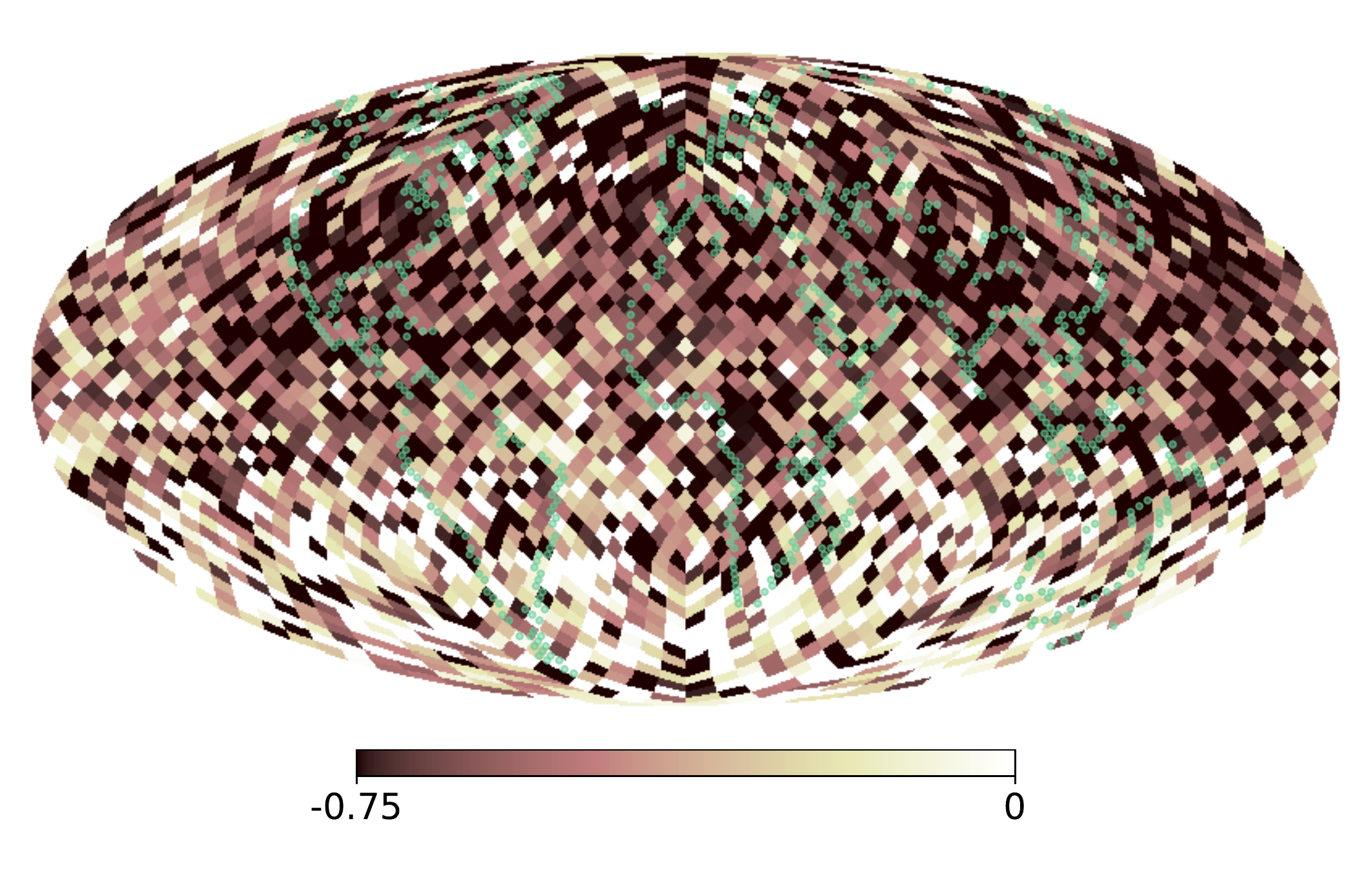}
   \includegraphics[width=0.32\linewidth]{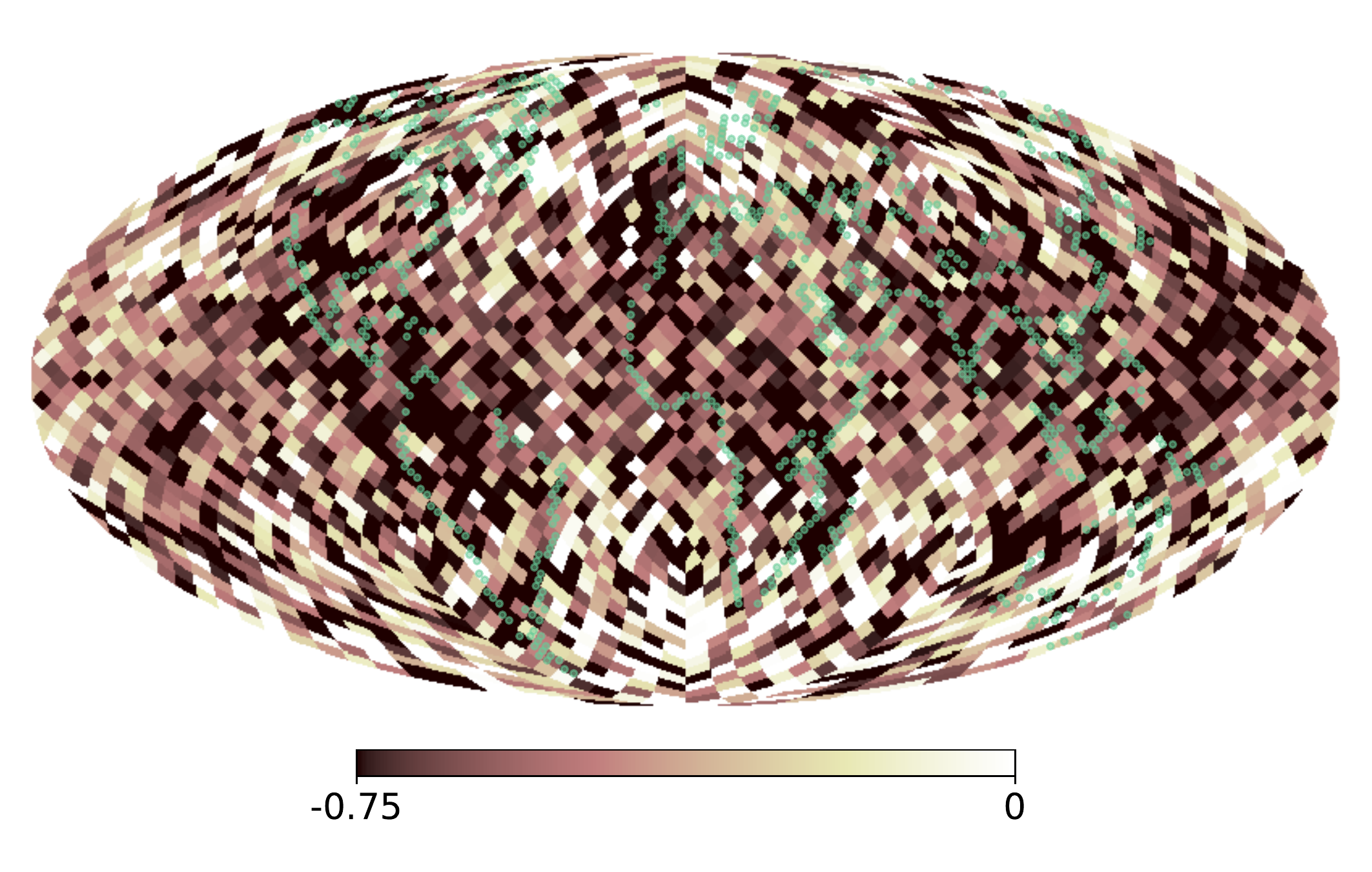}   
 \end{center}
 \caption{Randomized maps of the DSCOVR data (see Figure \ref{fig:ran_dysotdscovr}) for three different dates in January, May, and September in 2016 (from left to right).
 \label{fig:ran_dysotdscovr}}
 \end{figure*}

\rev{The advantage of the randomized map is that one can compile the uncertainty into a single map. The drawback is that the randomized map becomes non-intuitive when the map is less reliable. Figure \ref{fig:ran_dysotdscovr} displays the randomized maps for the DSCOVR data analyzed in Section 4. In this case, one find that the randomized maps are too noizy to recognize the structures. We suggest to use the percentile maps for such less reliable cases. }

\section{Comparison of the L2 and GP regularizations}\label{ss:testsot}

Assuming an indenpendent Gaussian as the spatial kernel as follows:
\begin{eqnarray}
\label{eq:l2l2}
 K_S = \alpha I \mbox{\,\, (L2)},
\end{eqnarray}
 we obtain the Bayesian interpretation of the Tikhonov (L2) regularization as the point estimate at a maximum a posteriori (MAP) in the case of an independent Gaussian prior of $\av$ \citep[Appendix C in][]{2012ApJ...755..101F,tarantola}. We then obtain the posterior as
\begin{eqnarray}
p(\av|\dv) &\propto& \Ng (\dv| W \av ,\sigma^2 I ) \Ng (\av| 0 , \sigma^2 K_S) \\
&\propto&  \exp{\left\{ - \frac{1}{2 \sigma^2} \left( ||\dv - W \av||^2_2 + \av^T \Pi_S \av \right)   \right\}},
\end{eqnarray}
where $|| \cdot ||^2_2$ is the L2 norm and $\Pi_S = K_S^{-1} = \alpha^{-1} I $ is the precision matrix\footnote{We note that we use a normalized form of the covariance by $\sigma^2$ because we assume an indenpedent Gaussian as the observational noise.}. The maximization of the posterior of $p(\av|\dv)$
is equivalent to the minimization of the cost function,
\begin{eqnarray}
\label{eq:costsot}
Q =||\dv - W \av||^2_2 + \alpha^{-1} ||\av||^2_2.
\end{eqnarray}
Adopting $\alpha = \lambda^{-2}$, Equation (\ref{eq:costsot}) is identical to that \cite{2011ApJ...739L..62K} proposed as the cost function of the original spin-orbit tomography.

As derived in Appendix \ref{apss:dervatami}, the MAP solution which minimizes Equation (\ref{eq:costsot}) is expressed as follows:
\begin{eqnarray}
\label{eq:normalsol}
\av^\mathrm{MAP}  &=& ( W^T W + \Pi_S )^{-1} W^T \dv.
\end{eqnarray}
We again stress that Equation (\ref{eq:normalsol}) is the mapping solution based on the point estimate. 

\begin{figure*}[]
 \begin{center}
    \includegraphics[width=0.35\linewidth]{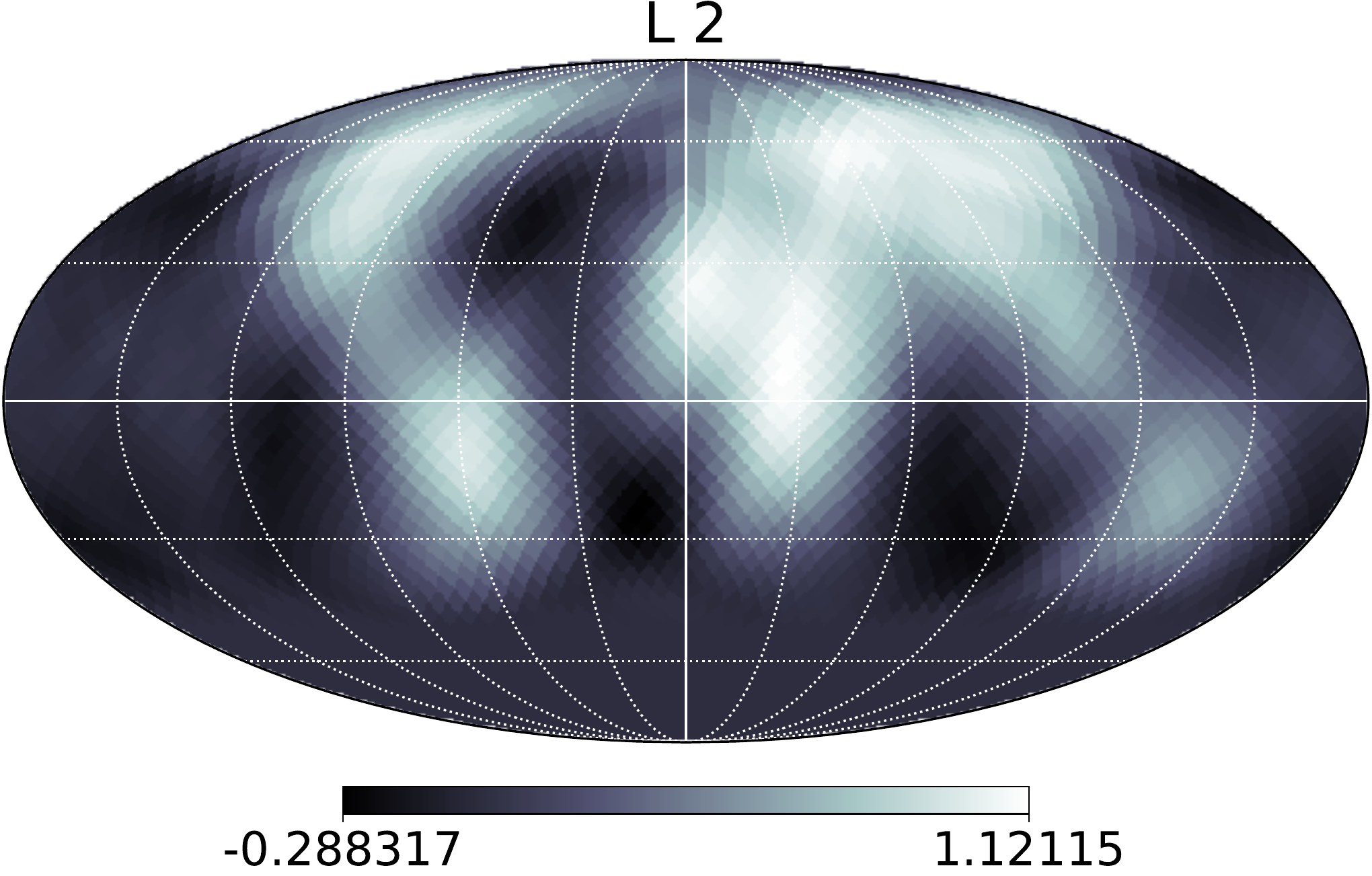}
    \includegraphics[width=0.35\linewidth]{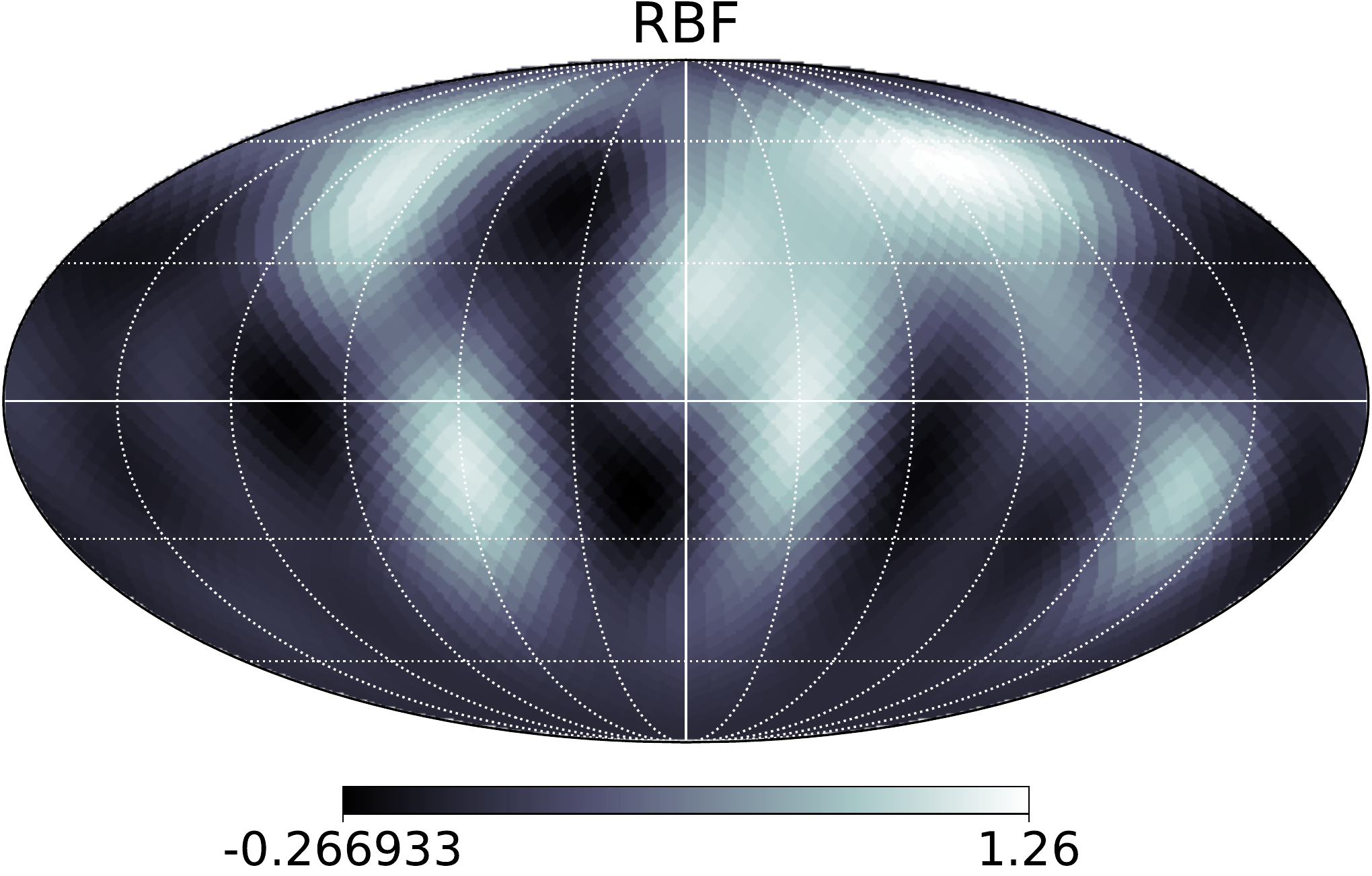}
    
        \includegraphics[width=0.35\linewidth]{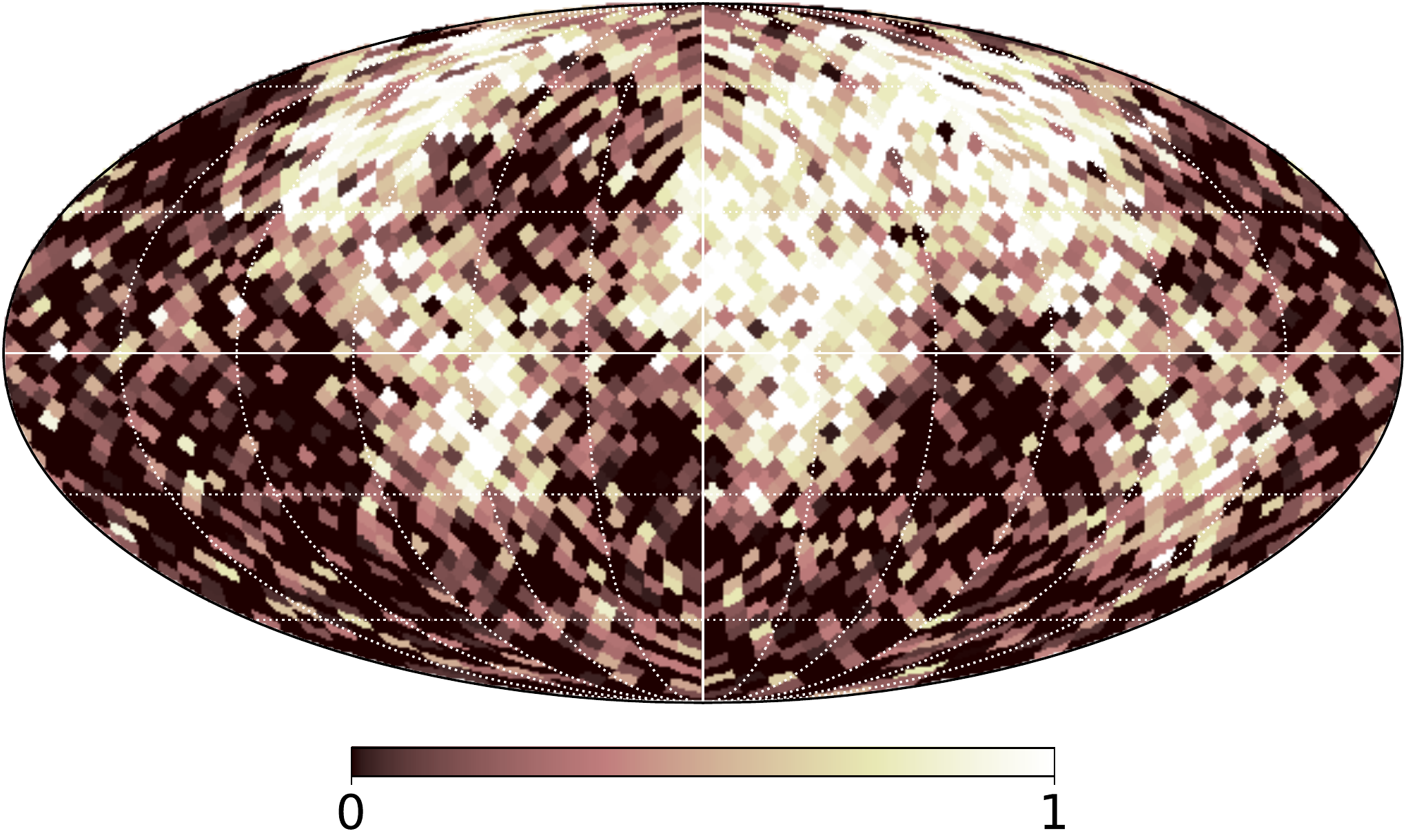}
    \includegraphics[width=0.35\linewidth]{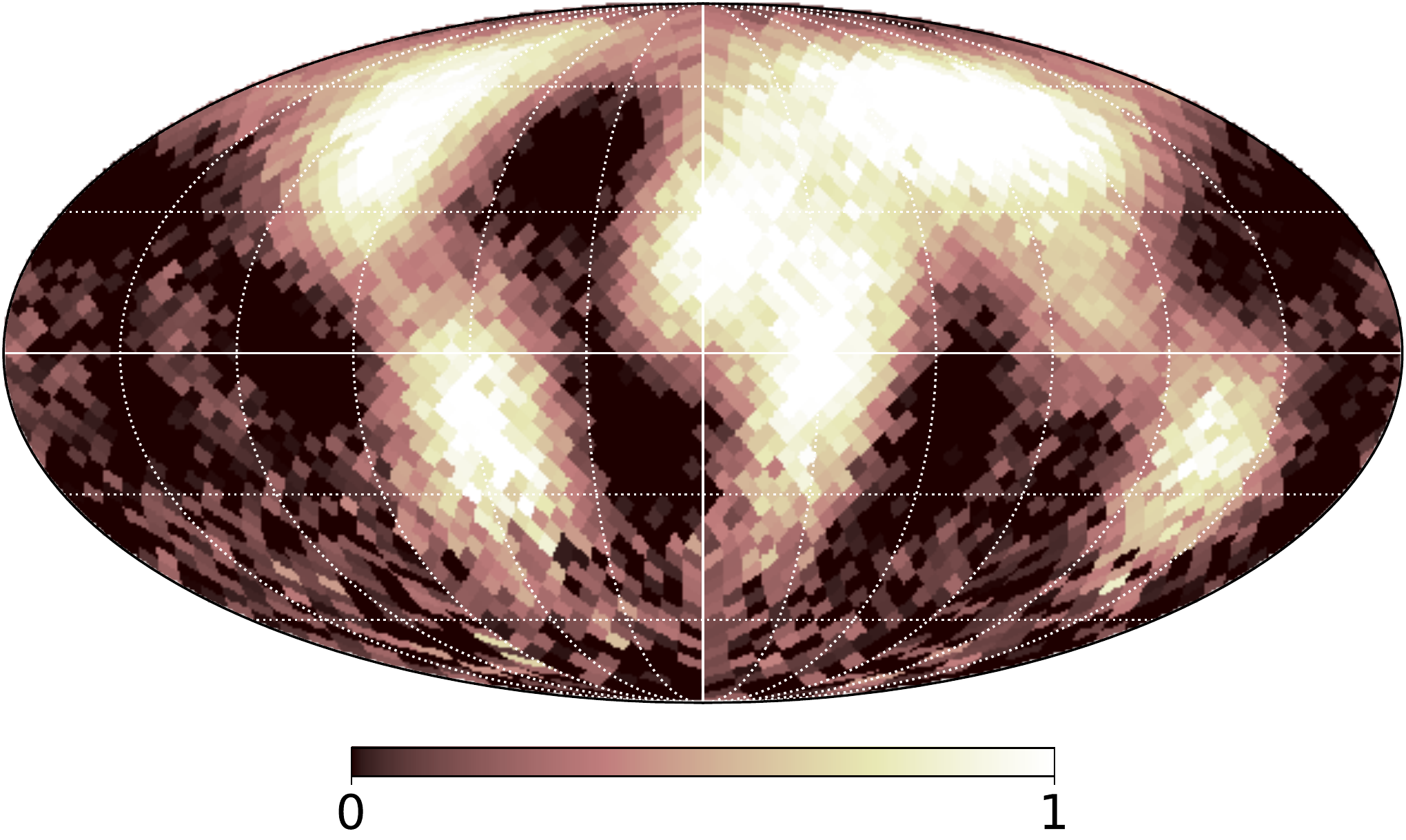}
 \end{center}
 \caption{Comparison of the mean (top) and randomized (bottom) maps of the L2 (left) regularization and RBF kernel (right), assuming a static geography. \label{fig:sot_compL2}}
\end{figure*}

Let us compare the L2 results with the model using the RBF kernel (GP modeling) as applied in the main body. The results of the point estimates are not siginificantly different from the GP modeling as shown in the map retrieved by the mean/MAP solution (top panels in Figure \ref{fig:sot_compL2}).  Recalling that the L2 case is an extreme situation of $\gamma \to 0$ in the GP kernel, we find that the spatial correlation scale $\gamma$ has little effect on the point estimate. However, this is not the case based on Bayesian statistics. The geography $\av$ can be sampled from the posterior distribution of Equation (\ref{eq:posteriorbayes}). As shown in the bottom panel, large difference can be seen in the randomized map. The L2 regularization provides an extremely noisy map, whereas the GP model provides smooth maps similar to the mean solution. The reason why the MAP solution in the L2 case exhibits a smooth structure is simply because it is averaged out. In contrast, the posterior distribution of the modeling by RBF provides a map similar to the mean map for each realization. The randomized maps show that the GP modeling generates better maps as realization of the posterior than those generated by the L2 case. In this sense, we prefer the GP modeling of the spatial correlation originally proposed by \cite{2018AJ....156..146F} over the L2 regularization when we adopt a Bayesian perspective. 

\bibliography{ref}{}

\begin{thebibliography}{}
\expandafter\ifx\csname natexlab\endcsname\relax\def\natexlab#1{#1}\fi
\providecommand{\url}[1]{\href{#1}{#1}}
\providecommand{\dodoi}[1]{doi:~\href{http://doi.org/#1}{\nolinkurl{#1}}}
\providecommand{\doeprint}[1]{\href{http://ascl.net/#1}{\nolinkurl{http://ascl.net/#1}}}
\providecommand{\doarXiv}[1]{\href{https://arxiv.org/abs/#1}{\nolinkurl{https://arxiv.org/abs/#1}}}

\bibitem[{{Aizawa} {et~al.}(2020){Aizawa}, {Kawahara}, \&
  {Fan}}]{2020arXiv200403941A}
{Aizawa}, M., {Kawahara}, H., \& {Fan}, S. 2020, arXiv e-prints,
  arXiv:2004.03941.
\newblock \doarXiv{2004.03941}

\bibitem[{{Berdyugina} \& {Kuhn}(2019)}]{2019AJ....158..246B}
{Berdyugina}, S.~V., \& {Kuhn}, J.~R. 2019, \aj, 158, 246,
  \dodoi{10.3847/1538-3881/ab2df3}

\bibitem[{Bishop(2006)}]{bishop2006pattern}
Bishop, C.~M. 2006, Pattern recognition and machine learning (springer)

\bibitem[{{Cowan} {et~al.}(2009){Cowan}, {Agol}, {Meadows}, {Robinson},
  {Livengood}, {Deming}, {Lisse}, {A'Hearn}, {Wellnitz}, {Seager},
  {Charbonneau}, \& {EPOXI Team}}]{2009ApJ...700..915C}
{Cowan}, N.~B., {Agol}, E., {Meadows}, V.~S., {et~al.} 2009, \apj, 700, 915,
  \dodoi{10.1088/0004-637X/700/2/915}

\bibitem[{{Fan} {et~al.}(2019){Fan}, {Li}, {Li}, {Bartlett}, {Jiang}, {Natraj},
  {Crisp}, \& {Yung}}]{2019ApJ...882L...1F}
{Fan}, S., {Li}, C., {Li}, J.-Z., {et~al.} 2019, \apjl, 882, L1,
  \dodoi{10.3847/2041-8213/ab3a49}

\bibitem[{{Farr} {et~al.}(2018){Farr}, {Farr}, {Cowan}, {Haggard}, \&
  {Robinson}}]{2018AJ....156..146F}
{Farr}, B., {Farr}, W.~M., {Cowan}, N.~B., {Haggard}, H.~M., \& {Robinson}, T.
  2018, \aj, 156, 146, \dodoi{10.3847/1538-3881/aad775}

\bibitem[{{Ford} {et~al.}(2001){Ford}, {Seager}, \&
  {Turner}}]{2001Natur.412..885F}
{Ford}, E.~B., {Seager}, S., \& {Turner}, E.~L. 2001, \nat, 412, 885,
  \dodoi{10.1038/35091009}

\bibitem[{Foreman-Mackey(2016)}]{corner}
Foreman-Mackey, D. 2016, The Journal of Open Source Software, 24,
  \dodoi{10.21105/joss.00024}

\bibitem[{{Foreman-Mackey} {et~al.}(2013){Foreman-Mackey}, {Hogg}, {Lang}, \&
  {Goodman}}]{2013PASP..125..306F}
{Foreman-Mackey}, D., {Hogg}, D.~W., {Lang}, D., \& {Goodman}, J. 2013, \pasp,
  125, 306, \dodoi{10.1086/670067}

\bibitem[{{Fujii} \& {Kawahara}(2012)}]{2012ApJ...755..101F}
{Fujii}, Y., \& {Kawahara}, H. 2012, \apj, 755, 101,
  \dodoi{10.1088/0004-637X/755/2/101}

\bibitem[{{Fujii} {et~al.}(2010){Fujii}, {Kawahara}, {Suto}, {Taruya},
  {Fukuda}, {Nakajima}, \& {Turner}}]{2010ApJ...715..866F}
{Fujii}, Y., {Kawahara}, H., {Suto}, Y., {et~al.} 2010, \apj, 715, 866,
  \dodoi{10.1088/0004-637X/715/2/866}

\bibitem[{{G{\'o}rski} {et~al.}(2005){G{\'o}rski}, {Hivon}, {Banday},
  {Wandelt}, {Hansen}, {Reinecke}, \& {Bartelmann}}]{2005ApJ...622..759G}
{G{\'o}rski}, K.~M., {Hivon}, E., {Banday}, A.~J., {et~al.} 2005, \apj, 622,
  759, \dodoi{10.1086/427976}

\bibitem[{Hunter(2007)}]{Hunter:2007}
Hunter, J.~D. 2007, Computing In Science \& Engineering, 9, 90,
  \dodoi{10.1109/MCSE.2007.55}

\bibitem[{{Jiang} {et~al.}(2018){Jiang}, {Zhai}, {Herman}, {Zhai}, {Hu}, {Su},
  {Natraj}, {Li}, {Xu}, \& {Yung}}]{2018AJ....156...26J}
{Jiang}, J.~H., {Zhai}, A.~J., {Herman}, J., {et~al.} 2018, \aj, 156, 26,
  \dodoi{10.3847/1538-3881/aac6e2}

\bibitem[{Jones {et~al.}(2001)Jones, Oliphant, Peterson, {et~al.}}]{scipy}
Jones, E., Oliphant, T., Peterson, P., {et~al.} 2001, {SciPy}: Open source
  scientific tools for {Python}.
\newblock \url{http://www.scipy.org/}

\bibitem[{Kailath(1980)}]{kailath1980linear}
Kailath, T. 1980, Linear systems, Vol. 156 (Prentice-Hall Englewood Cliffs, NJ)

\bibitem[{{Kawahara}(2016)}]{2016ApJ...822..112K}
{Kawahara}, H. 2016, \apj, 822, 112, \dodoi{10.3847/0004-637X/822/2/112}

\bibitem[{{Kawahara}(2020)}]{Kawahara2020}
---. 2020, \apj, 894, 58, \dodoi{10.3847/1538-4357/ab87a1}

\bibitem[{{Kawahara} \& {Fujii}(2010)}]{2010ApJ...720.1333K}
{Kawahara}, H., \& {Fujii}, Y. 2010, \apj, 720, 1333,
  \dodoi{10.1088/0004-637X/720/2/1333}

\bibitem[{{Kawahara} \& {Fujii}(2011)}]{2011ApJ...739L..62K}
---. 2011, \apjl, 739, L62, \dodoi{10.1088/2041-8205/739/2/L62}

\bibitem[{{Luger} {et~al.}(2019){Luger}, {Bedell}, {Vanderspek}, \&
  {Burke}}]{2019arXiv190312182L}
{Luger}, R., {Bedell}, M., {Vanderspek}, R., \& {Burke}, C.~J. 2019, arXiv
  e-prints, arXiv:1903.12182.
\newblock \doarXiv{1903.12182}

\bibitem[{{Nakagawa} {et~al.}(2020){Nakagawa}, {Kodama}, {Ishiwatari},
  {Kawahara}, {Suto}, {Takahashi}, {Hashimoto}, {Kuramoto}, {Nakajima},
  {Takehiro}, \& {Hayashi}}]{2020arXiv200611437N}
{Nakagawa}, Y., {Kodama}, T., {Ishiwatari}, M., {et~al.} 2020, arXiv e-prints,
  arXiv:2006.11437.
\newblock \doarXiv{2006.11437}

\bibitem[{{Oakley} \& {Cash}(2009)}]{2009ApJ...700.1428O}
{Oakley}, P.~H.~H., \& {Cash}, W. 2009, \apj, 700, 1428,
  \dodoi{10.1088/0004-637X/700/2/1428}

\bibitem[{Pedregosa {et~al.}(2011)Pedregosa, Varoquaux, Gramfort, Michel,
  Thirion, Grisel, Blondel, Prettenhofer, Weiss, Dubourg, Vanderplas, Passos,
  Cournapeau, Brucher, Perrot, \& Duchesnay}]{scikit-learn}
Pedregosa, F., Varoquaux, G., Gramfort, A., {et~al.} 2011, Journal of Machine
  Learning Research, 12, 2825

\bibitem[{Platnick {et~al.}(2003)Platnick, King, Ackerman, Menzel, Baum,
  Ri{\'e}di, \& Frey}]{platnick2003modis}
Platnick, S., King, M.~D., Ackerman, S.~A., {et~al.} 2003, IEEE Transactions on
  Geoscience and Remote Sensing, 41, 459

\bibitem[{Rasmussen(2003)}]{rasmussen2003gaussian}
Rasmussen, C.~E. 2003, in Summer School on Machine Learning, Springer, 63--71

\bibitem[{{Schwartz} {et~al.}(2016){Schwartz}, {Sekowski}, {Haggard},
  {Pall{\'e}}, \& {Cowan}}]{2016MNRAS.457..926S}
{Schwartz}, J.~C., {Sekowski}, C., {Haggard}, H.~M., {Pall{\'e}}, E., \&
  {Cowan}, N.~B. 2016, \mnras, 457, 926, \dodoi{10.1093/mnras/stw068}

\bibitem[{{Tarantola}(2005)}]{tarantola}
{Tarantola}, A. 2005, {Inverse Problem Theory and Methods for Model Parameter
  Estimation} (the Society for Industrial and Applied Mathematics)

\bibitem[{{van der Walt} {et~al.}(2011){van der Walt}, {Colbert}, \&
  {Varoquaux}}]{2011CSE....13b..22V}
{van der Walt}, S., {Colbert}, S.~C., \& {Varoquaux}, G. 2011, Computing in
  Science and Engineering, 13, 22, \dodoi{10.1109/MCSE.2011.37}

\end{thebibliography}
\bibliographystyle{aasjournal}
\end{document}